\documentclass[aps,prd,preprintnumbers,groupedaddress,nofootinbib,amssymb,notitlepage,eqsecnum]{revtex4-2}
\usepackage{here}
\usepackage{comment}
\usepackage[dvipdfmx]{graphicx}
\usepackage{amsmath,amsthm,amssymb}
\usepackage{bm}
\usepackage{color}
\usepackage{comment}
\usepackage{simplewick}
\usepackage{physics}
\usepackage{ulem}

\usepackage{amsfonts}
\usepackage{dcolumn}
\usepackage{hyperref}
\allowdisplaybreaks[1]
\usepackage{stackengine}

\newcommand{\be}{\begin{equation}}  
\newcommand{\ee}{\end{equation}}
\newcommand{\ba}{\begin{eqnarray}}
\newcommand{\ea}{\end{eqnarray}}

\newcommand{\rd}{{\rm d}}

\newcommand{\bem}{\begin{bmatrix}}
\newcommand{\eem}{\end{bmatrix}}
\newcommand{\Mpl}{M_{\rm Pl}}

\newcommand{\ri}{{\rm i}}

\allowdisplaybreaks

\begin{document}

\newcommand{\newc}{\newcommand}
\newc{\rH}{{\rm H}}
\newcommand{\rBH}{r_{s}}
\newcommand{\rc}{r_{c}}
\newcommand{\rh}{r_{h}}
\newcommand{\Xh}{X_{h}}
\makeatletter
\newsavebox{\@brx}
\newcommand{\llangle}[1][]{\savebox{\@brx}{\(\m@th{#1\langle}\)}%
  \mathopen{\copy\@brx\mkern2mu\kern-0.9\wd\@brx\usebox{\@brx}}}
\newcommand{\rrangle}[1][]{\savebox{\@brx}{\(\m@th{#1\rangle}\)}%
  \mathclose{\copy\@brx\mkern2mu\kern-0.9\wd\@brx\usebox{\@brx}}}
\makeatother

\preprint{WUCG-24-07}

\title{Proving the absence of large one-loop corrections to the power spectrum of curvature perturbations in transient ultra-slow-roll inflation within the path-integral approach}

\author{Ryodai Kawaguchi$^1$\footnote{{\tt ryodai0602@fuji.waseda.jp}}}
\author{Shinji Tsujikawa$^1$\footnote{{\tt tsujikawa@waseda.jp}}}
\author{Yusuke Yamada$^2$\footnote{{\tt y-yamada@aoni.waseda.jp}}}

\affiliation{${}^1$
Department of Physics, Waseda University, 3-4-1 Okubo, Shinjuku, Tokyo 169-8555, Japan}

\affiliation
{${}^2$
Waseda Institute for Advanced Study, Waseda University, 1-21-1 Nishi Waseda, Shinjuku, Tokyo 169-0051, Japan}

\begin{abstract}
We revisit one-loop corrections to the power spectrum of 
curvature perturbations $\zeta$ in an inflationary scenario containing a transient ultra-slow-roll (USR) period.
In Ref.~\cite{Kristiano:2022maq}, it was argued 
that one-loop corrections to the power spectrum 
of $\zeta$ can be larger than the tree-level one within the parameter region generating the seeds of primordial 
black holes during the USR epoch, which implies 
the breakdown of perturbation theory.
We prove that this is not the case by using   
a master formula for one-loop corrections to 
the power spectrum obtained in Ref.~\cite{Pimentel:2012tw}. 
We derive the same formula within the 
path-integral formalism, which is simpler than the original derivation in \cite{Pimentel:2012tw}. 
To show the smallness of one-loop corrections, 
the consistency relations and the effective constancy 
of tree-level mode functions of $\zeta$ for super-Hubble 
modes play essential roles, with which the master formula 
gives a simple expression for one-loop corrections.
For concreteness, we provide a reduced set 
of interactions including the leading-order one, 
while establishing the consistency relations in a 
self-consistent manner.
We also show how the consistency relations of 
various operators hold explicitly, which plays 
a key role in proving the absence of 
large one-loop corrections.
\end{abstract}

\date{\today}


\maketitle
\tableofcontents
\section{Introduction}

The inflationary paradigm was originally proposed 
to address several drawbacks 
in the big-bang cosmology--such as 
the flatness and horizon problems \cite{Starobinsky:1980te,Sato:1980yn,Kazanas:1980tx,Guth:1980zm,Linde:1981mu,Albrecht:1982wi}. 
Moreover, in this paradigm, the origins of Cosmic Microwave Background (CMB) temperature anisotropies 
and large-scale structures of the Universe 
can be explained by stretching quantum vacuum fluctuations 
over the super-Hubble scales during inflation \cite{Hawking:1982cz,Starobinsky:1982ee,Guth:1982ec,Mukhanov:1981xt,Bardeen:1983qw}.
The standard inflationary scenario is based on 
a single canonical scalar field $\phi$ dubbed 
``inflaton'' with a slowly varying potential $V(\phi)$, 
which is given by the action
\be
{\cal S}=\int \rd^4 x \sqrt{-g} \left[ 
\frac{\Mpl^2}{2}R-\frac{1}{2} 
\partial_{\mu} \phi \partial^{\mu} \phi
-V(\phi) \right]\,,
\label{action0}
\ee
where $g$ is a determinant of the metric tensor 
$g_{\mu \nu}$, $\Mpl$ is the reduced Planck mass, 
$R$ is the Ricci scalar, and 
$\partial_{\mu} \phi=\partial \phi/\partial x^{\mu}$ 
with a four-dimensional coordinate $x^{\mu}$. 
To address the evolution of cosmological perturbations during inflation, the commonly 
used line element is of the following 
Arnowitt-Deser-Misner (ADM) form \cite{Arnowitt:1962hi}
\be
\rd s^2=-N^2 \rd t^2+h_{ij}
\left( \rd x^i+N^i \rd t \right)
\left( \rd x^j+N^j \rd t \right)\,,
\label{ADM}
\ee
where $N$ is the lapse, $N^i$ is the shift, and 
\be
h_{ij}=a^2(t)e^{2\zeta (t,x^i)} \delta_{ij}\,.
\ee
Here, $a$ is the scale factor that is a function 
of the cosmic time $t$, and $\zeta$ is 
the curvature perturbation depending on 
both $t$ and the spatial coordinate $x^i$. 
A spatially-flat 
Friedmann-Lema\^{i}tre-Robertson-Walker (FLRW)
background corresponds to $N=1$, $N^i=0$,  
and $\zeta=0$. During slow-roll (SR) inflation 
driven by the inflaton potential $V(\phi)$, 
the scale factor exhibits quasi-exponential growth 
and hence the Hubble expansion rate 
$H \equiv \dot{a}/a$ is nearly constant 
(where a dot represents the derivative 
with respect to $t$).
To address the flatness and horizon problems, 
we require that the number of e-foldings 
$N=\ln (a_f/a_i)$, where $a_i$ and $a_f$ 
are the scale factors at the beginning 
and the end of inflation, is at least larger than 65. 

Besides the metric perturbations $\alpha=N-1$, 
$N_i=\partial_i \psi$, and $\zeta$, the inflaton field 
has a perturbation $\delta \phi=\phi-\phi_0$, 
where $\phi_0$ is the background value.
By choosing the unitary gauge 
\be
\delta \phi=0\,,
\label{unitary}
\ee
and using the Hamiltonian and momentum constraints 
to eliminate $\alpha$ and $\psi$, 
the linearized curvature perturbation 
$\zeta_k$ in Fourier space 
obeys \cite{Bardeen:1980kt,Kodama:1984ziu,Mukhanov:1990me,Bassett:2005xm}
\be
\left( \epsilon a^2 \zeta_k' \right)'
+\epsilon a^2 k^2 \zeta_k=0\,,
\label{zetaeq}
\ee
where 
\be
\epsilon \equiv -\frac{\dot{H}}{H^2}
\ee
is the first SR parameter, $k=|{\bm k}|$ is an absolute 
value of the comoving wavenumber ${\bm k}$, and
a prime represents the derivative with 
respect to the conformal time 
$\tau=\int a^{-1}\rd t$.
After the Hubble radius crossing ($k<aH$), 
the second term on the left-hand side of 
Eq.~(\ref{zetaeq}) is negligible 
and hence the resulting solution is given by 
\be
\zeta_k=c_1+c_2 \int \frac{\rd \tau}
{\epsilon a^2}\,.
\label{zetaso}
\ee
where $c_1$ and $c_2$ are constants. 
Since $\epsilon$ is nearly constant in standard 
slow-roll inflation, the second term on the 
right-hand side of Eq.~(\ref{zetaso}) decays 
quickly. Therefore, $\zeta_k$ approaches a 
constant value $c_1$ after the Hubble 
radius crossing. During SR inflation, 
the Hubble expansion rate slowly varies over time 
and hence the power spectrum of curvature perturbations 
slightly deviates from the scale-invariant 
Harrison-Zeldovich spectrum.

The CMB temperature anisotropies correspond to the wavelength of perturbations that crossed the Hubble radius about 
60 e-foldings before the end of inflation. 
The primordial power spectrum of curvature perturbations 
${\cal P}_{\zeta}$ computed in the framework of 
single-field SR inflation is consistent 
with the CMB temperature anisotropies 
observed by the WMAP \cite{WMAP:2003elm} 
and Planck \cite{Planck:2013pxb} satellites. 
The amplitude of the power spectrum of curvature perturbations is constrained to be of the order
${\cal P}_{\zeta}=2\times 10^{-9}$, 
with the scalar spectral index 
$n_s=0.965 \pm 0.004$ at the pivot wavenumber 
$k_0=0.002$~Mpc$^{-1}$ (68\,\% CL) \cite{Planck:2018vyg,Planck:2018jri}. 

Since the perturbations with shorter 
wavelengths (wavenumber $k \gg k_0$) have been less 
constrained from observations, there is a possibility for breaking a nearly scale-invariant 
power spectrum. If such small-scale modes are 
enhanced in some non-standard inflationary scenarios, 
they can be responsible for the seeds of 
primordial black holes (PBHs) \cite{zel1967hypothesis,Hawking:1971ei,Carr:1974nx,Carr:1975qj}. 
In the presence of high-density regions that exceed a certain threshold, PBHs may form in the radiation-dominated epoch 
after inflation (see \cite{Sasaki:2018dmp,Carr:2020xqk,Green:2020jor,Carr:2020gox,Villanueva-Domingo:2021spv,
Carr:2021bzv,Escriva:2022duf,Karam:2022nym,Ozsoy:2023ryl,Choudhury:2024aji} for recent reviews).
Unlike astrophysical black holes, PBHs can have a variety of masses depending on the epoch when 
curvature perturbations are subject to growth during inflation. PBHs may play important roles as the source for dark matter \cite{Chapline:1975ojl,Meszaros:1975ef} 
and as the possible origins of binary black holes detected by gravitational waves  \cite{Bird:2016dcv,Sasaki:2016jop,Clesse:2016vqa,Wang:2016ana} and supermassive black holes at the center of galaxies \cite{Bean:2002kx}.

So far, a large number of inflationary models 
realizing the enhancement of small-scale curvature perturbations have been proposed to generate the seeds of PBHs. 
One of the most well-known scenarios is 
a transient ultra-slow-roll (USR) model, 
in which the inflaton potential has a very 
flat region (USR region) in which 
$V_{,\phi} \equiv \rd V/\rd \phi$ almost vanishes \cite{Garcia-Bellido:2017mdw,Kannike:2017bxn,Germani:2017bcs,Ezquiaga:2017fvi,Motohashi:2017kbs}. 
On the spatially-flat FLRW background, the 
gravitational and scalar-field 
equations of motion are given by 
\be
3\Mpl^2H^2=\frac{1}{2}\dot{\phi}^2+V(\phi)\,,
\qquad 
2\Mpl^2 \dot{H}=-\dot{\phi}^2\,,
\qquad 
\ddot{\phi}+3H \dot{\phi}+V_{,\phi}=0\,.
\label{backeq}
\ee
Here and in the following, we omit the subscript ``0'' 
for the background inflaton field.
Since $V_{,\phi} \simeq 0$ in the USR regime, 
we have $\ddot{\phi}+3H \dot{\phi} \simeq 0$ 
and hence the inflaton velocity decreases as 
$\dot{\phi} \propto a^{-3}$. 
This rapid decrease of $\dot{\phi}$ toward 0 
leads to the almost de-Sitter background with 
$\epsilon \propto \dot{\phi}^2 \propto a^{-6}$.
In this regime, the second term on 
the left-hand side of Eq.~(\ref{zetaso}) increases after 
the Hubble radius crossing, which results in 
the enhancement of curvature perturbations 
$\zeta_k$.
So long as the small-scale power spectrum 
${\cal P}_{\zeta}$ is amplified by more than $10^7$ 
times, it is possible to produce a sufficient 
amount of PBHs responsible for dark matter.

On the other hand, it was pointed out by 
Kristiano and Yokoyama \cite{Kristiano:2022maq,Kristiano:2023scm} (see also Refs.~\cite{Cheng:2021lif,Inomata:2022yte}) that the strong amplification of small-scale curvature perturbations can modify the power spectrum on scales relevant to CMB. 
Since one-loop corrections to the large-scale 
power spectrum may exceed the tree-level contribution, 
this spoils the perturbative 
control of underlying theory~\cite{Kristiano:2022maq}. 
Follow-up papers \cite{Riotto:2023hoz,Choudhury:2023vuj,Choudhury:2023jlt,Riotto:2023gpm,Choudhury:2023rks,Firouzjahi:2023aum,Motohashi:2023syh,Choudhury:2023hvf,Firouzjahi:2023ahg,Franciolini:2023agm,Tasinato:2023ukp,Cheng:2023ikq,Fumagalli:2023hpa,Maity:2023qzw,Tada:2023rgp,Jackson:2023obv,Firouzjahi:2023bkt,Davies:2023hhn,Iacconi:2023ggt,Firouzjahi:2024psd,Inomata:2024lud,Choudhury:2024ybk,Ballesteros:2024zdp,Kristiano:2024vst,Kristiano:2024ngc} claim various arguments, which may support the original one or may not, or improve some technical points. However, the issue of one-loop corrections within the transient USR model is unsettled and yet controversial.

\subsection{What is a problem?}

Before going to the quantitative discussion, it would be worthwhile to illustrate the heart of the controversy. In the work of Kristiano and Yokoyama~\cite{Kristiano:2022maq}, it was argued that super-Hubble perturbations may be affected by a late-time event through quantum corrections by short modes that experience 
the late-time event inside the Hubble radius, 
even if the super-Hubble modes exit the Hubble radius 
well before the late-time event takes place. 
The late-time event we refer to is the short-time USR phase 
with SR-USR-SR phase transitions. The USR stage can 
cause the strong enhancement of curvature perturbations 
for wavelengths smaller than the CMB scale, 
with the generation of CMB-scale perturbations during 
the first SR phase.

It seems possible that large one-loop corrections are generated up to super-Hubble modes simply because the couplings of curvature perturbations may be controlled by the second SR parameter 
\be
\eta \equiv \frac{\dot{\epsilon}}{H \epsilon}\,.
\ee
While $\epsilon=-\dot{H}/H^2$ is negligibly small, 
$\eta$ reaches the value $-6$ during the USR epoch. 
Furthermore, there is a cubic interaction proportional to $\dot{\eta}$, 
which may become very large if 
the phase transition occurs suddenly.

On the other hand, corrections to super-Hubble perturbations (without volume suppression) seem rather peculiar 
in the following sense. 
Suppose the first slow-roll phase continues for 100 e-foldings 
and consider the mode that exits the 
Hubble radius almost at the beginning of the 
first SR inflation. In such a case, at the USR phase transition, the momentum of the corresponding mode is roughly 
$k\sim e^{-100}H_i$, where $H_i$ is the Hubble 
expansion rate when the mode exits the horizon. 
As the momentum is quite small, such a perturbation 
cannot be distinguished from the zero mode 
$\zeta_0$ of $\zeta$. 
However, the zero mode is eliminated by a coordinate transformation, since 
$a^2(t)e^{2\zeta(t,x^i)}$ in Eq.~(\ref{ADM}) 
is invariant under the changes 
\be
\zeta(t,x^i) \to \zeta(t,x^i)+\delta\zeta\,,\quad
{\rm and} \quad
a(t)\to a(t)e^{-\delta\zeta}\,.
\ee
Therefore, the constant zero mode of $\zeta$ is unphysical and should decouple from the shorter wavelength perturbations. 
Now, a mode having $k\sim e^{-100}H_i$ should be approximately a zero mode and would decouple from all the dynamics, which is the physical reason why we expect the conservation of super-Hubble 
curvature perturbations. If this is the case, there would be no large one-loop corrections to sufficiently long-wavelength modes whatever late-time events happen (as long as the zero mode is constant at leading order).

The above physical reasoning is based on the symmetry 
of spacetime mentioned above, and therefore, 
it seems that symmetry is a key to proving the constancy 
of super-Hubble curvature perturbation. 
In quantum field theory, 
consistency relations first shown by Maldacena~\cite{Maldacena:2002vr} manifest the intuitive property of $\zeta$. They follow from the hidden conformal 
symmetry of the FLRW background~\cite{Creminelli:2012ed,Hinterbichler:2012nm,Assassi:2012zq,Hinterbichler:2013dpa,Kundu:2015xta}. 
In literature, consistency relations or corresponding conformal symmetry play crucial roles in proving the constancy of $\zeta$ on super-Hubble scales \cite{Pimentel:2012tw,Assassi:2012et}. 
In the context of the transient USR model, the authors of Ref.~\cite{Tada:2023rgp} used the same argument to show that the large one-loop corrections to super-Hubble curvature perturbations are absent by improving the work of Fumagalli~\cite{Fumagalli:2023hpa}. 
However, due to the incorrect argument regarding the regularization of ultra-violet (UV) divergences, the results of 
Ref.~\cite{Tada:2023rgp} were criticized 
in Ref.~\cite{Firouzjahi:2023bkt}.\footnote{We will comment on the regularization issue later.} 

In the proof of the $\zeta$-constancy for 
super-Hubble perturbations given 
in Ref.~\cite{Pimentel:2012tw}, detailed information is not necessary and can be applied to any models satisfying
\begin{itemize}
    \item Consistency relations for all operators appearing in cubic interactions,
    \item The super-Hubble curvature perturbation under consideration is constant at the tree level.
\end{itemize}
The latter condition is not independent of the former, 
but we have just distinguished it for technical reasons. 
In particular, the derivation of the formula in Ref.~\cite{Pimentel:2012tw} does not rely on particular inflationary models. It should be applied to the transient USR model as long as the above properties are satisfied. 
Nevertheless, it would be important to note that the second requirement may not be satisfied e.g., by the modes that exit the Hubble radius near the USR phase transition.
In such a case, our result may not be applied directly, 
and it requires more detailed consideration.

Another advantage of the master formula in~Ref.~\cite{Pimentel:2012tw} is its independence from the Lagrangian we use. 
Indeed, the controversy also originates from the subtlety of 
time-total-derivative interactions and associated terms proportional to the free equations 
of motion (EOMs), which can be treated 
also by performing a field redefinition~\cite{Maldacena:2002vr,Seery:2006tq,Arroja:2011yj,Burrage:2011hd,Braglia:2024zsl,Kawaguchi:2024lsw}.
At the tree level, their contributions can be incorporated by the field redefinition but may become subtle at one-loop order. 
More importantly, such terms turn out to be crucial 
in proving the consistency relations of some operators 
with various momenta.
Since the effective constancy of $\zeta$ can be proven only 
when the consistency relations hold~\cite{Pimentel:2012tw}, the subtle interaction terms must be treated very carefully.
Nevertheless, on the formal level, without specifying the concrete Lagrangian we use, the master formula tells us that as long as the consistency relations and the constancy of free mode functions hold for the modes under consideration, we expect that one-loop corrections to the super-Hubble modes are small, contrary to the claim in Ref.~\cite{Kristiano:2022maq}.
In this sense, if we assume the above-mentioned requirements that would be true even in the transient USR model, we do not need to discuss further, and simply conclude that there are no large 
one-loop corrections. Nevertheless, it would be worth showing the absence of large one-loop corrections explicitly. 

\subsection{What do we discuss?}

In this work, we revisit one-loop corrections to 
the power spectrum of curvature perturbations
in the transient USR model on the basis 
of Ref.~\cite{Pimentel:2012tw}, which proves the time-independence of the super-Hubble mode of $\zeta$. 
The key observation made in Ref.~\cite{Pimentel:2012tw} 
is that various contributions to the 
two-point correlation function 
$\langle\zeta_k\zeta_k\rangle$ from 
cubic interactions as well as quartic ones can be summarized into 
a simple formula. 
Taking the super-Hubble limit $k \ll aH$ of 
an external $\zeta_k$ and using the consistency relations, one can show that the two-point function acquires no time-dependence, namely, it is constant over time.\footnote{The all-order proof of the constancy can be found in Ref.~\cite{Senatore:2012ya} and another all-order proof is given in Ref.~\cite{Assassi:2012et}, but provided that constancy of $\zeta$ holds at tree level in both of the proofs.} As we will review, the soft limit of the external momentum leads to a simple formula that is similar to the one given 
in Ref.~\cite{Tada:2023rgp} but is a more general form~\cite{Pimentel:2012tw}. 
It turns out that the leading order one-loop corrections can be evaluated as a total derivative form in the loop momentum depending only on the small momentum mode, which finally shows 
the smallness of one-loop corrections. 
Note that a consistency relation has been examined in 
Refs.~ \cite{Kristiano:2023scm,Motohashi:2023syh,Tada:2023rgp,Namjoo:2024ufv,Kristiano:2024vst} within the transient USR model, but it is of the bispectrum of $\zeta$ without derivatives. To prove 
the above-mentioned result, we need to check other sets 
of three-point functions as will be clear from the master formula. We also comment on the regularization of the UV modes used 
in Ref.~\cite{Tada:2023rgp} but criticized 
in Ref.~\cite{Firouzjahi:2023bkt}.

We should emphasize that our results are mostly based on the earlier work~\cite{Pimentel:2012tw}. 
Nevertheless, our work contains several new results: 
First, we will derive the master formula within 
the path-integral formalism, which seems simpler than 
the operator formalism used in Ref.~\cite{Pimentel:2012tw}. Secondly, we particularly focus on a subset of cubic interactions that possibly give rise to large one-loop corrections due to unsuppressed couplings during the USR epoch. We also check whether the master formula of Ref.~\cite{Pimentel:2012tw} yields the same conclusion within the transient USR model by examining the consistency relations of relevant cubic interactions. 
It will turn out that one should include quartic interactions that must exist 
in the Lagrangian for consistency relations 
or the symmetry of the zero mode of $\zeta$ to hold. 
With an appropriate subset of interactions we can show 
the consistency relations that are essential to the final result, and we find no significant enhancement of the power spectrum 
due to one-loop corrections. It is also worth emphasizing that the same results should be derived within the operator formalism by just following the discussion in Ref.~\cite{Pimentel:2012tw}. 
The only difference with our result would be rather technical. For instance, as we will discuss, the EOM terms play crucial roles within the path-integral formalism, whereas the boundary terms do in the operator formalism and the EOM terms automatically vanish for interaction picture field operators. 
So long as the consistency relations of all relevant correlators are satisfied, the master formula for one-loop corrections applies and it will give us a simple 
answer.\footnote{We note that the absence of large one-loop corrections 
is also shown in Ref.~\cite{Inomata:2024lud}, where the author uses 
a spatially flat gauge. 
Since the method is quite different from what we will discuss, 
it is not easy to compare the results directly, 
but they would be related to each other.}

The rest of this work is organized as follows. 
In Sec.~\ref{Setup}, we revisit the dynamics of 
linear perturbations in the transient USR model 
and the cubic-order action of curvature perturbations.
In Sec.~\ref{master}, we derive the master formula 
for one-loop corrections to the power spectrum of 
curvature perturbations within the path-integral formalism. 
This formula is rather independent of 
the Lagrangian of a scalar field we consider. 
In Sec.~\ref{AppMaldacena}, we analytically compute the three-point correlation $\langle\zeta^3\rangle$ in the squeezed limit for the transient USR model. 
This analysis reveals that, in the limit of small $\epsilon$, we need only three 
leading-order terms to reproduce the consistency relation of $\langle\zeta^3\rangle$. 
Based on such an observation, we consider a reduced Lagrangian that consists of 
the aforementioned leading-order interactions. 
Adding quartic interactions necessary for consistency relations of three-point functions including spatial derivatives, we prove the consistency relations within the reduced Lagrangian system in Sec.~\ref{effectiveUSR}, justifying the master formula. 
Then, we explicitly evaluate one-loop corrections 
to the power spectrum of 
super-Hubble curvature perturbations in Sec.~\ref{oneloopresult}, which 
shows that there are no corrections at the leading order in the momentum expansion. Sec.~\ref{conclusion} is devoted to conclusions. 

We have several Appendices: In Appendix \ref{mostgeneralpathintegral}, 
we discuss the path-integral formulation with time-derivative interactions. 
In Appendix.~\ref{Bterminpathintegral}, we briefly show that the boundary terms in the Lagrangian do not contribute to the correlation functions with our prescription of the path-integral formalism. We give some detail of the equality of \eqref{one-loop-cut-Fourier} in Appendix~\ref{proofid}. In Appendix.~\ref{renormalization}, we briefly comment on the structure of UV divergences within the transient USR model and the standard SR models. 
We also give some details of one-loop corrections within our reduced system and discuss some subtlety associated with the time-ordering of operators in Appendix~\ref{explicitform}, which however does not affect our results.

\section{Transient ultra-slow-roll inflation}
\label{Setup}

In the transient USR inflationary model which belongs to 
the action (\ref{action0}), we review the dynamics of 
linear perturbations and the cubic-order action of 
curvature perturbations relevant to the computation 
of one-loop corrections performed later in Sec.~\ref{master}. 
We consider the 1st SR stage followed by the transient 
USR regime, which finally connects to the 2nd SR phase.  
The field equations of motion on the 
spatially-flat FLRW background with the 
line element $\rd s^2=-\rd t^2+a^2(t) 
\delta_{ij}\rd x^i \rd x^j$ are given by (\ref{backeq}). 
As we already mentioned in Introduction, 
the USR stage can be realized by 
an inflaton potential with a very flat region. 
During the SR stages, the SR parameters $\epsilon=-\dot{H}/H^2$ 
and $\eta=\dot{\epsilon}/(H \epsilon)$ are
much smaller than 1, whereas in the USR regime, 
we have $\epsilon \propto a^{-6}$ and 
$\eta \simeq -6$. The time dependence of $\eta$ 
may be quantified as 
\be
\eta(t)=\Delta\eta \left[ 
\theta(t-t_s)-\theta(t-t_e)\right]\,,\qquad
\Delta\eta=-6\,,
\label{eta}
\ee
where 
\begin{align}
\theta(t)=\left\{\begin{array}{ll}1&
\text{for }t>0\\ 1/2&\text{for }t=0\\ 0&\text{for }t<0
\end{array} \right.
\label{Hevi}
\end{align}
is the Heaviside step function, and 
$t_s$ and $t_e$ are the instants at which 
the USR period starts and ends, respectively.
Here we assume that $\eta=0$ in two SR 
regimes, which allows us to approximate $\epsilon$ to be constant. Integrating the relation $\eta=\dot{\epsilon}/(H \epsilon)$ 
with neglect of the time-dependence of $H$,  
it follows that 
\be
\epsilon(t)=\begin{cases}\epsilon_1 \hspace{2.9cm} (\text{1st SR stage}) \\ 
\epsilon_1 e^{\Delta\eta H(t-t_s)} \hspace{1.3cm} (\text{USR stage})\\
\epsilon_2=\epsilon_1 e^{\Delta\eta H(t_e-t_s)} \hspace{0.4cm} (\text{2nd SR stage})\end{cases}\,,
\label{epsilon}
\ee
where $\epsilon_1$ and $\epsilon_2$ are constants.
Due to the rapid decrease of $\epsilon$ during the 
USR regime, $\epsilon_2$ is smaller than $\epsilon_1$.

\subsection{Second-order action and enhancement 
of small-scale linear perturbations}
\label{linear}

To study the evolution of cosmological perturbations
for the transient USR model 
mentioned above, we take into account a 
Gibbons-Hawking-York boundary term to 
the action (\ref{action0}), such that 
\be
{\cal S}=\int_{\cal V} \rd^4 x \sqrt{-g} \left[ 
\frac{\Mpl^2}{2}R-\frac{1}{2} 
\partial_{\mu} \phi \partial^{\mu} \phi
-V(\phi) \right]+{\cal{S}}_{\rm GHY}\,,
\label{action}
\ee
where
\be
{\cal{S}}_{\rm GHY}=\Mpl^2\oint_{\partial{\cal{V}}} {\rm d}^3 y \sqrt{|h|}\, \varepsilon_b K \,.
\label{Gibbons-Hawking-York} 
\ee
Here, $h$ is a determinant of the 3-space metric $h_{ij}$ appearing in Eq.~(\ref{ADM}), $K=K_{ij}h^{ij}$ is the trace of an extrinsic curvature $K_{ij}=(\dot{h}_{ij}
-\nabla^{(3)}_i N_j-\nabla^{(3)}_j N_i)/(2N)$ with 
$\nabla^{(3)}_i$ being the 3-space covariant 
derivative operator, and $\varepsilon_b$ is a parameter 
that is $-1$ at temporal boundaries and $+1$ at spatial boundaries.
The Gibbons-Hawking-York action ${\cal{S}}_{\rm GHY}$ is required for boundary value problems to be well defined. 

To study the evolution of linear perturbations, 
we expand the action up to quadratic order in 
perturbed variables. 
We choose the unitary gauge (\ref{unitary}) and 
consider the metric perturbations in the ADM 
line element (\ref{ADM}) as 
$N=1+\alpha$ and $N_i=\partial_i \psi$, 
so that 
\be
{\rm d}s^2=-[(1+\alpha)^2-a^{-2}e^{-2\zeta}(\partial\psi)^2] {\rm d}t^2+2\partial_i\psi {\rm d}t{\rm d}x^i+a^2e^{2\zeta}\delta_{ij}{\rm d}x^i{\rm d}x^j\,,
\label{perturbedmetric}
\ee
where $(\partial\psi)^2 \equiv (\partial_i\psi)(\partial_i\psi)
=(\partial_{x^1}\psi)^2+(\partial_{x^2}\psi)^2
+(\partial_{x^3}\psi)^2$.
We can rewrite the action~\eqref{action} 
in the form of the $3+1$ ADM decomposition as 
\ba
&&
{\cal{S}}=\int_{\cal{V}} {\rm d} t {\rm d}^3 x
\sqrt{h}\left[\frac{\Mpl^2}{2}N\left(R^{(3)}+K^{ij}K_{ij}-K^2\right)+\frac{1}{2N}\dot{\phi}^2-NV(\phi) \right] \notag\\
&&
\hspace{2.7cm}+\Mpl^2\int_{\cal{V}} {\rm d} t {\rm d}^3 x\frac{\rm d}{{\rm d}t}\left(\sqrt{h}K\right)+(\text{spatial boundary terms})\,,
\label{action3+1}
\ea
where $R^{(3)}$ is the 3-dimensional Ricci scalar 
of hypersurfaces foliating ${\cal{V}}$. 
Expanding the action up to quadratic order 
in perturbations and eliminating the 
non-dynamical perturbations $\alpha$ and $\psi$ 
on account of their constrained equations of motion, the resulting second-order action 
for $\zeta$ is expressed in the form 
\be
{\cal{S}}^{(2)}=\Mpl^2\int{\rm d}\tau{\rm d}^3x\biggl[\epsilon a^2\zeta'^2-\epsilon a^2 (\partial\zeta)^2-\frac{\rm d}{{\rm d}\tau}f_2\biggr]\,,
\label{secondorderactionforzeta} 
\ee
where we recall that a prime represents the derivative 
with respect to $\tau=\int a^{-1} \rd t$, and 
we have kept the temporal boundary term
\be
f_2=9a^3H\zeta^2-\frac{a}{H}(\partial\zeta)^2 \,,
\label{F2}
\ee
just for completeness.\footnote{The spatial translation 
invariance implies that there are no spatial boundary contributions to the action, whereas we would have initial and final spacelike surfaces, which can be regarded as temporal boundary surfaces. Indeed, some surface terms can contribute to the correlation functions 
when the perturbations are quantized.}
Varying Eq.~(\ref{secondorderactionforzeta}) 
with respect to $\zeta$, the linear curvature 
perturbation obeys
\be
\left(\epsilon  a^2 \zeta' \right)' 
-\epsilon a^2 \partial^2 \zeta=0\,.
\label{eom}
\ee
We decompose the curvature perturbation 
in the real space coordinate ${\bm x}=(x^1,x^2,x^3)$
in terms of the Fourier modes $\zeta_k(\tau)$, as
\begin{equation}
\zeta(\tau,{\bm{x}})=\frac{1}{(2\pi)^{3}}\int 
\rd^{3}{k}\,\tilde{\zeta} (\tau,{\bm{k}})
e^{\ri{\bm{k}}\cdot{\bm{x}}}\,,
\qquad 
\tilde{\zeta} (\tau,{\bm{k}})=\zeta_k(\tau)\hat{a}_{\bm{k}}
+\bar{\zeta}_k(\tau)\hat{a}^{\dagger}_{-{\bm{k}}}\,,
\label{RFourier}
\end{equation}
where $\hat{a}_{\bm{k}}$ and $\hat{a}^{\dagger}_{\bm{k}}$ 
are the annihilation and creation operators, respectively, 
satisfying the commutation relations
\begin{equation}
\left[\hat{a}_{{\bm{k}}_{1}},
\hat{a}^{\dagger}_{{\bm{k}}_{2}}\right]
=(2\pi)^{3}\delta^{(3)}({\bm{k}}_{1}+{\bm{k}}_{2})\,,
\qquad\left[\hat{a}_{{\bm{k}}_{1}},\hat{a}_{{\bm{k}}_{2}}\right]
=\left[\hat{a}^{\dagger}_{{\bm{k}}_{1}},
\hat{a}^{\dagger}_{{\bm{k}}_{2}}\right]=0\,.
\end{equation}
The canonical commutation relation $[{\zeta}(\tau,\bm x),{\pi}_\zeta(\tau,\bm y)]=\ri\delta^{(3)}(\bm x-\bm y)$, 
 where ${\pi}_\zeta$ is a canonical momentum of $\zeta$, 
yields the normalization condition on the mode function $\zeta_k$, as
\be
\zeta_k \bar{\zeta}_k{}'-\zeta_k' \bar{\zeta}_k=\frac{\ri}{2a^2\epsilon \Mpl^2},\label{normalizationzeta}
\ee
where a bar represents the complex conjugate.
From \eqref{eom}, the mode function in 
Fourier space satisfies Eq.~(\ref{zetaeq}), 
which is expressed in the form 
\be
\zeta_k''+\frac{(z^2)'}{z^2}\zeta_k'
+k^2\zeta_k=0\,,
\label{modeeom}
\ee
where $z\equiv a \Mpl \sqrt{2\epsilon}$. 
Introducing a canonically normalized field 
$u_k=z\zeta_k$, Eq.~(\ref{modeeom}) reduces to
\be
u_k''+\left( k^2-\frac{\nu^2
-1/4}{\tau^2} \right)u_k=0\,,
\label{ueq}
\ee
with
\be
\nu^2=\tau^2\frac{z''}{z}+\frac{1}{4} \simeq 
\left(\frac{3+\eta}{2}\right)^2-\frac{1}{2}\eta'\tau\,,\label{defnu}
\ee
where we assumed that the first slow-roll parameter $\epsilon$ is negligibly small and the background spacetime is given by 
a quasi de Sitter background, so that $\tau \simeq -1/(aH)$.

Since we are considering the case in which 
$\eta$ is given by \eqref{eta}, we can deal 
with $\nu$ as constants in each stage. 
This enables us to solve Eq.~\eqref{ueq} analytically.
During the two SR regimes, we have $\eta \simeq 0$, 
whereas, in the USR stage, $\eta \simeq -6$. 
Hence it follows that $\nu^2=9/4$ in both SR and 
USR regions. Imposing the Bunch-Davies vacuum condition 
$u_k (\tau \to -\infty)=e^{-\ri k\tau}/\sqrt{2k}$ 
in the asymptotic past, the solutions to 
curvature perturbations can be expressed in the form
\ba
&&
\zeta_k(\tau)=\ri \frac{H}{\Mpl}\frac{1}{\sqrt{4\epsilon(\tau)k^3}}\xi(k\tau,k\tau_s,k\tau_e) \label{zetak}\,,\\
&&
\xi(x,y,z)={\cal{A}}_k(y,z)(1+\ri x)e^{-\ri x}
+{\cal{B}}_k(y,z)(1-\ri x)e^{+\ri x}\,,
\label{xik}
\ea
where $\tau_s$ and $\tau_e$ are the conformal times 
at which the USR stage starts and ends, 
respectively, and $x=k\tau$, $y=k\tau_s$, and $z=k\tau_e$.
The coefficients ${\cal{A}}_k$ and ${\cal{B}}_k$, 
which depend on $y$ and $z$, can be determined 
by matching the solutions of 
$\zeta_k$ at $\tau=\tau_s$ and $\tau=\tau_e$. 
From Eq.~\eqref{modeeom}, we impose the 
following matching conditions 
\be
\zeta_k(\tau_{i-})=\zeta_k(\tau_{i+}), \hspace{1cm}\text{and} \hspace{1cm} z^2(\tau_{i-})\zeta_k'(\tau_{i-})=z^2(\tau_{i+})\zeta_k'(\tau_{i+}), \label{matching}
\ee
where $i=s,e$, and the subscripts $\pm$ represent 
$\tau_{i\pm}=\tau_i\pm 0$, respectively.
Substituting Eqs.~\eqref{zetak} and \eqref{xik} 
and their $\tau$-derivatives into Eq.~\eqref{matching}, the coefficients in each epoch are determined to be \cite{Byrnes:2018txb}
\begin{itemize}
\item 1st SR stage
\be
{\cal{A}}_k=1\,,\qquad
{\cal{B}}_k=0 \,, 
\ee
\item USR stage
\be
{\cal{A}}_k=1+\frac{3\ri (1+y^2)}{2y^3}\,,
\qquad
{\cal{B}}_k=\frac{3\ri (-\ri+y)^2 e^{-2\ri y}}{2y^3}\,, 
\ee
\item 2nd SR stage
\ba
&&
{\cal{A}}_k=\frac{1}{4y^3z^3}
\Biggl[-9(-\ri+y)^2(\ri+z)^2 e^{2\ri(z-y)}
+\left(3\ri+3\ri y^2+2y^3\right)\left(-3\ri-3\ri z^2+2z^3\right)\Biggr]\,,\label{2ndSRAB0}\\
&&
{\cal{B}}_k=\frac{3\ri}{4y^3z^3}
\Biggl[(-\ri+y)^2
\left(3\ri+3\ri z^2+2z^3\right)e^{-2\ri y}
-(-\ri+z)^2 \left(3\ri+3\ri y^2+2y^3\right)e^{-2\ri z}
\Biggr]\,.
\label{2ndSRAB}
\ea
\end{itemize}

For long-wavelength perturbations, we take the 
$\tau$-derivatives of the above solutions and 
pick up the leading-order terms with respect to 
small values of $k$.
This manipulation leads to 
\ba
&&
\zeta_k'(\tau)\simeq \ri\frac{H}{\Mpl}\frac{1}{\sqrt{4\epsilon_1 k^3}}k^2\tau\,,\hspace{4.5cm} (\text{1st SR stage})\,,
\\
&&
\zeta_k'(\tau)\simeq \ri\frac{H}{\Mpl}\frac{1}{\sqrt{4\epsilon(\tau)k^3}}\frac{6+k_s^5\tau^5}{5k_s^2\tau^2}k^2\tau\,,\hspace{2.8cm} (\text{USR stage})\,,
\\
&&
\zeta_k'(\tau)\simeq \ri\frac{H}{\Mpl}\frac{1}{\sqrt{4\epsilon_2 k^3}}\left(\frac{k_s^3}{k_e^3}-\frac{6k_e^3\tau}{5k_s^2}+\frac{6k_s^3\tau}{5k_e^2}\right)k^2\tau\,,\hspace{1.0cm} (\text{2nd SR stage})\,,
\ea
where $k_s$ and $k_e$ are comoving wavenumbers that exit the Hubble horizon at $\tau=\tau_s$ and $\tau=\tau_e$, i.e., $k_s=a(\tau_s)H(\tau_s)\simeq-1/\tau_s$ and 
$k_e=a(\tau_e)H(\tau_e)\simeq-1/\tau_e$, respectively. 
For finite values of $k$, we have the time 
dependence $|\zeta_k'(\tau)| \propto a^{-1}$ 
during the two SR stages and 
$|\zeta_k'(\tau)| \propto a^{4}$ during the 
USR stage. So long as $k$ is not much smaller 
than $k_s$, there is the enhancement of 
$|\zeta_k(\tau)|$ proportional to $a^3$ 
during the USR period.
In the limit that $k/k_s \to 0$, however, 
$\zeta_k'(\tau)$ vanishes in all three periods.

\begin{figure}[t]
\centering
\includegraphics[width=10cm]{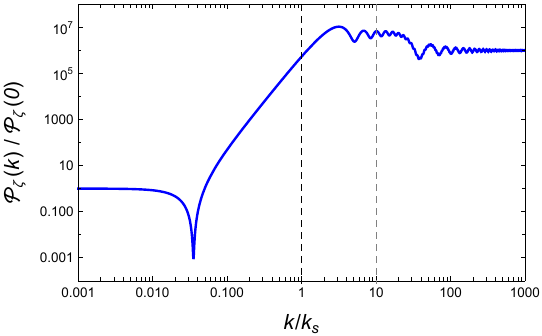}
\caption{\label{powerspectrum} 
The power spectrum of curvature perturbations 
in the transient USR model for $\tau_s/\tau_e=10.0041$ 
and $\epsilon_1=0.01$. 
We normalize ${\cal P}_{\zeta}(k)$ and $k$ by  
the power spectrum ${\cal P}_{\zeta}(0)$ 
in the long-wavelength limit and 
$k_s$ corresponding to the wavenumber at $\tau=\tau_s$, 
respectively. 
The increased power for small-scale perturbations
($k \gtrsim k_s$) arises from the enhancement 
of $|\zeta_k|$ during the transient USR period.
}
\end{figure}

The tree-level power spectrum ${\cal P}_{\zeta} (\tau_*,k)$ at a particular moment 
$\tau=\tau_*$ is defined by the following 
vacuum expectation value
\be
\langle 0| \tilde{\zeta} (\tau_*, {\bm k}_1) \tilde{\zeta} (\tau_*, {\bm k}_2) | 0 \rangle
=\frac{2\pi^2}{k_1^3}{\cal P}_{\zeta} (\tau_*, k_1)\,(2\pi)^3 \delta^{(3)} ({\bm k}_1+{\bm k}_2)
\,,
\ee
where $|0\rangle$ is the Bunch-Davies vacuum state.
By using Eq.~\eqref{RFourier}, we obtain
\be
{\cal P}_{\zeta} (\tau_*, k)=\frac{k^3}{2\pi^2}|\zeta_k(\tau_*)|^2=\frac{H^2}{8\pi^2\Mpl^2\epsilon_1}|{\cal{A}}_k(\tau_*)+{\cal{B}}_k(\tau_*)|^2\,.
\label{analyticpowerspectrum}
\ee
To compute the power spectrum at the end of 
inflation ($\tau_* \to 0$), we substitute
the coefficients (\ref{2ndSRAB0}) and 
(\ref{2ndSRAB}) into 
Eq.~(\ref{analyticpowerspectrum}).
In Fig.~\ref{powerspectrum}, we plot an example 
of the power spectrum normalized by the one 
in the large-scale limit versus the normalized 
wavenumber $k/k_s$. 
In this plot, the ratio $\tau_s/\tau_e$ is about $10$ and hence 
the number of e-foldings acquired during 
the USR period corresponds to 2.3.
We observe that small-scale curvature perturbations ($k \gtrsim k_s$) are subject to strong enhancement compared to 
large-scale modes ($k \ll k_s$). 
This property arises from the growth $|\zeta_k| \propto a^3$ 
during the USR period for small-scale modes.

\subsection{The effective constancy of curvature 
perturbations for super-Hubble modes}
\label{effconst}

In the subsequent sections, we often exploit the fact that 
sufficiently long-wavelength perturbations such as those 
relevant to CMB temperature anisotropies behave 
as effectively constant in time even during the USR phase. 
It is worth quantifying it more explicitly to confirm 
how the effective constancy is valid. 
This amounts to expanding the solution (\ref{zetak}) 
with respect to the small momentum $k$, yielding
\be
\zeta_k(\tau)=\ri \frac{H}{\Mpl} 
\frac{1}{\sqrt{4\epsilon_1 k^3}} 
\left[ 1+r_k (\tau) \right]\,,
\ee
where the next-to-leading order terms 
in $r_k(\tau)$ are given by 
\begin{align}
r_k(\tau)=
\left\{\begin{array}{l} 
\dfrac{1}{2}k^2 \tau^2 \qquad (\text{for }\tau<\tau_s)\,,\\ 
-\dfrac{2}{5}(k\tau_s)^2\left(\dfrac{\tau_s}{\tau}\right)^3 \qquad (\text{for }\tau_s <\tau<\tau_e)\,, \\ 
\dfrac{2}{5}(k\tau_s)^2 \left(\dfrac{\tau}{\tau_s}\right)^3\left(\dfrac{\tau_s}{\tau_e}\right)^6\qquad 
(\text{for }\tau>\tau_e)\,.
\end{array}\right.\label{defrk}
\end{align}
This shows that $r_k$ has the scale-factor dependence $r_k \propto a^{-2}$, $r_k \propto a^{3}$, and $r_k \propto a^{-3}$ 
during the 1st SR, USR, and 2nd SR stages, 
respectively.

Here, we notice that the next-to-leading order correction is universally bounded for $\forall\tau>\tau_s$ as
\be
|r_k(\tau)|<|r_k(\tau_e)|=\frac{2}{5}e^{-2\Delta N_{k}
+3\Delta N_{\rm USR}}\,,
\label{rkbound}
\ee
where $\Delta N_k \equiv -\log (k\tau_s)=
-\log [a(\tau_k)/a(\tau_s)]$ is the number of 
e-foldings between the time $\tau_k$ of Hubble radius 
crossing of the $k$-mode and the time $\tau_s$, 
and similarly, 
$\Delta N_{\rm USR}=\log [a(\tau_e)/a(\tau_s)]$ 
is the number of e-foldings acquired during the USR era.  

The above bound is crucial for our later discussion in the sense that despite the presence of a growing mode during the USR phase, if $k$ is sufficiently small, namely, $\Delta N_k$ is large enough, the mode function $\zeta_k$ is approximately constant. 
For the CMB modes ($k \ll k_s$) with 
$\Delta N_k \gg (3/2)\Delta N_{\rm USR}$, taking the typical values $\Delta N_k=40$ and 
$\Delta N_{\rm USR}=3$ relevant to 
the seed generation of PBHs, we have that 
$|r_k(\tau)|<6 \times 10^{-32}$ for $\forall \tau>\tau_s$.
Thus, we can neglect the time dependence of $\zeta_k$ for the CMB modes satisfying the condition $\Delta N_k \gg 
(3/2)\Delta N_{\rm USR}$. In the subsequent discussion, we often approximate a super-Hubble mode function by constant for $\tau>\tau_s$. 
We should remember that it is correct up to 
$r_k(\tau)$, which is bounded by \eqref{rkbound}.

One may wonder if the time derivative of $\zeta_k(\tau)$ 
is small particularly during the USR phase. 
For $\tau_s<\tau<\tau_e$, 
the $k$-expansion of $\zeta'_k(\tau)$ yields
\ba
\left|\frac{\tau\zeta_k'(\tau)}{\zeta_k(\tau)}\right|
\simeq \frac65 (k\tau_s)^2\left(\frac{\tau_s}{\tau}\right)^3<\frac65 e^{-2\Delta N_{k}+3\Delta N_{\rm USR}}\,,
\ea
or equivalently
\ba
|\zeta_k'(\tau)|<\frac{3|r_k(\tau_e)|}{\tau}|\zeta_k(\tau)|\,.
\ea
Once again, we find a similar bound, and as is clear from the above discussion, we may neglect $\zeta_k'(\tau)$ for a typical CMB mode thanks to the smallness of $r_k(\tau_e)$. 
Thus we showed that, even during the USR phase, the time 
derivative of $\zeta_k$ sufficiently outside 
the Hubble horizon is negligible.

\subsection{Cubic-order action}
\label{thirdorderaction}

For the calculation of one-loop corrections to 
the power spectrum of $\zeta$, 
we need to expand the action (\ref{action}) 
higher than second order in perturbations.
The cubic-order action was first derived by 
Maldacena~\cite{Maldacena:2002vr} in the context 
of inflationary non-Gaussianities. 
Keeping the boundary terms associated with the 
$\tau$-derivatives and using the Hamiltonian and momentum constraints, the cubic-order action 
following from (\ref{action}) can be expressed 
in the form 
\be
{\cal S}^{(3)}= 
\Mpl^2 \int  \rd\tau\rd^3 x 
\Biggl[ a^2 \epsilon^2 \zeta \zeta'^2
+a^2 \epsilon^2 \zeta (\partial \zeta)^2
-a \epsilon \left( 2-\frac{1}{2} \epsilon 
\right) \zeta' \partial_i \zeta 
\partial_i \chi 
+\frac{a^2}{2}\epsilon\eta'\zeta^2\zeta'
+\frac{\epsilon}{4} \partial^2 \zeta (\partial \chi)^2
+2f(\zeta) D_2 \zeta
-\frac{\rm d}{{\rm d} \tau}f_3
\Biggr]\,,
\label{S3JY}
\ee
where 
\ba
& &
\chi=a \epsilon \partial^{-2} \zeta'\,,\qquad 
D_2 \zeta
=\frac{\rd}{\rd \tau} \left( a^2 \epsilon \zeta' 
\right) -a^2 \epsilon \partial^2 \zeta\,,\label{EOMopdef}\\
& &
f(\zeta)=\frac{1}{4}\eta\zeta^2+\frac{1}{aH} \zeta \zeta'
-\frac{1}{4a^2 H^2} \left[ (\partial \zeta)^2 
-\partial^{-2}(\partial_i \partial_j (\partial_i \zeta 
\partial_j \zeta)) \right] 
+\frac{1}{2a^2 H} \left[ (\partial_i \zeta) (\partial_i \chi)
-\partial^{-2}(\partial_i \partial_j (\partial_i \zeta 
\partial_j \chi)) \right]\,,
\label{fzetaJY} \\
& &
f_3= 9a^3 H\zeta^3
+\frac{1}{2}a^2\epsilon\eta\zeta^2\zeta'
+\frac{\epsilon a}{H}\left(1-\frac{\epsilon}{2}\right)\zeta\zeta'^2
\notag\\
& &
\hspace{0.8cm}
-(1-\epsilon)\frac{a}{H}\zeta(\partial\zeta)^2
+\frac{\epsilon}{2H^2}\zeta\zeta'\partial^2\zeta
+\frac{1}{4aH^3}\left[ (\partial\zeta)^2\partial^2\zeta
-2H\zeta\partial_i\partial_j\chi\left(\partial_i\partial_j(\zeta-H\chi)\right)\right]\,.
\label{F3JY}
\ea
Note that the boundary terms associated with the 
spatial derivatives of $\zeta$ have been dropped 
after the integration by parts.
The first five contributions to (\ref{S3JY}) 
correspond to the bulk terms. The contributions 
$2f(\zeta)D_2 \zeta$, which we call the EOM terms, 
are proportional to the linear perturbation equation 
of $\zeta$. The last contributions $-\rd f_3/\rd \tau$ 
correspond to the boundary terms. 
In the operator formalism, we cannot generally 
ignore the boundary terms, while the EOM terms 
may be neglected \cite{Seery:2005wm,Chen:2006nt,Arroja:2011yj,Burrage:2011hd}.
In Ref.~\cite{Kawaguchi:2024lsw}, 
it was shown in the context of slow-roll inflation that 
the path-integral formalism allows us to neglect 
the boundary terms, while the EOM terms bring 
contributions to cosmological correlation functions.
As we will show in Appendix.~\ref{Bterminpathintegral}, 
the latter property in the path-integral 
formalism holds in general, including the 
case of transient USR inflation.

\section{Master formula for one-loop corrections 
within the path-integral formalism}
\label{master}

In this section, we prove the master formula of one-loop corrections to the two-point correlation function of $\zeta$ by using the path-integral formalism. In Sec.~\ref{Setup}, we considered the 
theory given by~(\ref{action}) and derived the 
cubic-order action~(\ref{S3JY}) of $\zeta$. 
In this section, we will consider the case 
in which more general interacting Lagrangians are present.
Indeed, the master formula itself can be derived without 
any details of the Lagrangian and background field configurations 
such as the time dependence of slow-roll parameters. 
While the same formula was obtained
in Ref.~\cite{Pimentel:2012tw} within the operator formalism, 
we recognize that the path-integral approach is simpler than the former in its derivation. 

\subsection{In-in path-integral formalism}\label{ininpath1}

In this subsection, we briefly review the 
in-in path-integral formalism and introduce 
necessary tools for proving the master formula of one-loop corrections. 
This is also known as the 
Schwinger-Keldysh path-integral formalism, see 
Refs.~\cite{Weinberg:2005vy,Chen:2017ryl} 
for further details. 
Readers who are familiar with this 
formalism may directly go 
to the next subsection~\ref{ininpath}.

Within the in-in path-integral scheme, 
we evaluate the expectation values of quantum operators rather than the transition amplitudes given as $S$-matrices. Accordingly, there are two fields, the $+$ and $-$ fields, which correspond to the ``ket'' and ``bra'' fields, respectively. 
The generating functional for interacting systems can formally be written as
\be
Z[J^{+},J^{-}]={\cal{N}}\exp\left[\ri{\cal{S}}_{\rm int}\left[\frac{\delta}{\ri\delta J^+}\right]-\ri{\cal{S}}_{\rm int}\left[\frac{\delta}{-\ri\delta J^-}\right]\right]Z_0[J^+,J^-]\,,
\label{generatingfunctional}
\ee
where $\cal{N}$ is a normalization factor, 
$J^{\pm}$ are external source fields, ${\cal{S}}_{\rm int}$ 
is an interacting part of the action and $Z_0$ is 
a generating functional for the free field, 
which is given by
\be
Z_0[J^+,J^-]={\cal{N}}_0\int{\cal{D}}\zeta^+{\cal{D}}\zeta^- \delta(\zeta^+(\tau_f,\bm{x})-\zeta^-(\tau_f,\bm{x}))\exp
\left[\ri{\cal{S}}^{(2)}[\zeta^+]-\ri{\cal{S}}^{(2)}[\zeta^-]
+\ri\int^{\tau_f}_{-\infty}{\rm d}\tau\int{\rm d}^3x(J^+ \zeta^+-J^-\zeta^-)\right]\,,
\ee
where $\tau_f$ is the time on the final slice. We have a few remarks about the path-integral formulation. We will deal with interactions that contain time derivatives of $\zeta$. As we will clarify in Appendix~\ref{mostgeneralpathintegral}, the canonical momentum $\pi_\zeta$ is no longer linearly related to $\zeta'$ and therefore, there are non-trivial additional terms that are not present 
in the original Lagrangian. Such additional terms may have a singular factor $\delta^{(4)}(0)$, but we generally prove that they are necessary to cancel unphysical singular contributions associated with derivative interactions. For notational simplicity, we make the additional terms implicit and simply neglect the unphysical singular contributions since they should be canceled by the terms with the singular coefficient. The second remark is about the prescription for the final time $\tau_f$. We take $\tau_f$ to be later than the time at which we evaluate operators' expectation values. As we will 
prove in Appendix~\ref{Bterminpathintegral}, we can neglect the boundary interactions within this prescription. A simple explanation of this fact is due to causality since the boundary interactions are at $\tau_f$, which cannot affect the operators in their past.

From the above generating functional, the expectation values of $\zeta$ fields can 
be calculated by the functional derivatives with respect to $J^{\pm}$.
For example, the three-point correlation function of the 
$\zeta^+$ field can be obtained by
\be
\ev{\zeta^+(\tau_*,\bm{x}_1)\zeta^+(\tau_*,\bm{x}_2)\zeta^+(\tau_*,\bm{x}_3)}=\frac{\delta}{\ri\delta J^+(\tau_*,\bm{x}_1)}\frac{\delta}{\ri\delta J^+(\tau_*,\bm{x}_2)}\frac{\delta}{\ri\delta J^+(\tau_*,\bm{x}_3)}Z[J^+,J^-]\biggr|_{J^{\pm}=0}\,.
\ee
In practice, we can use diagrammatic calculations of 
the correlation functions 
with a set of Feynman rules, which we will briefly describe
below (see Ref.~\cite{Chen:2017ryl} for the derivation).

\begin{itemize}

\item Free propagators 

Since there are two types of the $\zeta$ fields ($+$ and $-$), 
we have four free propagators given below.
\ba
& &
-\ri\Delta_{++}(\tau_1,\bm{x}_1;\tau_2,\bm{x}_2)
=\frac{\delta}{\ri\delta J^+(\tau_1,\bm{x}_1)}\frac{\delta}{\ri\delta J^+(\tau_2,\bm{x}_2)}Z_0[J^+,J^-]\biggr|_{J^{\pm}=0}=\bra{0}T\{\zeta(\tau_1,\bm{x}_1)\zeta(\tau_2,\bm{x}_2)\}
\ket{0}
\notag\\
& &
=\theta(\tau_1-\tau_2)\int\frac{{\rm d}^3k}{(2\pi)^3}\zeta_k(\tau_1)\bar{\zeta}_k(\tau_2)e^{\ri\bm{k}\cdot(\bm{x}_1-\bm{x}_2)}+\theta(\tau_2-\tau_1)\int\frac{{\rm d}^3k}{(2\pi)^3}\bar{\zeta}_k(\tau_1)\zeta_k(\tau_2)e^{-\ri\bm{k}\cdot(\bm{x}_1-\bm{x}_2)}\,,
\label{Delpp}\\
& &
-\ri\Delta_{+-}(\tau_1,\bm{x}_1;\tau_2,\bm{x}_2)
= \frac{\delta}{\ri\delta J^+(\tau_1,\bm{x}_1)}\frac{\delta}{-\ri\delta J^-(\tau_2,\bm{x}_2)}Z_0[J^+,J^-]\biggr|_{J^{\pm}=0}=\bra{0}\zeta(\tau_2,\bm{x}_2)\zeta(\tau_1,\bm{x}_1)\ket{0}
\notag\\
&&=\int\frac{{\rm d}^3k}{(2\pi)^3}\bar{\zeta}_k(\tau_1)\zeta_k(\tau_2)e^{-\ri\bm{k}\cdot(\bm{x}_1-\bm{x}_2)}\,,
\\
& &
-\ri\Delta_{-+}(\tau_1,\bm{x}_1;\tau_2,\bm{x}_2)
= \frac{\delta}{-\ri\delta J^-(\tau_1,\bm{x}_1)}\frac{\delta}{\ri\delta J^+(\tau_2,\bm{x}_2)}Z_0[J^+,J^-]\biggr|_{J^{\pm}=0}=\bra{0}\zeta(\tau_1,\bm{x}_1)\zeta(\tau_2,\bm{x}_2)\ket{0}
\notag\\
&&= \int\frac{{\rm d}^3k}{(2\pi)^3}\zeta_k(\tau_1)\bar{\zeta}_k(\tau_2)e^{\ri\bm{k}\cdot(\bm{x}_1-\bm{x}_2)}\,,
\\
& &
-\ri\Delta_{--}(\tau_1,\bm{x}_1;\tau_2,\bm{x}_2)
=\frac{\delta}{-\ri\delta J^-(\tau_1,\bm{x}_1)}\frac{\delta}{-\ri\delta J^-(\tau_2,\bm{x}_2)}Z_0[J^+,J^-]\biggr|_{J^{\pm}=0}=\bra{0}\bar{T}\{\zeta(\tau_1,\bm{x}_1)\zeta(\tau_2,\bm{x}_2)\}
\ket{0}
\notag\\
&&= \theta(\tau_1-\tau_2)\int\frac{{\rm d}^3k}{(2\pi)^3}\bar{\zeta}_k(\tau_1)\zeta_k(\tau_2)e^{-\ri\bm{k}\cdot(\bm{x}_1-\bm{x}_2)}+\theta(\tau_2-\tau_1)\int\frac{{\rm d}^3k}{(2\pi)^3}\zeta_k(\tau_1)\bar{\zeta}_k(\tau_2)e^{\ri\bm{k}\cdot(\bm{x}_1-\bm{x}_2)}\,,
\label{Delmm}
\ea
where $\zeta_k$ is the mode function of $\zeta$ 
in Fourier space defined in Eq.~\eqref{RFourier} 
and the solution to $\zeta_k$ is given by Eq.~\eqref{zetak} 
in the transient USR model, 
and $T\,(\bar{T})$ denotes the (anti-)time-ordered symbol.
For the correlation function of real fields to be real, 
the Heaviside step function needs to be defined as Eq.~(\ref{Hevi}). 
The propagators
in Fourier space corresponding to 
Eqs.~(\ref{Delpp})-(\ref{Delmm}) are 
given, respectively, by  
\ba
&&
G_{++}(k;\tau_1,\tau_2)=\zeta_k(\tau_1)\bar{\zeta}_k(\tau_2)\theta(\tau_1-\tau_2)+\bar{\zeta}_k(\tau_1)\zeta_k(\tau_2)\theta(\tau_2-\tau_1)\,,
\label{G++}
\\
&&
G_{+-}(k;\tau_1,\tau_2)=\bar{\zeta}_k(\tau_1)\zeta_k(\tau_2)\,,
\label{G+-}
\\
&&
G_{-+}(k;\tau_1,\tau_2)=\zeta_k(\tau_1)\bar{\zeta}_k(\tau_2)\,,
\label{G-+}
\\
&&
G_{--}(k;\tau_1,\tau_2)=\bar{\zeta}_k(\tau_1)\zeta_k(\tau_2)\theta(\tau_1-\tau_2)+\zeta_k(\tau_1)\bar{\zeta}_k(\tau_2)\theta(\tau_2-\tau_1)\,.
\label{G--}
\ea
In Fig.~\ref{figpropagators}, we show the diagrammatic rules for propagators, 
where the black and white dots denote the $+$ and $-$ fields, respectively. 
Later, we use the following formulas of the (anti-)time-ordered propagator,
\begin{align}
D_2 G_{\pm\pm}(k;\tau,\tau_*)=
\mp\frac{\ri\delta(\tau-\tau_*)}{2\Mpl^2}\,,
\label{EOMtimeprop}
\end{align}
and
\begin{align}
D_2 G_{\pm\mp}(k;\tau,\tau_*)=0\,,
\end{align}
where $D_2$ is the EOM derivative operator 
defined in~Eq.~\eqref{EOMopdef}.

\begin{figure}[h]
\centering
\includegraphics[width=10cm]{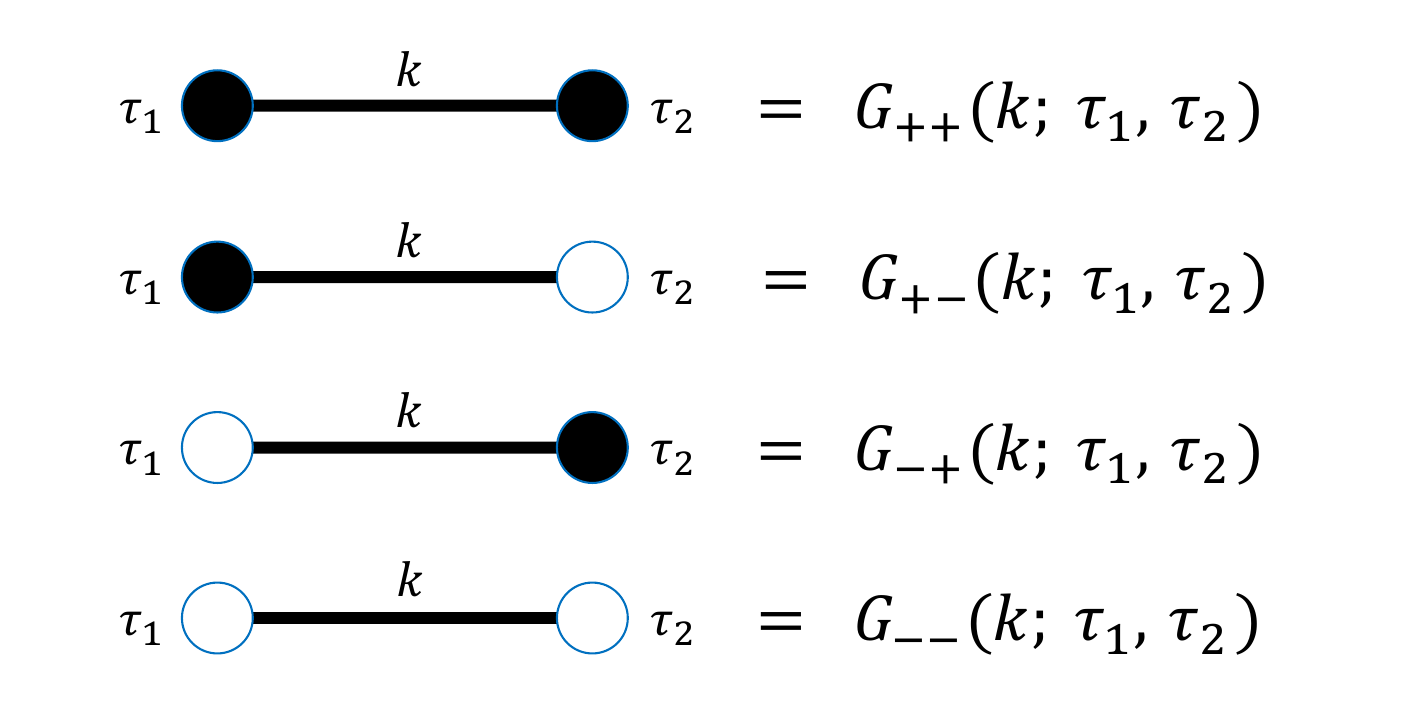}
\caption{\label{figpropagators} 
From top to bottom, we show the propagators corresponding to 
\eqref{G++},\eqref{G+-},\eqref{G-+}, and \eqref{G--}, respectively. 
The black and white dots denote the $+$ and $-$ type $\zeta$ 
fields, respectively.
}
\end{figure}

%
\item Vertices 

Let us consider a general form of the interacting Lagrangian,
\be
{\cal{L}}_{\rm int}(\zeta)=\lambda(\tau){\cal{D}}_1\zeta{\cal{D}}_2
\zeta{\cal{D}}_3\zeta\,,
\ee
where $\lambda$ is a function of $\tau$, 
${\cal{D}}_i$'s are derivative operators including 
the identity operator. 
At first order in $\lambda(\tau)$, the correction to the generating 
functional in Eq.~\eqref{generatingfunctional} from 
the $+$ field is 
\be
\ri\int^{\tau_f}_{-\infty}{\rm d}\tau
\int{\rm d}^3 x\, {\cal{L}}_{\rm int}
\left(\frac{\delta}{\ri\delta J^+(\tau,\bm{x})}\right)Z_0[J^+,J^-]\,.
\ee
By further applying the functional derivative with respect to the remaining 
$J^{\pm}$ and taking the limit $J^\pm\to0$ at the end, we obtain 
the three-point correlation function with (external) legs 
labeled by $A, B, C$ as
\ba
(\text{3-point vertex by the $+$ field}) &=& 
\ri\int^{\tau_f}_{-\infty}{\rm d}\tau\lambda(\tau)\left({\cal{D}}_1 G_{+s_A}(k_A;\tau,\tau_A)\right)\left({\cal{D}}_2 
G_{+s_B}(k_B;\tau,\tau_B)\right)
\left({\cal{D}}_3 G_{+s_C}(k_C;\tau,\tau_C)\right) 
\nonumber \\
& &
+(\text{5 perms})\,,
\label{+vertex}
\ea
where $s_{A,B,C}$ are either $+$ or $-$, and
$k_{A,B,C}$ and $\tau_{A,B,C}$ are the momenta and times 
of each propagator. Note that we have just applied the functional derivatives 
$-\ri s_{i}\delta/\delta J^{s_i}(\bm k_i,\tau_i)$ ($i=A, B, C$) to the residual 
source functions and have taken the limit $J^\pm\to0$, and therefore 
the points $A, B, C$ could be connected to 
each other by propagators. 
In Eq.~(\ref{+vertex}), ``5 perms'' means summing up all 
the possible combinations of 
$(1,2,3)$ and $(A,B,C)$.
We note that the derivative operators ${\cal{D}}_i$ 
act on the propagators.

Similarly, the three-point correlation induced 
by the $-$ field is given by
\ba
(\text{3-point vertex by the $-$ field})
&=& 
-\ri\int^{\tau_f}_{-\infty}{\rm d}\tau\lambda(\tau)\left({\cal{D}}_1 
G_{-s_A}(k_A;\tau,\tau_A)\right)
\left({\cal{D}}_2 G_{-s_B}(k_B;\tau,\tau_B)\right)
\left({\cal{D}}_3 G_{-s_C}(k_C;\tau,\tau_C)\right)
\nonumber \\
&&+(\text{5 perms})\,.
\label{-vertex}
\ea
The Feynman diagrams representing the three-point vertices generated 
from the $+$ and $-$ fields are shown in Fig.~\ref{fig3pointvertex}.
Again, the $+$ and $-$ fields 
are represented by the black and white dots, respectively.

\begin{figure}[h]
\centering
\includegraphics[width=18cm]{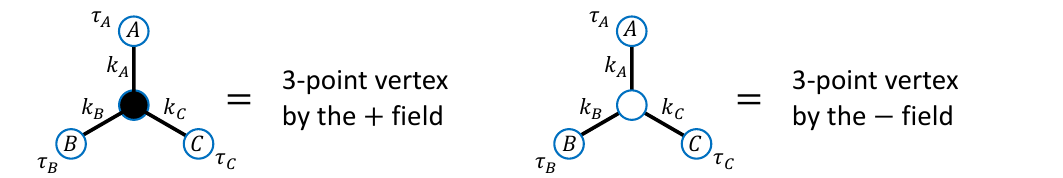}
\caption{\label{fig3pointvertex}
3-point vertices corresponding to 
\eqref{+vertex} (left) and \eqref{-vertex} (right), 
respectively.
}
\end{figure}

\item Feynman rules
\begin{enumerate}
  \item Determine the number of interaction vertices, namely the order 
  of perturbations and the number of external points corresponding to 
  the correlation functions we evaluate. 
  In most cases, we consider the correlation functions of operators at equal time. 
  In such a case, external legs can be either $+$ or $-$ as equal-time correlators 
  would have no distinction between (anti-)time-ordered or non-time-ordered ones.
  \item Draw vertices and here one has to implement both black and white vertices. 
  If a diagram consists of multiple interaction vertices, each vertex 
  can be $+$ or $-$.
  \item Connect all the external points and vertices with propagators.
  Assign the momentum to propagators so that it is conserved at each vertex.
  \item The calculations are carried out according to the rules 
  in Figs.~\ref{figpropagators} and \ref{fig3pointvertex}.
  For the diagrams including loops, we should integrate 
  over the loop momentum.
\end{enumerate}
Note that some of the diagrams can be related to their complex conjugations. 
In particular, when we evaluate real correlation functions, we expect some 
reduction in the number of possible diagrams to compute. 
See e.g.,~Ref.~\cite{Chen:2017ryl}.
\end{itemize}

Before deriving the master formula in the path-integral approach,  
we would like to give a few comments on the differences between 
the path-integral and operator formalisms, which 
was also noted in our previous work~\cite{Kawaguchi:2024lsw}.

The first remark is that, in the path-integral formalism, the operator 
insertion done by a source functional derivative $\delta/(\pm \ri\delta J^\pm)$ yields the expectation value of a Heisenberg picture operator. 
This is because the source function is the one for a Heisenberg picture 
operator but not for an interaction picture one.  
On the other hand, in the operator formalism, one has to be 
careful particularly if one considers the insertion of operators with time derivatives simply because $\zeta_I'(\tau)\neq\zeta_H'(\tau)$, 
where the subscripts $I$ and $H$ denote the interaction picture and 
Heisenberg picture operators, respectively. 
This difference leads to an ``effective Hamiltonian''~\cite{Pimentel:2012tw}. 
In the path-integral formalism, such an effective Hamiltonian 
does not appear. It may seem peculiar, but 
there is another crucial difference, which is discussed below.

In the operator formalism, the interaction terms which are proportional to 
the free EOM should automatically vanish for perturbation 
theory in the interacting picture. 
This is not the case for path-integral formalism since 
the (anti-)time-ordered propagator gives non-vanishing contributions 
to correlation functions. 
Crudely speaking, in the path integral, the derivatives acting 
on operators are multiplied to propagators, for instance, $\langle T\zeta'(\tau_1)\zeta(\tau_2)\rangle_{\text{path-int}}=\partial_{\tau_1}G_{++}(\tau_1,\tau_2)$. On the other hand, 
in operator formalism, one would evaluate the expectation value 
of the operator with a time derivative directly. 
This is why EOM terms vanish in the operator formalism, 
whereas not in the path-integral formalism. 
Therefore, one should not mix the result in each formulation, and 
can compare only the final result. 
Nevertheless, each formalism is self-consistent. 

\subsection{Master formula}
\label{ininpath}

Now, we are ready to show the master formula of 
one-loop corrections under the in-in path-integral formalism.
At one-loop order, only cubic- or quartic-order 
operators contribute to the two-point correlation 
functions. Therefore, we are only concerned with 
the interactions
\be
{\cal{L}}^{(3)}=\sum_{A=1}^{n}\lambda_3^{(A)}(\tau)
\prod_{i=1}^{3}{\cal{D}}_i^{(A)}\zeta\,,\label{cubicgen}
\ee
and
\be
{\cal{L}}^{(4)}=\sum_{A=1}^{m}\lambda_4^{(A)}(\tau)
\prod_{i=1}^{4}{\cal{D}}_i^{(A)}\zeta\,,\label{quarticgen}
\ee
where $A$ represents the label of each interaction, 
$n,m$ denote the total number of different terms 
in the Lagrangian, $\lambda_{3,4}^{(A)}$ are the 
time-dependent coupling constants, and ${\cal{D}}_i^{(A)}$'s are the 
derivative operators acting on the $i$-th $\zeta$ 
of the $A$-th term. 
We also have particular quartic interactions that play important roles 
in proving consistency relations including operators with spatial derivatives~\cite{Senatore:2012wy,Pimentel:2012tw}. 
For instance, when we have cubic interactions
\begin{align}
\mathcal{L}^{(3)}\supset\lambda_3^A\zeta'(\partial\zeta)^2
+\lambda_3^B\zeta (\partial\zeta)^2\,,
\label{cubicex}
\end{align}
where $\lambda_3^{A,B}$ are coupling constants, 
there should be quartic interactions
\begin{align}
\mathcal{L}^{(4)}\supset-2\lambda_4^A\zeta\zeta'(\partial\zeta)^2
-\lambda_4^B\zeta^2(\partial\zeta)^2\,.
\label{quarex}
\end{align}
These quartic terms must appear to realize a rescaling symmetry 
$x^i \to e^{-\delta\zeta} x^i$ and 
$\zeta(\tau,x^i)\to \zeta(\tau,x^i)+\delta\zeta$ 
with a constant $\delta\zeta$, 
which follows from the definition of 
$\zeta$ in the line element 
$\rd s^2=-\rd t^2+a^2 e^{2\zeta} 
\delta_{ij}\rd x^i \rd x^j$.\footnote{The shift of the zero mode of
$\zeta$ is equivalent to rescaling of the scale factor $a(t)$ as well.} 
Therefore, the nonlinear expression of operators reproducing 
the interactions (\ref{cubicex}) and (\ref{quarex}) 
should be of the form 
\begin{align}
\lambda_3^A\zeta'\frac{1}{e^{2\zeta}}(\partial\zeta)^2
-\frac{1}{2e^{2\zeta}}\lambda_3^B(\partial\zeta)^2\,.
\end{align}
We will consider particular quartic interactions like (\ref{quarex}) 
associated with the cubic interactions with spatial derivatives.
The one-loop corrections consist of contractions of the two external legs 
with either two $\mathcal{L}^{(3)}$ or a single 
$\mathcal{L}^{(4)}$, which we will discuss separately. 

\subsubsection{Cubic interactions}

\begin{figure}[h]
\centering
\includegraphics[width=9cm]{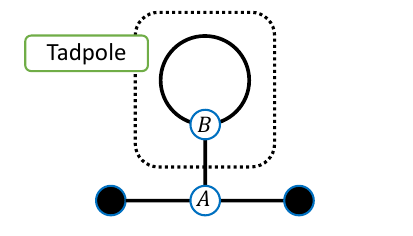}
\caption{\label{fig_tadpole}
Schematic figure of non-1PI three-point correlation 
functions with a tadpole diagram}
\end{figure}

\begin{figure}[h]
\centering
\includegraphics[width=15cm]{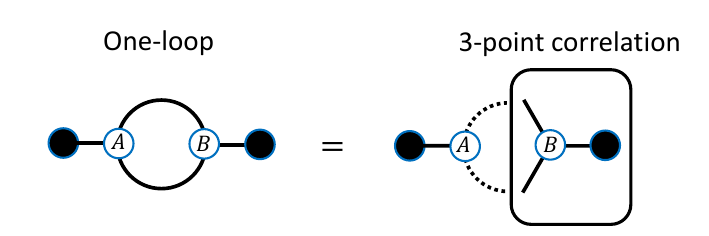}
\caption{\label{figone-loop-cut}
Schematic figure of Eq.~\eqref{one-loop-cut}.
The dotted lines denote uncontracted legs of vertex $a$.
}
\end{figure}

By using the generating functional (\ref{generatingfunctional}) of 
one-particle-irreducible (1PI) diagrams, the one-loop corrections from 
two cubic vertices associated with $\mathcal{L}^{(3)}$ 
can be generally written as
\ba
&&
\ev{\zeta(\tau_*,\bm{x}_1)\zeta(\tau_*,\bm{x}_2)}_{\text{1-loop,1PI}}^{(3)}=\left(\frac{\delta}{\ri\delta J^+(\tau_*,\bm{x}_1)}\right)\left(\frac{\delta}{\ri\delta J^+(\tau_*,\bm{x}_2)}\right)Z[J^+,J^-]\biggr|_{J^\pm=0}^{\text{1-loop,1PI}}
\notag\\
&&
\hspace{0.2cm}
=\frac{1}{2}\left(\frac{\delta}{\ri\delta J^+(\tau_*,\bm{x}_1)}\right)\left(\frac{\delta}{\ri\delta J^+(\tau_*,\bm{x}_2)}\right)
\left(\ri\int^{\tau_f}_{-\infty}{\rm d}\tau\int{\rm d}^3 x
\left[{\cal{L}}^{(3)}
\left(\frac{\delta}{\ri\delta J^+(\tau,\bm{x})}\right)-{\cal{L}}^{(3)} \left(\frac{\delta}{-\ri\delta J^-(\tau,\bm{x})}\right)\right]\right)^2 \nonumber \\
&&
\hspace{0.5cm}
\times Z_0[J^+,J^-]\biggr|_{J^\pm=0}^{\text{1-loop,1PI}}
\notag\\
&&
\hspace{0.2cm}
=\frac{1}{2}\left(\frac{\delta}{\ri\delta J^+(\tau_*,\bm{x}_1)}\right)\uwave{\left(\frac{\delta}{\ri\delta J^+(\tau_*,\bm{x}_2)}\right)}\ri\int^{\tau_f}_{-\infty}{\rm d}\tau\int{\rm d}^3 x\left[\sum_{A=1}^{n}\lambda_3^{(A)}\left(\prod_{i=1}^3\uwave{{\cal{D}}_i^{(A)}\frac{\delta}{\ri\delta J^+(\tau,\bm{x})}}-\prod_{i=1}^3\uwave{{\cal{D}}_i^{(A)}\frac{\delta}{-\ri\delta J^-(\tau,\bm{x})}}\right)\right]
\notag\\
&&
\hspace{0.5cm}
\times Z^{(1)} [J^+,J^-]\biggr|_{J^\pm=0}^{\text{1-loop,1PI}}\,,
\ea
where
\begin{align}
Z^{(1)}[J^+,J^-]=\ri\int^{\tau_f}_{-\infty}{\rm d}\tau'\int{\rm d}^3 y\left[{\cal{L}}^{(3)}\left(\frac{\delta}{\ri\delta J^+(\tau',\bm{y})}\right)-{\cal{L}}^{(3)}\left(\frac{\delta}{-\ri\delta J^-(\tau',\bm{y})}\right)\right]
Z_0[J^+,J^-]\,.
\end{align}
Here, $Z^{(1)}[J^+, J^-]$ is the generating functional with the first-order correction by cubic interactions. Notice also that a functional derivative of an external leg $\bm x_2$ and one of the functional derivatives in cubic interactions at $\bm x$ (indicated by wavy underlines) yield propagators connecting $\bm x$ and $\bm x_2$, and therefore
\ba
\hspace{-1.0cm}
&&\ev{\zeta(\tau_*,\bm{x}_1)\zeta(\tau_*,\bm{x}_2)}_{\text{1-loop,1PI}}^{(3)}
\notag\\
\hspace{-1.0cm}
&&
=\frac{1}{2}\dotuline{\uuline{\left(\frac{\delta}{\ri\delta J^+(\tau_*,\bm{x}_1)}\right)}}\ri\int^{\tau_f}_{-\infty}{\rm d}\tau\int{\rm d}^3 x\Biggl[\sum_{A=1}^{n}\lambda_3^{(A)}\sum_{i=1}^3\Biggl({\cal{D}}_i^{(A)}(-\ri\Delta_{++}(\tau,\bm{x};\tau_*,\bm{x}_2))\uuline{\prod_{j\ne i}{\cal{D}}_j^{(A)}\frac{\delta}{\ri\delta J^+(\tau,\bm{x})}}
\notag
\\
\hspace{-1.0cm}
&&\qquad\qquad\qquad\qquad\qquad\qquad\qquad\qquad
-{\cal{D}}_i^{(A)}(-\ri\Delta_{-+}(\tau,\bm{x};\tau_*,\bm{x}_2))\dotuline{\prod_{j\ne i}{\cal{D}}_j^{(A)}\frac{\delta}{-\ri\delta J^-(\tau,\bm{x})}}\Biggr)\Biggr]\dotuline{\uuline{Z^{(1)}[J^+,J^-]\Bigr|_{J^\pm=0}}}.
\ea
Now, we further notice that the remaining functional 
derivatives acting on $Z^{(1)}[J^{+}, J^{-}]$ bring us 
the expectation values of tree-level three-point functions, 
which are indicated by the double or dotted underlines. 
Thus, we obtain the final expression
\ba
&&\ev{\zeta(\tau_*,\bm{x}_1)\zeta(\tau_*,\bm{x}_2)}_{\text{1-loop,1PI}}^{(3)}
\notag\\
&&
\hspace{0.2cm}
=\ri\int^{\tau_f}_{-\infty}{\rm d}\tau\int{\rm d}^3 x\Biggl[\sum_{A=1}^{n}\lambda_3^{(A)}\sum_{i=1}^3\Biggl({\cal{D}}_i^{(A)}(-\ri\Delta_{++}(\tau,\bm{x};\tau_*,\bm{x}_2))\uuline{\ev{\zeta^+(\tau_*,\bm{x}_1)\prod_{j\ne i}{\cal{D}}_j^{(A)}\zeta^+(\tau,\bm{x})}_{\rm tree}}
\notag
\\
&&
\hspace{3.0cm}
-{\cal{D}}_i^{(A)}(-\ri\Delta_{-+}(\tau,\bm{x};\tau_*,\bm{x}_2))\dotuline
{\ev{\prod_{j\ne i}{\cal{D}}_j^{(A)}\zeta^-(\tau,\bm{x})\zeta^+(\tau_*,\bm{x}_1)}_{\rm tree}}\Biggr)\Biggr]\,,
\label{one-loop-cut}
\ea
where $\ev{}_{\rm tree}$ indicates that the three-point functions are 
evaluated at tree level (namely, at first order in cubic interactions), 
and no loop diagram forms. In the above discussion, we have put aside the 
non-1PI contribution such as
\begin{align}
\ev{\zeta^+(\tau_*,\bm{x}_1)\prod_{j\ne i}{\cal{D}}_j^{(A)}\zeta^+(\tau,\bm{x})} \supset {\cal D}^{(A)}_{j_1}\Delta_{++}(\tau_*,\bm x_1;\tau,\bm x)\mathcal{D}^{(A)}_{j_2}\ev{\zeta^+(\tau,\bm x)}_{\rm 1st}\,,
\end{align}
where 
$j_1,j_2\neq i$, $\ev{\zeta^+(\tau,\bm x)}_{\rm 1st}$ denotes the (one-loop) tadpole terms arising from the first-order correction with cubic interactions. 
The non-1PI diagram refers to the one shown in Fig.~\ref{fig_tadpole}. The contributions from non-1PI diagrams contain UV divergences which have to be regulated in a covariant way as discussed in Ref.~\cite{Pimentel:2012tw}. 
Thus, all the one-loop diagrams from cubic interactions are either 
the 1PI ones or the non-1PI ones, and we concentrate on 
the 1PI contributions in the following.
The formula~\eqref{one-loop-cut} can be diagrammatically 
understood as Fig.~\ref{figone-loop-cut}.


The Fourier transformation of the above formula is 
\ba
&&
P_\zeta^{(\text{1-loop})}(\tau_*,k_1) \equiv
\llangle\tilde{\zeta}(\tau_*,\bm{k}_1)
\tilde{\zeta}(\tau_*,-\bm{k}_1)
\rrangle_{\text{1-loop,1PI}}^{(3)}
\notag\\
&&
=\frac{\ri}{2} \int^{\tau_f}_{-\infty}{\rm d}\tau\Biggl[\sum_{A=1}^{n}\lambda_3^{(A)}
\sum_{(i_1,i_2,i_3)}
\Biggr({\cal{D}}_{i_1}^{(A)}G_{++}(k_1;\tau,\tau_*)
\int\frac{{\rm d}^3 k}{(2\pi)^3}\llangle
\tilde{\zeta}^+(\tau_*,\bm{k}_1){\cal{D}}_{i_2}^{(A)}\tilde{\zeta}^+(\tau,\bm{k}){\cal{D}}_{i_3}^{(A)}\tilde{\zeta}^+(\tau,-\bm{k}_1-\bm{k})\rrangle_{{\rm tree}}
\notag\\
&&
\hspace{2.0cm}
-{\cal{D}}_{i_1}^{(A)}G_{-+}(k_1;\tau,\tau_*)\int\frac{{\rm d}^3 k}{(2\pi)^3}\llangle{\cal{D}}_{i_2}^{(A)}\tilde{\zeta}^-(\tau,\bm{k}){\cal{D}}_{i_3}^{(A)}\tilde{\zeta}^-(\tau,-\bm{k}_1-\bm{k})\tilde{\zeta}^+(\tau_*,\bm{k}_1)\rrangle_{{\rm tree}}\Biggr)\Biggr]
\notag\\
&&
=\frac{\ri}{2} \int^{\tau_*}_{-\infty}{\rm d}\tau\Biggl[\sum_{A=1}^{n}\lambda_3^{(A)}
\sum_{(i_1,i_2,i_3)}\Biggr({\cal{D}}_{i_1}^{(A)}G_{++}(k_1;\tau,\tau_*)\int\frac{{\rm d}^3 k}{(2\pi)^3}\llangle
\tilde{\zeta}^+(\tau_*,\bm{k}_1){\cal{D}}_{i_2}^{(A)}\tilde{\zeta}^+(\tau,\bm{k}){\cal{D}}_{i_3}^{(A)}\tilde{\zeta}^+(\tau,-\bm{k}_1-\bm{k})\rrangle_{{\rm tree}}
\notag\\
&&
\hspace{2.0cm}
-{\cal{D}}_{i_1}^{(A)}G_{-+}(k_1;\tau,\tau_*)\int\frac{{\rm d}^3 k}{(2\pi)^3}\llangle{\cal{D}}_{i_2}^{(A)}\tilde{\zeta}^-(\tau,\bm{k}){\cal{D}}_{i_3}^{(A)}\tilde{\zeta}^-(\tau,-\bm{k}_1-\bm{k})\tilde{\zeta}^+(\tau_*,\bm{k}_1)\rrangle_{{\rm tree}}\Biggr)\Biggr],
\label{one-loop-cut-Fourier}
\ea
where $\llangle  \rrangle$ is the expectation value without 
the momentum $\delta$-function, i.e., 
\be
\ev{X(\bm{k}_1)X(\bm{k}_2)\cdots}=(2\pi)^3\delta^{(3)}(\bm{k}_1+\bm{k}_2+\cdots)\llangle X(\bm{k}_1)X(\bm{k}_2)\cdots \rrangle\,,
\ee
and $(i_1,i_2,i_3)$ runs over all the possible permutations of $(1,2,3)$.
In the second equality of Eq.~(\ref{one-loop-cut-Fourier}), we have changed the upper limit of $\tau$-integration to be $\tau_*$. 
In Appendix~\ref{proofid}, we will prove this equality.\footnote{Technically speaking, the upper limit of $\tau$-integration is infinitesimally greater than $\tau_*$ 
such that the possible contribution from $\delta$-function can be included. 
See Apppendix~\ref{proofid}.}

We discuss whether the CMB modes that are supposed to exit the Hubble 
radius well before the USR transition can be amplified during the USR phase. 
We formally separate the $\bm k$-integration as 
\be
\int \frac{\rd^3 k}{(2\pi)^3}=\int_{k>k_{\rm min}}
\frac{\rd^3 k}{(2\pi)^3}+\int_{k<k_{\rm min}}\frac{\rd^3k}{(2\pi)^3}\,.
\label{kintdiv}
\ee
We choose $k_{\rm min}$ to be sufficiently larger than $k_1$ 
but much smaller than the momentum that exits the Hubble horizon 
at the beginning of the USR phase, i.e.,
\be
k_1\ll k_{\rm min}\ll k_s=-\tau_s^{-1}\,.
\label{krange}
\ee
Then, we can take the squeezing limit within the loop momentum, 
so that the consistency relations hold as will be discussed later. 
The second integral on the right-hand side of Eq.~(\ref{kintdiv}) does
not give significant contributions. Therefore, only the first integral 
would give possibly large contributions to the power spectrum, 
and hereafter, we consider the momentum integration 
in the range $k>k_{\rm min}$.

We also separate the time-integration region as 
\be
\int_{-\infty}^{\tau_0} \rd \tau
+\int_{\tau_0}^{\tau_*}\rd \tau\,,
\label{intsum}
\ee
where $\tau_0$ is an arbitrary time before the USR transition but after 
the $k_1$ mode exits the Hubble horizon, and we will assume 
it to be the time slightly before the USR period starts, $\tau_0\simeq \tau_s$. 
Since the time region $-\infty<\tau<\tau_0$ is before the onset 
of the USR, there should be no large enhancement effects. 
Thus, we are only concerned with the second integral in Eq.~(\ref{intsum}) hereafter. 

From the above argument for the ranges of integrals, the 1PI one-loop 
corrections to the power spectrum associated with cubic vertices 
can be summarized as
\ba
&&
P_\zeta^{(\text{1-loop})}(\tau_*,k_1)
\notag\\
&&
\simeq \frac{\ri}{2} \int^{\tau_*}_{\tau_0}{\rm d}\tau\Biggl[\sum_{A=1}^{n}\lambda_3^{(A)}
\sum_{(i_1,i_2,i_3)}\Biggr({\cal{D}}_{i_1}^{(A)}G_{++}(k_1;\tau,\tau_*)\int_{k>k_{\rm min}}\frac{{\rm d}^3 k}{(2\pi)^3}\llangle
\tilde{\zeta}^+(\tau_*,\bm{k}_1){\cal{D}}_{i_2}^{(A)}\tilde{\zeta}^+(\tau,\bm{k}){\cal{D}}_{i_3}^{(A)}\tilde{\zeta}^+(\tau,-\bm{k}_1-\bm{k})\rrangle_{\rm tree}
\notag\\
&&
\hspace{2.0cm}
-{\cal{D}}_{i_1}^{(A)}G_{-+}(k_1;\tau,\tau_*)\int_{k>k_{\rm min}}\frac{{\rm d}^3 k}{(2\pi)^3}\llangle{\cal{D}}_{i_2}^{(A)}\tilde{\zeta}^-(\tau,\bm{k}){\cal{D}}_{i_3}^{(A)}\tilde{\zeta}^-(\tau,-\bm{k}_1-\bm{k})\tilde{\zeta}^+(\tau_*,\bm{k}_1)\rrangle_{\rm tree}\Biggr)\Biggr].\label{premaster}
\ea

Let us consider another crucial property of $\zeta$, 
consistency relations. If they hold, this implies that, 
for sufficiently small $k_1$, the three-point function 
is approximated as
\begin{align}
\llangle
\tilde{\zeta}(\tau_*,\bm{k}_1){\cal{D}}_{i_2}^{(A)}\tilde{\zeta}(\tau,\bm{k}){\cal{D}}_{i_3}^{(A)}\tilde{\zeta}(\tau,-\bm{k}_1-\bm{k})\rrangle_{\rm tree}\underset{k_1\to 0}{\longrightarrow}
-P_\zeta^{(\rm tree)}(\tau_*,k_1)\frac{1}{k^3}\frac{\rd \left(k^3\llangle{\cal{D}}_j^{(A)}\tilde{\zeta}(\tau,\bm{k})
{\cal{D}}_k^{(A)}\tilde{\zeta}(\tau,-\bm{k})\rrangle\right)}
{\rd \log k},\label{MCRlikerelations}
\end{align}
where $P_\zeta^{(\rm tree)}(\tau_*,k_1)$ 
is the tree-level power spectrum defined by 
\be
P_\zeta^{(\rm tree)}(\tau_*,k_1) \equiv 
\llangle\tilde{\zeta}(\tau_*,\bm k_1)
\tilde{\zeta}(\tau_*,-\bm k_1)\rrangle_{\rm tree}\,,\label{Pzetadef}
\ee
which differs from the dimensionless one by 
a factor $k_1^3/(2\pi^2)$.  
Recall that we did not introduce the same factor 
for $P^{(\text{1-loop})}_\zeta$ in Eq.~\eqref{one-loop-cut-Fourier} either. 
Note that Eq.~(\ref{MCRlikerelations}) corresponds to 
the leading expression in $k_1$. Here, we have assumed 
the unusual forms of consistency relations since the 
super-Hubble mode $\tilde{\zeta}(\tau_*,\bm k_1)$ corresponds to 
the time $\tau_*$ different from $\tau$ appearing in the other operators. 
Nevertheless, the effective constancy of $\tilde{\zeta}$ discussed in Sec.~\ref{effconst} justifies this since it just follows from the property of the tree-level mode function. Furthermore, we have assumed that the consistency relations of three-point functions consisting 
of the $+$ and $-$ fields yield the same result. 
This is rather nontrivial, but we will show it 
within our reduced system later.

What is the consequence of consistency relations? 
To perform the momentum integration properly, 
we use dimensional regularization, which results in\footnote{Strictly speaking, we need to include the corrections from quartic interactions. For now, we assume such contributions are implicitly included.}
\ba
\hspace{-0.3cm}
&&
\frac{P_\zeta^{(\text{1-loop})}(\tau_*,k_1)}
{P_\zeta^{(\rm tree)}(\tau_*,k_1)}
\notag\\
\hspace{-0.3cm}
&&
\simeq 
-\frac{\ri}{2} \int^{\tau_*}_{\tau_0}{\rm d}\tau \left[\sum_{s=\pm}s\sum_{A=1}^{n}\lambda_3^{(A)}\sum_{(i_1,i_2,i_3)}{\cal{D}}_{i_1}^{(A)}G_{s+}(k_1;\tau,\tau_*)\int^{k_{\rm max}}_{k_{\rm min}}\frac{{\rm d}^{p}k}{(2\pi)^{p}}\frac{1}{k^{p}}\frac{{\rm d}\left(\mu^{-\delta}k^{p}
\llangle {\cal{D}}_{i_2}^{(A)}\tilde{\zeta}^{s}(\tau,\bm{k}){\cal{D}}_{i_3}^{(A)}\tilde{\zeta}^{s}(\tau,-\bm{k})\rrangle\right)}{{\rm d}\log k}
\right]
\notag\\
\hspace{-0.3cm}
&&=-\ri \int^{\tau_*}_{\tau_0}{\rm d}\tau\Biggl[\sum_{s=\pm}s\sum_{A=1}^{n}\lambda_3^{(A)}\sum_{(i_1,i_2,i_3)}{\cal{D}}_{i_1}^{(A)}G_{s+}(k_1;\tau,\tau_*)\frac{1}
{(4\pi)^{p/2}\Gamma\left(p/2 \right)}\left(\mu^{-\delta}k^{p}
\llangle {\cal{D}}_{i_2}^{(A)}\tilde{\zeta}^s(\tau,\bm{k}){\cal{D}}_{i_3}^{(A)}\tilde{\zeta}^s(\tau,-\bm{k})\rrangle\right)\Biggr]^{k_{\rm max}}_{k_{\rm min}},\nonumber\\
\label{ratio-one-loop-tree}
\ea
where $p \equiv 3+\delta$, 
$\mu$ is the renormalization scale introduced to keep 
the correct dimensionality, $s$ denotes the sign of the fields, 
and we used $P_\zeta^{(\rm tree)}(\tau_*,k_1)
=\llangle \tilde{\zeta}^+(\tau_*,\bm k_1)\tilde{\zeta}^-(\tau_*,-\bm k_1)\rrangle_{\rm tree}=\llangle \tilde{\zeta}^+(\tau_*,\bm k_1)\tilde{\zeta}^+(\tau_*,-\bm k_1)\rrangle_{\rm tree}$. 
In the last equality, we dropped the contribution from $k_{\rm max}$, 
which does not contribute when we apply dimensional regularization. 
In Ref.~\cite{Tada:2023rgp}, it was argued that the 
$\ri \varepsilon$ prescription 
cuts off the contribution from UV modes, and such an argument was 
criticized in Ref.~\cite{Firouzjahi:2023bkt}.\footnote{Indeed, 
the argument in Ref.~\cite{Tada:2023rgp} cannot be applied if 
we consider a manifestly unitary $\ri \varepsilon$-prescription 
suggested in Ref.~\cite{Baumgart:2020oby},
where the interacting Hamiltonian 
is modified as $H_I(t)\to H_I(t)e^{\varepsilon t}$ 
such that it is turned off in the asymptotic past 
$t\to-\infty$, while the time evolution operator 
$U_{\varepsilon} = T\exp(-\ri\int^t_{-\infty} 
\rd t'H_I(t')e^{\varepsilon t'})$ is manifestly unitary.} 
Nevertheless, the claim that $k_{\rm max}\to\infty$ does not 
give contributions is correct for the following reasons: We define the momentum integration in spatial dimension $3+\delta$ and take ${\rm Re}\ \delta$ to be negative such that $k_{\rm max}\to\infty$ is convergent (under which the dependence on 
$k_{\rm max}$ disappears). 
Then, we analytically continue the result to the limit 
$\delta\to 0$. Hence only logarithmic divergence 
may generally appear as the terms are proportional 
to the negative powers of $\delta$.\footnote{An interesting discussion about the relation between dimensional regularization and cut-off regularization can be found e.g., 
in Ref.~\cite{Branchina:2022jqc}, which would clarify 
the above comments regarding the analytic continuation.} 
Thus, when Eq.~\eqref{MCRlikerelations} holds, 
the 1PI one-loop corrections to the power spectrum are summarized as \eqref{ratio-one-loop-tree}. As anticipated from the above expression, we can readily show that the 1PI one-loop corrections are small assuming that Eq.~\eqref{ratio-one-loop-tree} holds.

One may wonder whether the limit $\delta \to0$ 
leads to singular terms, namely, whether logarithmic 
divergence exists or not.
The answer is no, and there are no $1/\delta$ poles. 
For the UV divergent cases, $1/\delta$ is associated with 
the $\Gamma$ function with a negative integer argument that appears when we perform momentum integration within dimensional regularization. However, unlike the usual case, the momentum integration is 
performed as that of total derivative terms and therefore, 
there is no room for the $\Gamma$ function of a negative integer 
to appear in the limit $\delta\to0$. 
The mode function behaves continuously for $\delta\to 0$, 
which may be understood as follows: The dimensional regularization 
leads to $\nu \simeq (3+\eta)/2 \to (3+\delta+\eta)/2$ 
in Eq.~\eqref{defnu}.
Therefore, it only leads to 
a small shift of the second slow-roll parameter $\eta$. 
Some powers of the scale factor and momentum can also change, 
but such factors do not give any $1/\delta$ singularity. 
Thus, as it is consistent with Ref.~\cite{Pimentel:2012tw}, 
we can safely neglect the UV divergences of 1PI one-loop corrections. 
Nevertheless, we will give a few comments on the UV divergences 
in Appendix~\ref{renormalization}. 

\subsubsection{Quartic interactions}

Let us also consider one-loop corrections arising from 
quartic interactions of the form \eqref{quarticgen}. 
They are given by 
\begin{align}
&\ev{\zeta(\tau_*,\bm{x}_1)\zeta(\tau_*,\bm{x}_2)}_{\text{1-loop,1PI}}^{(4)}\nonumber\\
&=\left(\frac{\delta}{\ri\delta J^+(\tau_*,\bm{x}_1)}\right)\left(\frac{\delta}
{\ri\delta J^+(\tau_*,\bm{x}_2)}\right)
\left(\ri\int^{\tau_*}_{\tau_0}{\rm d}\tau\int{\rm d}^3 x\left[{\cal{L}}^{(4)}\left(\frac{\delta}{\ri\delta J^+(\tau,\bm{x})}\right)-{\cal{L}}^{(4)}\left(\frac{\delta}{-\ri\delta J^-(\tau,\bm{x})}\right)\right]\right)
Z_0[J^+,J^-]\biggr|_{J^\pm=0}\nonumber\\
&=\left(\frac{\delta}{\ri\delta J^+(\tau_*,\bm{x}_1)}\right)\ri\int^{\tau_*}_{\tau_0} 
\rd \tau\int \rd^3x\Biggl[\sum_{A=1}^{m}\lambda_4^{(A)}
\sum_{i=1}^4\Biggl({\cal{D}}_i^{(A)}(-\ri\Delta_{++}(\tau,\bm{x};\tau_*,\bm{x}_2))\prod_{j\ne i}{\cal{D}}_j^{(A)}\frac{\delta}{\ri\delta J^+(\tau,\bm{x})}\nonumber\\
&\hspace{5.4cm}-{\cal{D}}_i^{(A)}(-\ri\Delta_{-+}(\tau,\bm{x};\tau_*,\bm{x}_2))\prod_{j\ne i}{\cal{D}}_j^{(A)}\frac{\delta}{-\ri\delta J^-(\tau,\bm{x})}\Biggr)\Biggr]Z_0[J^+,J^-]
\biggr|_{J^\pm=0}\nonumber\\
&=\ri\int^{\tau_*}_{\tau_0} \rd \tau\int \rd^3x
\Biggl[\sum_{A=1}^{m}\lambda_4^{(A)}\sum_{i=1}^4\Biggl({\cal{D}}_i^{(A)}(-\ri\Delta_{++}(\tau,\bm{x};\tau_*,\bm{x}_2))\ev{\zeta^+(\bm x_1,\tau_*)\prod_{i\neq j}{\cal D}_j^{(A)}\zeta^+(\bm x,\tau)}\nonumber\\
&\hspace{2.9cm}-{\cal{D}}_i^{(A)}(-\ri\Delta_{-+}(\tau,\bm{x};\tau_*,\bm{x}_2))\ev{\zeta^+(\bm x_1,\tau_*)\prod_{i\neq j}{\cal D}_j^{(A)}\zeta^-(\bm x,\tau)}\Biggr)\Biggr]\,\nonumber\\
&\simeq\frac{\ri}{2}\int^{\tau_*}_{\tau_0} 
\rd \tau\int \rd^3x 
\nonumber \\
&~~~\times
\Biggl[\sum_{A=1}^{m}\lambda_4^{(A)}\sum_{(i_1,i_2,i_3,i_4)}\Biggl({\cal{D}}_{i_1}^{(A)}(-\ri\Delta_{++}(\tau,\bm{x};\tau_*,\bm{x}_2))\ev{\zeta^+(\bm x_1,\tau_*){\cal D}_{i_2}^{(A)}\zeta^+(\bm x,\tau)}\ev{{\cal D}_{i_3}^{(A)}\zeta^+(\bm x,\tau){\cal D}_{i_4}^{(A)}\zeta^+(\bm x,\tau)}\nonumber\\
&\hspace{0.9cm}-{\cal{D}}_{i_1}^{(A)}(-\ri\Delta_{-+}(\tau,\bm{x};\tau_*,\bm{x}_2))\ev{\zeta^+(\bm x_1,\tau_*){\cal D}_{i_2}^{(A)}\zeta^-(\bm x,\tau)}\ev{{\cal D}_{i_3}^{(A)}\zeta^-(\bm x,\tau){\cal D}_{i_4}^{(A)}\zeta^-(\bm x,\tau)}\Biggr)\Biggr]\,,
\end{align}
where the overall factor $1/2$ is a symmetric factor, 
$(i_1,i_2,i_3,i_4)$ runs over all the possible permutations 
of $(1,2,3,4)$, and, in the last equality, we used the fact that 
the four-point functions should be the products of the two-point function at one-loop order. Note that we understand the two-point function of the same spacetime points will be defined in a way 
that it is regulated.
The Fourier transform of the above contribution 
(with dimensional regularization) is
\ba
&&
\llangle\tilde{\zeta}(\tau_*,\bm{k}_1)\tilde{\zeta}(\tau_*,-\bm{k}_1)\rrangle^{(4)}_{\text{1-loop,1PI}}
\notag\\
&&
\hspace{0.2cm}
\simeq \frac{\ri}{2} 
\int^{\tau_*}_{\tau_0}{\rm d}\tau \nonumber \\
&& \qquad \times
\Biggl[\sum_{A=1}^{m}\lambda_4^{(A)}\int\frac{\mu^{-\delta}{\rm d}^{p}k}{(2\pi)^{p}}
\sum_{(i_1,i_2,i_3,i_4)}\Biggl({\cal{D}}_{i_1}^{(A)}G_{++}(k_1;\tau,\tau_*)\llangle \tilde{\zeta}^+(\tau_*,\bm{k}_1){\cal{D}}_{i_2}^{(A)}\tilde{\zeta}^+(\tau,-\bm{k}_1)\rrangle\llangle{\cal{D}}_{i_3}^{(A)}\tilde{\zeta}^+(\tau,\bm{k}){\cal{D}}_{i_4}^{(A)}\tilde{\zeta}^+(\tau,-\bm{k})\rrangle
\notag\\
&&
\hspace{0.9cm}
-{\cal{D}}_{i_1}^{(A)}G_{-+}(k_1;\tau,\tau_*)\llangle
{\cal{D}}_{i_2}^{(A)}\tilde{\zeta}^-(\tau,-\bm{k}_1)\tilde{\zeta}^+(\tau_*,\bm{k}_1)\rrangle\llangle{\cal{D}}_{i_3}^{(A)}\tilde{\zeta}^-(\tau,\bm{k}){\cal{D}}_{i_4}^{(A)}\tilde{\zeta}^-(\tau,-\bm{k})\rrangle\Biggr)\Biggr]\,.
\label{1PIquarticloop}
\ea
Since we are interested in the contributions from 
short-wavelength modes ${\bm k}$ experiencing the USR phase inside 
the Hubble horizon to the power spectrum of a long-wavelength 
mode ${\bm k}_1$ associated with the CMB, 
we will further take the squeezing limit $k_1 \ll k$ later.
As shown later, such squeezing limit results in the contributions 
that restore consistency relations of 
operators with spatial derivatives.

\section{Consistency relation of 
$\langle\zeta\zeta\zeta\rangle$ in the transient USR model}
\label{AppMaldacena}

In the previous section, we showed that, if the consistency relations~\eqref{MCRlikerelations} and the effective constancy 
of super-Hubble perturbations hold, we can exploit the very 
simple formula~\eqref{ratio-one-loop-tree} to evaluate one-loop correction to the power spectrum of super-Hubble 
modes at leading order. 
If we want to prove the consistency relations of all the 
three-point functions that appear in interactions, it is rather cumbersome since the number of cubic interactions is not a few. Therefore, 
we consider a method to find a self-consistent set of interactions 
that keeps the symmetry of the system, namely, 
that shows the correct consistency relations.

Before going into details of the consistency relation, we give a few comments regarding the consistency relations and the presence of the USR phase. It is known that, in the USR phase being non-attractor inflation, the constancy of the scalar curvature perturbation can be violated 
by a growing mode even after the 
Hubble-radius crossing. 
This fact often causes some confusion that the consistency relation may not hold in the models including the USR phase. 
However, this is not the case. We are concerned with the long-wavelength modes that exit the 
Hubble radius during the slow-roll phase with 
an inflationary attractor. 
The growing solution of such super-Hubble modes is significantly suppressed as we have quantitatively shown in Sec.~\ref{effconst}, and accordingly, the violation of the consistency relation due to the growing mode should also be suppressed.  Therefore, the long-wavelength modes of our interest satisfy the consistency relations.

We will construct a reduced Lagrangian by 
the following strategy: First, we compute the three-point function 
\be
\llangle\tilde{\zeta}^3\rrangle \equiv
\llangle \tilde{\zeta}^+(\tau_*,\bm k_1)\tilde{\zeta}^+(\tau_*,\bm k_2)\tilde{\zeta}^+(\tau_*,\bm k_3)\rrangle\,,
\ee
with all the cubic interactions in the original action. 
It will turn out that at leading order of the squeezing limit 
and slow-roll limit $\epsilon \to 0$ (while $\eta\neq 0$), 
only a few interactions give the leading contributions, 
and we will identify them as leading interactions 
in the transient USR model. 
Then, we will find a reduced Lagrangian.

In this section, as a first step to find the reduced Lagrangian, we explicitly compute the three-point function $\llangle\tilde{\zeta}^3\rrangle$
and look for the set of interaction terms that are necessary to show the consistency relations for arbitrary wavenumbers 
$\bm k_2\sim -\bm k_3$ in the squeezing limit $|\bm k_1|\to 0$.

Before entering the detailed computations, we give a few technical remarks. One is that when a correlation function is evaluated 
at the same moment, choosing either $+$ or $-$ fields 
as external fields does not give differences. 
Therefore, we take all the external fields to be the 
$+$ field for simplicity. Another is that the boundary terms in the interaction Lagrangian give no contribution under our prescription 
given in Ref.~\cite{Kawaguchi:2024lsw} and 
in Appendix~\ref{Bterminpathintegral}. 
Therefore, we are only concerned with the bulk and EOM terms.

\begin{figure}[h]
\centering
\includegraphics[width=19cm]{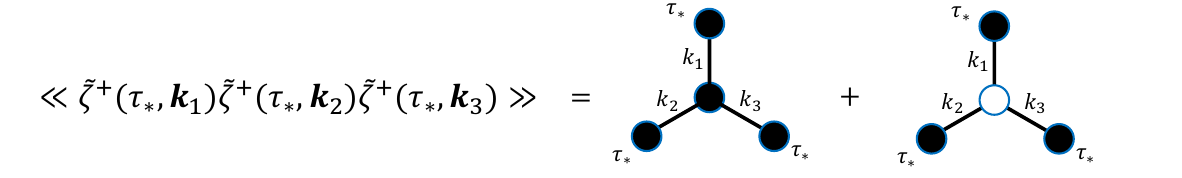}
\caption{\label{figbispectrum}
The diagram corresponding to the bispectrum 
$\llangle \tilde{\zeta}^+(\tau_*,\bm k_1)\tilde{\zeta}^+(\tau_*,\bm k_2)\tilde{\zeta}^+(\tau_*,\bm k_3)\rrangle$.
}
\end{figure}

We consider a single scalar-field theory given 
by the action (\ref{action}) and compute the three-point correlation 
function of $\zeta$ for the diagram illustrated 
in Fig.~\ref{figbispectrum}. 
As we showed in Eq.~(\ref{S3JY}), the cubic-order action of $\zeta$ 
consists of the five bulk contributions, EOM terms, and boundary terms. 
Since the boundary terms do not contribute to  
the three-point function in the in-in path-integral formalism, 
we will compute $\llangle\tilde{\zeta}^3\rrangle$ 
arising from each of the bulk and EOM terms.
In the following, we will denote the contributions to the 
three-point function from the $+$ and $-$ vertices, as 
$\llangle\tilde{\zeta}^3\rrangle_{+}$ and 
$\llangle\tilde{\zeta}^3\rrangle_{-}$, respectively. 

\vspace{0.2cm}

The contributions arising from each of the bulk terms 
in Eq.~(\ref{S3JY}) are given by 
\begin{itemize}
\item ${\cal L}_{1,{\rm bulk}}=\Mpl^2a^2\epsilon^2\zeta\zeta'^2$
\ba
&&
\llangle\tilde{\zeta}^3\rrangle_+=2\ri\Mpl^2\Biggl(\int^{\tau_f}_{\tau_*}{\rm d}\tau a^2\epsilon^2\zeta'_{k_1}(\tau)\zeta'_{k_2}(\tau)\zeta_{k_3}(\tau)\bar{\zeta}_{k_1}(\tau_*)\bar{\zeta}_{k_2}(\tau_*)\bar{\zeta}_{k_3}(\tau_*)
\notag
\\
&&
\hspace{2.5cm}
+\int^{\tau_*}_{-\infty}{\rm d}\tau a^2\epsilon^2\bar{\zeta}'_{k_1}(\tau)\bar{\zeta}'_{k_2}(\tau)\bar{\zeta}_{k_3}(\tau){\zeta}_{k_1}(\tau_*){\zeta}_{k_2}(\tau_*){\zeta}_{k_3}(\tau_*)\Biggr)
+(\text {2 perms})\,,
\\
&&
\llangle\tilde{\zeta}^3\rrangle_-=-2\ri\Mpl^2\Biggl(\int^{\tau_f}_{-\infty}{\rm d}\tau a^2\epsilon^2\zeta'_{k_1}(\tau)\zeta'_{k_2}(\tau)\zeta_{k_3}(\tau)\bar{\zeta}_{k_1}(\tau_*)\bar{\zeta}_{k_2}(\tau_*)\bar{\zeta}_{k_3}(\tau_*)\Biggr)+(\text {2 perms})\,,
\\
&&
\llangle\tilde{\zeta}^3\rrangle=\llangle\tilde{\zeta}^3\rrangle_+ +\llangle\tilde{\zeta}^3\rrangle_-
=4\Mpl^2\int^{\tau_*}_{-\infty}{\rm d}\tau a^2\epsilon^2{\rm Im}\left(\zeta'_{k_1}(\tau)\zeta'_{k_2}(\tau)\zeta_{k_3}(\tau)\bar{\zeta}_{k_1}(\tau_*)\bar{\zeta}_{k_2}(\tau_*)\bar{\zeta}_{k_3}(\tau_*)\right)+(\text {2 perms})\,,
\ea
where we used the symbol ``2 perms'' for the symmetric terms 
with respect to the wavenumbers $\bm{k}_1, \bm{k}_2, \bm{k}_3$. 

\item ${\cal L}_{2,{\rm bulk}}=
\Mpl^2a^2\epsilon^2\zeta(\partial\zeta)^2$
\ba
&&
\llangle\tilde{\zeta}^3\rrangle_+=-2\ri\Mpl^2\left(\bm{k}_1\cdot\bm{k}_2+\bm{k}_2\cdot\bm{k}_3+\bm{k}_3\cdot\bm{k}_1\right)\Biggl(\int^{\tau_f}_{\tau_*}{\rm d}\tau a^2\epsilon^2\zeta_{k_1}(\tau)\zeta_{k_2}(\tau)\zeta_{k_3}(\tau)\bar{\zeta}_{k_1}(\tau_*)\bar{\zeta}_{k_2}(\tau_*)\bar{\zeta}_{k_3}(\tau_*)
\notag
\\
&&
\hspace{7.0cm}
+\int^{\tau_*}_{-\infty}{\rm d}\tau a^2\epsilon^2\bar{\zeta}_{k_1}(\tau)\bar{\zeta}_{k_2}(\tau)\bar{\zeta}_{k_3}(\tau){\zeta}_{k_1}(\tau_*){\zeta}_{k_2}(\tau_*){\zeta}_{k_3}(\tau_*)\Biggr)\,,
\\
&&
\llangle\tilde{\zeta}^3\rrangle_-=2\ri\Mpl^2\left(\bm{k}_1\cdot\bm{k}_2+\bm{k}_2\cdot\bm{k}_3+\bm{k}_3\cdot\bm{k}_1\right)\int^{\tau_f}_{-\infty}{\rm d}\tau a^2\epsilon^2\zeta_{k_1}(\tau)\zeta_{k_2}(\tau)\zeta_{k_3}(\tau)\bar{\zeta}_{k_1}(\tau_*)\bar{\zeta}_{k_2}(\tau_*)\bar{\zeta}_{k_3}(\tau_*)\,,
\\
&&
\llangle\tilde{\zeta}^3\rrangle
=-4\Mpl^2\left(\bm{k}_1\cdot\bm{k}_2+\bm{k}_2\cdot\bm{k}_3+\bm{k}_3\cdot\bm{k}_1\right)
\int^{\tau_*}_{-\infty}{\rm d}\tau a^2\epsilon^2{\rm Im}\left(\zeta_{k_1}(\tau)\zeta_{k_2}(\tau)\zeta_{k_3}(\tau)\bar{\zeta}_{k_1}(\tau_*)\bar{\zeta}_{k_2}(\tau_*)\bar{\zeta}_{k_3}(\tau_*)\right)\,.
\ea
\item ${\cal L}_{3,{\rm bulk}}=-\Mpl^2 a\epsilon 
\left(2-\dfrac{\epsilon}{2} \right) \zeta'
\partial_i \zeta \partial_i \chi
=-\Mpl^2a^2\epsilon^2 
\left(2-\dfrac{\epsilon}{2}\right)\zeta'
\partial_i \zeta \partial_i \partial^{-2}\zeta'
\simeq 
-2\Mpl^2a^2\epsilon^2\zeta'
\partial_i \zeta \partial_i \partial^{-2}\zeta'$
\ba
&&
\llangle\tilde{\zeta}^3\rrangle_+=
-2\ri\Mpl^2\frac{\bm{k}_2\cdot\bm{k}_3}{k_3^2}\Biggl(\int^{\tau_f}_{\tau_*}{\rm d}\tau a^2\epsilon^2\zeta'_{k_1}(\tau)\zeta_{k_2}(\tau)\zeta'_{k_3}(\tau)\bar{\zeta}_{k_1}(\tau_*)\bar{\zeta}_{k_2}(\tau_*)\bar{\zeta}_{k_3}(\tau_*)
\notag
\\
&&
\hspace{3.8cm}
+\int^{\tau_*}_{-\infty}{\rm d}\tau a^2\epsilon^2\bar{\zeta}'_{k_1}(\tau)\bar{\zeta}_{k_2}(\tau)\bar{\zeta}'_{k_3}(\tau){\zeta}_{k_1}(\tau_*){\zeta}_{k_2}(\tau_*){\zeta}_{k_3}(\tau_*)\Biggr)
+(\text {5 perms})\,,
\\
&&
\llangle\tilde{\zeta}^3\rrangle_-=2\ri\Mpl^2\frac{\bm{k}_2\cdot\bm{k}_3}{k_3^2}\Biggl(\int^{\tau_f}_{-\infty}{\rm d}\tau a^2\epsilon^2\zeta'_{k_1}(\tau)\zeta_{k_2}(\tau)\zeta'_{k_3}(\tau)\bar{\zeta}_{k_1}(\tau_*)\bar{\zeta}_{k_2}(\tau_*)\bar{\zeta}_{k_3}(\tau_*)\Biggr)+(\text {5 perms})\,,
\\
&&
\llangle\tilde{\zeta}^3\rrangle
=-4\Mpl^2\frac{\bm{k}_2\cdot\bm{k}_3}{k_3^2}\int^{\tau_*}_{-\infty}{\rm d}\tau a^2\epsilon^2{\rm Im}\left(\zeta'_{k_1}(\tau)\zeta_{k_2}(\tau)\zeta'_{k_3}(\tau)\bar{\zeta}_{k_1}(\tau_*)\bar{\zeta}_{k_2}(\tau_*)\bar{\zeta}_{k_3}(\tau_*)\right)+(\text {5 perms})\,.
\ea
\item ${\cal{L}}_{4,{\rm bulk}}=
\Mpl^2\dfrac{a^2}{2}\epsilon\eta'\zeta^2\zeta'$
\ba
&&
\llangle\tilde{\zeta}^3\rrangle_+=\ri\Mpl^2\Biggl(\int^{\tau_f}_{\tau_*}{\rm d}\tau a^2\epsilon\eta'\zeta'_{k_1}(\tau)\zeta_{k_2}(\tau)\zeta_{k_3}(\tau)\bar{\zeta}_{k_1}(\tau_*)\bar{\zeta}_{k_2}(\tau_*)\bar{\zeta}_{k_3}(\tau_*)
\notag
\\
&&
\hspace{2.3cm}
+\int^{\tau_*}_{-\infty}{\rm d}\tau a^2\epsilon\eta'\bar{\zeta}'_{k_1}(\tau)\bar{\zeta}_{k_2}(\tau)\bar{\zeta}_{k_3}(\tau){\zeta}_{k_1}(\tau_*){\zeta}_{k_2}(\tau_*){\zeta}_{k_3}(\tau_*)\Biggr)
+(\text {2 perms})\,,
\\
&&
\llangle\tilde{\zeta}^3\rrangle_-=-\ri\Mpl^2\Biggl(\int^{\tau_f}_{-\infty}{\rm d}\tau a^2\epsilon\eta'\zeta'_{k_1}(\tau)\zeta_{k_2}(\tau)\zeta_{k_3}(\tau)\bar{\zeta}_{k_1}(\tau_*)\bar{\zeta}_{k_2}(\tau_*)\bar{\zeta}_{k_3}(\tau_*)\Biggr)+(\text {2 perms})\,,
\\
&&
\llangle\tilde{\zeta}^3\rrangle
=2\Mpl^2\int^{\tau_*}_{-\infty}{\rm d}\tau a^2\epsilon\eta'{\rm Im}\left(\zeta'_{k_1}(\tau)\zeta_{k_2}(\tau)\zeta_{k_3}(\tau)\bar{\zeta}_{k_1}(\tau_*)\bar{\zeta}_{k_2}(\tau_*)\bar{\zeta}_{k_3}(\tau_*)\right)+(\text {2 perms})\,.
\label{C12}
\ea
\item ${\cal L}_{5,{\rm bulk}}=
\Mpl^2 \dfrac{\epsilon}{4} \partial^2 \zeta 
(\partial \chi)^2=
\Mpl^2\dfrac{a^2}{4}\epsilon^3\partial^2
\zeta(\partial_i \partial^{-2}\zeta')^2$
\ba
&&
\llangle\tilde{\zeta}^3\rrangle_+=\frac{\ri}{2}\Mpl^2\frac{k_1^2}{k_2^2 k_3^2}\bm{k}_2\cdot\bm{k}_3\Biggl(\int^{\tau_f}_{\tau_*}{\rm d}\tau a^2\epsilon^3\zeta_{k_1}(\tau)\zeta'_{k_2}(\tau)\zeta'_{k_3}(\tau)\bar{\zeta}_{k_1}(\tau_*)\bar{\zeta}_{k_2}(\tau_*)\bar{\zeta}_{k_3}(\tau_*)
\notag
\\
&&
\hspace{4.2cm}
+\int^{\tau_*}_{-\infty}{\rm d}\tau a^2\epsilon^3\bar{\zeta}_{k_1}(\tau)\bar{\zeta}'_{k_2}(\tau)\bar{\zeta}'_{k_3}(\tau){\zeta}_{k_1}(\tau_*){\zeta}_{k_2}(\tau_*){\zeta}_{k_3}(\tau_*)\Biggr)
+(\text {2 perms})\,,
\\
&&
\llangle\tilde{\zeta}^3\rrangle_-=-\frac{\ri}{2}\Mpl^2\frac{k_1^2}{k_2^2 k_3^2}\bm{k}_2\cdot\bm{k}_3\Biggl(\int^{\tau_f}_{-\infty}{\rm d}\tau a^2\epsilon^3\zeta_{k_1}(\tau)\zeta'_{k_2}(\tau)\zeta'_{k_3}(\tau)\bar{\zeta}_{k_1}(\tau_*)\bar{\zeta}_{k_2}(\tau_*)\bar{\zeta}_{k_3}(\tau_*)\Biggr)+(\text {2 perms})\,,
\\
&&
\llangle\tilde{\zeta}^3\rrangle
=\Mpl^2\frac{k_1^2}{k_2^2 k_3^2}\bm{k}_2\cdot\bm{k}_3
\int^{\tau_*}_{-\infty}{\rm d}\tau a^2\epsilon^3{\rm Im}\left(\zeta_{k_1}(\tau)\zeta'_{k_2}(\tau)\zeta'_{k_3}(\tau)\bar{\zeta}_{k_1}(\tau_*)\bar{\zeta}_{k_2}(\tau_*)\bar{\zeta}_{k_3}(\tau_*)\right)+(\text {2 perms})\,.
\ea

For the EOM terms, the $-$ type vertex does not contribute 
to $\llangle\tilde{\zeta}^3\rrangle$  
due to the property $D_2 G_{-+}=0$. 
Each contribution arising from the 
EOM terms in Eq.~(\ref{S3JY}) is
\item ${\cal{L}}_{1,{\rm EOM}}=
\Mpl^2\dfrac{\eta}{2}\zeta^2D_2\zeta$
\be
\llangle\tilde{\zeta}^3\rrangle
=\frac{\eta(\tau_*)}{2}|\zeta_{k_1}(\tau_*)|^2|
\zeta_{k_2}(\tau_*)|^2+(\text {2 perms})\,,
\label{C16}
\ee
where we used the relation \eqref{EOMtimeprop}.

\item ${\cal{L}}_{2,{\rm EOM}}=
\Mpl^2\dfrac{2}{aH}\zeta\zeta'D_2\zeta$
\be
\llangle\tilde{\zeta}^3\rrangle
=\frac{1}{2a(\tau_*)H}(|\zeta_{k_1}(\tau_*)|^2)'|
\zeta_{k_2}(\tau_*)|^2+(\text {5 perms})\,.
\label{C17}
\ee
\item ${\cal{L}}_{3,{\rm EOM}}=
-\dfrac{\Mpl^2}{2a^2H^2}(\partial\zeta)^2D_2\zeta$
\be
\llangle\tilde{\zeta}^3\rrangle
=\frac{\bm{k}_1\cdot\bm{k}_2}{2a^2(\tau_*)H^2}|\zeta_{k_1}(\tau_*)|^2|\zeta_{k_2}(\tau_*)|^2+(\text {2 perms})\,.
\label{L3EOM}
\ee
\item ${\cal{L}}_{4,{\rm EOM}}=\dfrac{\Mpl^2}{2a^2H^2}\partial^{-2}(\partial_i\partial_j\left(\partial_i\zeta\partial_j\zeta\right))D_2\zeta$
\be
\llangle\tilde{\zeta}^3\rrangle
=-\frac{1}{2a^2(\tau_*)H^2}\frac{(\bm{k}_1\cdot\bm{k}_3)(\bm{k}_2\cdot\bm{k}_3)}{k_3^2}|\zeta_{k_1}(\tau_*)|^2|
\zeta_{k_2}(\tau_*)|^2+(\text {2 perms})\,.
\ee
\item ${\cal{L}}_{5,{\rm EOM}}=\dfrac{\Mpl^2}{a^2 H} 
(\partial_i \zeta)(\partial_i \chi)D_2 \zeta
=\Mpl^2\dfrac{\epsilon}{aH}(\partial_i\zeta)(\partial_i\partial^{-2}\zeta')D_2\zeta$
\be
\llangle\tilde{\zeta}^3\rrangle
=\dfrac{\epsilon(\tau_*)}{4a(\tau_*)H}\frac{\bm{k}_1\cdot\bm{k}_2}{k_2^2}|\zeta_{k_1}(\tau_*)|^2\left(|\zeta_{k_2}(\tau_*)|^2\right)'
+(\text {5 perms})\,.
\ee
\item ${\cal{L}}_{6,{\rm EOM}}
=-\dfrac{\Mpl^2}{a^2 H}\partial^{-2}(\partial_i \partial_j (\partial_i \zeta 
\partial_j \chi))D_2\zeta
=-\Mpl^2\dfrac{\epsilon}{aH}\partial^{-2}(\partial_i\partial_j(\partial_i
\zeta\partial_j\partial^{-2}\zeta'))D_2\zeta$
\be
\llangle\tilde{\zeta}^3\rrangle
=-\frac{\epsilon(\tau_*)}{4a(\tau_*)H}\frac{(\bm{k}_1\cdot\bm{k}_3)(\bm{k}_2\cdot\bm{k}_3)}{k_2^2k_3^2}|\zeta_{k_1}(\tau_*)|^2\left(|\zeta_{k_2}(\tau_*)|^2\right)'+(\text {5 perms})\,.
\label{L6EOM}
\ee
\end{itemize}

Let us consider the transient USR model in which  
$\eta$ and $\epsilon$ evolve according to 
Eqs.~(\ref{eta}) and (\ref{epsilon}), respectively. 
During the whole SR+USR+SR stages, $\epsilon$ is much smaller than 1, 
while $\eta$ exhibits the transition from the SR value $\eta \simeq 0$ 
to the USR value $\eta \simeq -6$. 
In the squeezed limit ($|{\bm k}_1| \ll 
|{\bm k}_2| \simeq |{\bm k}_3|$), 
the contributions to $\llangle\tilde{\zeta}^3\rrangle$ arising 
from Eqs.~(\ref{L3EOM})-(\ref{L6EOM}) are suppressed relative to 
those from Eqs.~(\ref{C16}) and (\ref{C17}). 
The first three bulk terms ${\cal L}_{1,{\rm bulk}}$, 
${\cal L}_{2,{\rm bulk}}$, ${\cal L}_{3,{\rm bulk}}$ and 
the last one ${\cal L}_{5,{\rm bulk}}$ 
give rise to contributions to $\llangle\tilde{\zeta}^3\rrangle$ 
of order $\epsilon^2$ and $\epsilon^3$, respectively, so that 
they are negligible relative to (\ref{C16}) and (\ref{C17}).
In the transient USR model, the derivative $\eta'$ is not 
small during the transition from the SR to USR stages and hence 
the contribution (\ref{C12}) from 
the fourth bulk term ${\cal L}_{4,{\rm bulk}}$ cannot be 
ignored.\footnote{In standard slow-roll inflation where $|\eta|$ 
is at most of the same order as $\epsilon$, 
the ${\cal L}_{4,{\rm bulk}}$ 
contributions to $\llangle\tilde{\zeta}^3\rrangle$ can be 
neglected relative to (\ref{C16}) and (\ref{C17}).}

In summary, in the context of transient USR inflation, 
the dominant contributions to $\llangle\tilde{\zeta}^3\rrangle$
come from the three terms \eqref{C12}, \eqref{C16}, 
and \eqref{C17}. For the computation of $\llangle\tilde{\zeta}^3\rrangle$ 
arising from these dominant contributions, we exploit 
the background Eq.~(\ref{eta}) for $\eta$ and 
use the approximations $\zeta_{k_1}'(\tau) \simeq 0$ 
after the Hubble radius crossing for the large-scale 
mode with $k_1=|{\bm k}_1|$, and $\zeta_{k_1}(\tau_*) \simeq 
\zeta_{k_1}(\tau_s) \simeq \zeta_{k_1}(\tau_e)$,   
where $\tau_s$ and $\tau_e$ correspond to conformal 
times at the start and end of the USR period, respectively. 
Picking up the leading-order terms around $k_1\to 0$, 
the squeezed-limit bispectrum is given 
by\footnote{Here, we use the inequality  
$|\zeta_{k}(\tau_*)|^2\ll|\zeta_{k_1}(\tau_*)|^2$ 
for the wavenumber $k$ much larger than $k_1$. 
When the transient USR period is present, 
the small-scale dimensionless power spectrum 
${\cal P}_\zeta(k)=k^3|\zeta_k|^2/(2\pi^2)$ can be 
enhanced by a factor of order $10^7$ relative to the 
value ${\cal P}_\zeta(k_1)
=k_1^3|\zeta_{k_1}|^2/(2\pi^2)$ on CMB scales, 
see Fig.~\ref{powerspectrum}. 
Provided that $k/k_1 \gg 10^{7/3}$, however, 
the inequality $|\zeta_{k}(\tau_*)|^2\ll|\zeta_{k_1}(\tau_*)|^2$ 
holds.}
\ba
&&
\llangle \tilde{\zeta}(\tau_*,\bm{k}_1)\tilde{\zeta}(\tau_*,\bm{k})\tilde{\zeta}(\tau_*,-\bm{k})\rrangle
\simeq
|\zeta_{k_1}(\tau_*)|^2
\Biggl[4\Mpl^2\frac{\epsilon_1\Delta\eta}{\tau_s^2H^2}{\rm Im}\left(\zeta_k'(\tau_s)\zeta_k(\tau_s)
\bar{\zeta}_k(\tau_*)\bar{\zeta}_k(\tau_*)\right)\theta(\tau_*-\tau_s) \nonumber
\\
&&
\hspace{6.3cm}
-4\Mpl^2\frac{\epsilon_2\Delta\eta}{\tau_e^2H^2}{\rm Im}\left(\zeta_k'(\tau_e)\zeta_k(\tau_e)\bar{\zeta}_k(\tau_*)\bar{\zeta}_k(\tau_*)\right)\theta(\tau_*-\tau_e) \nonumber
\\
&&
\hspace{6.4cm}
+\eta(\tau_*)|\zeta_k(\tau_*)|^2-\tau_*
\left(|\zeta_k(\tau_*)|^2\right)'\Biggr]\,.
\ea
By using the mode function~\eqref{zetak}, 
it follows that 
\ba
&&
\eta(\tau_*)|\zeta_k(\tau_*)|^2-\tau_*\left(|\zeta_k(\tau_*)|^2\right)'=-\frac{H^2}{4\epsilon(\tau_*)\Mpl^2k^2}\frac{\partial|\xi(x,y,z)|^2}{\partial x}\frac{{\rm d}x}{{\rm d}k}\biggr|_{(x,y,z)=(k\tau_*,k\tau_s,k\tau_e)}\,,\label{cube1}
\\
&&
4\Mpl^2\frac{\epsilon_1\Delta\eta}{\tau_s^2H^2}{\rm Im}\left(\zeta_k'(\tau_s)\zeta_k(\tau_s)\bar{\zeta}_k(\tau_*)\bar{\zeta}_k(\tau_*)\right)\theta(\tau_*-\tau_s)=-\frac{H^2}{4\epsilon(\tau_*)\Mpl^2k^2}\frac{\partial|\xi(x,y,z)|^2}{\partial y}\frac{{\rm d}y}{{\rm d}k}\biggr|_{(x,y,z)=(k\tau_*,k\tau_s,k\tau_e)}\,,\label{cube2}
\\
&&
-4\Mpl^2\frac{\epsilon_2\Delta\eta}{\tau_e^2H^2}{\rm Im}\left(\zeta_k'(\tau_e)\zeta_k(\tau_e)\bar{\zeta}_k(\tau_*)\bar{\zeta}_k(\tau_*)\right)\theta(\tau_*-\tau_e)=-\frac{H^2}{4\epsilon(\tau_*)\Mpl^2k^2}\frac{\partial|\xi(x,y,z)|^2}{\partial z}\frac{{\rm d}z}{{\rm d}k}\biggr|_{(x,y,z)=(k\tau_*,k\tau_s,k\tau_e)}\,,\label{cube3}
\ea
where $\xi(x,y,z)$ is defined in Eq.~\eqref{xik}.
Then, the three-point correlation function 
in the squeezed limit yields
\ba
&&
\llangle \tilde{\zeta}(\tau_*,\bm{k}_1)\tilde{\zeta}(\tau_*,\bm{k})\tilde{\zeta}(\tau_*,-\bm{k})\rrangle
\simeq 
-\frac{H^2}{4\epsilon(\tau_*)k^2\Mpl^2}|\zeta_{k_1}(\tau_*)|^2\frac{{\rm d}|\xi(k\tau_*,k\tau_s,k\tau_e)|^2}{{\rm d}k}
\notag\\
&&
\hspace{4.3cm}
=-\llangle \tilde{\zeta}(\tau_*,\bm{k}_1)\tilde{\zeta}(\tau_*,-\bm{k}_1)\rrangle\frac{1}{k^3}\frac{{\rm d}\left(k^3\llangle \tilde{\zeta}(\tau_*,\bm{k})\tilde{\zeta}(\tau_*,-\bm{k})\rrangle\right)}{{\rm d}\ln k}\,,
\label{zeta3consistency}
\ea
which is equivalent to the consistency relation originally derived by 
Maldacena in the context of slow-roll inflation~\cite{Maldacena:2002vr}.
We proved that the same consistency relation also holds in the context 
of transient USR inflation, as it is consistent with the numerical 
analysis in Ref.~\cite{Motohashi:2023syh,Tada:2023rgp}.\footnote{The results of Refs.~\cite{Motohashi:2023syh,Tada:2023rgp} seem to be based on the operator formalism, and they use total time-derivative Hamiltonian 
terms rather than the EOM terms.}
Note that the relation (\ref{zeta3consistency}) is a general result 
valid for the momentum ${\bm k}$ in the squeezed limit 
($|{\bm k}| \gg |{\bm k}_1|$). 
Moreover, it holds at any time $\tau_*$ during inflation 
after the large-scale mode ${\bm k}_1$ exits the Hubble horizon.
These properties are important to show 
the absence of large one-loop corrections to the power spectrum 
of long-wavelength perturbations.
As it is clear from the above proof, even in the presence of 
multiple USR stages, the three-point correlation function 
would be proportional to the change of $|\xi|^2$ 
with respect to $k$, which should finally translate into 
the consistency relation~(\ref{zeta3consistency}).

From the above discussion, we have found that the leading-order 
interactions in the transient USR model are those yielding 
\eqref{C12}, \eqref{C16}, and \eqref{C17}. 
In Sec.~\ref{effectiveUSR}, we study whether the reduced 
Lagrangian consisting of the leading-order terms proves 
consistency relations by themselves.

\section{Effective description of the transient 
USR model and other consistency relations}\label{effectiveUSR}

In this section, we consider a reduced Lagrangian that consists of 
leading-order terms, which gives the correct consistency relations of 
three-point correlation functions. 
In the literature, it is often the case that only the interaction 
term ${\cal L}_{1,{\rm bulk}}=
\Mpl^2 a^2\epsilon\eta'\zeta^2\zeta'/2$ is 
taken into account for computing 
quantum corrections to the power spectrum. 
However, this is inadequate since it cannot prove the consistency relation of the simplest three-point function. 
Indeed, from our analytic results in Sec.~\ref{AppMaldacena}, 
we showed that a reduced cubic action  which results in the correct 
consistency relation consists of the terms 
${\cal L}_{4,{\rm bulk}}$, 
${\cal L}_{1,{\rm EOM}}$, and ${\cal L}_{2,{\rm EOM}}$, i.e.,
\footnote{In Refs.~\cite{Motohashi:2023syh,Tada:2023rgp}, it was 
shown that the leading bulk term~${\cal L}_{1,{\rm bulk}}$ and the boundary terms ${\cal L}_\partial=[-(a\epsilon/H)\zeta\zeta'^2
-(a^2\epsilon\eta/2)\zeta^2\zeta']'$ lead to the correct consistency relation of $\langle\zeta^3\rangle$ for a very wide range of momentum modes. Since they use the operator formalism, we are not able to find the one-to-one correspondence between our results and theirs. Nevertheless, we suspect that their sets of interactions would be the self-consistent reduced Hamiltonian within the operator formalism. It would be worth proving the consistency relations corresponding to their reduced Hamiltonian interactions as we will do in the path integral formalism, which would justify the application of the master formula within the operator formalism self-consistently.}
\be
\tilde{\cal{S}}^{(3)}=\Mpl^2\int{\rm d}\tau{\rm d}^3 x
\left( \frac{a^2}{2}\epsilon\eta'\zeta^2\zeta'+\frac{\eta}{2}\zeta^2D_2 \zeta+\frac{2}{aH}\zeta\zeta'D_2\zeta \right)\,,
\label{cubicL}
\ee
where
\begin{align}
D_2 \zeta
=\frac{\rd}{\rd \tau} \left( a^2 \epsilon \zeta' 
\right) -a^2 \epsilon \partial^2 \zeta.
\end{align}
We emphasize that the reduced Lagrangian (\ref{cubicL}) is yet a candidate for a self-consistent reduced Lagrangian. 
This is because we have not yet proven the consistency relations of 
three-point functions that appear in the interactions. 
As we will show below, the result in Sec.~\ref{AppMaldacena} immediately 
shows the consistency relation of $\llangle\zeta^2\zeta' \rrangle$ that corresponds to the first term in the reduced 
interaction Lagrangian \eqref{cubicL}. 
To use the master formula~\eqref{ratio-one-loop-tree}, we need to prove consistency relations of the three-point functions $\llangle \zeta^2D_2 \zeta\rrangle$ and $\llangle \zeta\zeta'D_2\zeta\rrangle$ as well. 
Here, it is required to include quartic interactions, 
since the $D_2$ operator contains two spatial derivatives. 
To find necessary quartic interactions, we need to consider 
the following nonlinear completion of the Lagrangian 
that respects the trivial symmetry associated with 
the zero mode of curvature perturbations:
\begin{align}
\tilde{\cal{L}}^{(3)}\supset \Mpl^2
\left( -\frac{a^2 \epsilon\eta}{2}\zeta^2\partial^2 \zeta
-\frac{2 a\epsilon}{H}\zeta\zeta'\partial^2\zeta 
\right)\to \Mpl^2 \left(-\frac{a^2\epsilon\eta}{4}
e^{-2\zeta}\partial^2 \zeta+\frac{a\epsilon}{H}\zeta'
e^{-2\zeta}\partial^2\zeta\right)\,.
\end{align}
One would be able to reproduce the cubic interaction by expanding 
the exponential factor, while recovering the shift symmetry 
of $\zeta$ compensated by coordinate rescalings $\zeta \to \zeta+\delta \zeta$ and 
$x^i \to e^{-\delta \zeta} x^i$ with a constant $\delta \zeta$. 
Accordingly, we should have the following quartic action
\begin{align}
{\cal S}^{(4)} \supset 
\Mpl^2\int{\rm d}\tau{\rm d}^3x
\left( \frac{a^2\epsilon\eta}{3}\zeta^3\partial^2\zeta
+\frac{2a\epsilon}{H}\zeta^2\zeta'\partial^2\zeta \right)\,.
\label{reducedS4}
\end{align}
Note that quartic interactions may cause the UV divergences 
at one-loop order. However, the UV divergences appear only 
in the forms respecting the original symmetry of the system. 
The local Lagrangian respecting the residual diffeomorphism 
on the FLRW cosmological background is summarized as the effective field theory of inflation (EFToI)~\cite{Cheung:2007st}, in which 
all perturbative UV divergences should be renormalized by the interactions of the EFToI. As shown in Ref.~\cite{Pimentel:2012tw}, 
the counterterms to tadpole UV divergences are also related to the quadratic UV divergent 
terms that appear due to quartic interactions. 
If we absorb the tadpole counterterm within the EFToI, we automatically introduce quadratic counterterms that cancel the UV divergent quadratic terms associated with quartic interactions. This should be the case as long as perturbative corrections do not spoil the original symmetry of the Lagrangian.\footnote{The series of terms in the EFToI are spatial derivative expansions, and the contractions of multiple derivative interactions can lead to UV divergent new derivative interactions as is usual in non-renormalizable theory. Such terms cannot be absorbed into 
lower-order terms in the EFToI but by higher-order ones. 
In any case, the UV divergent terms would be absorbed into 
the EFToI Lagrangian.}

The cubic-order action (\ref{cubicL}) consists of the interactions 
of the three products $\zeta^2 \zeta'$, 
$\zeta^2 (D_2 \zeta)$, and $\zeta \zeta'(D_2 \zeta)$.
In the expression of one-loop corrections to the power spectrum 
arising from cubic-order vertices~\eqref{premaster}, 
they correspond to the expectation values 
(see Appendix \ref{explicitform} for details):
\be
\llangle\zeta_L\zeta_S\zeta_S\rrangle,\hspace{0.5cm}
\llangle\zeta_L\zeta_S'\zeta_S\rrangle,\hspace{0.5cm}\llangle\zeta_L D_2\zeta_S\zeta_S\rrangle,\hspace{0.5cm}\llangle\zeta_L D_2\zeta_S\zeta'_S\rrangle\,,
\ee
where we omitted the $\pm$ signs of fields. 
Here and in the following, $\zeta_L$ is the large-scale mode 
with the wavenumber ${\bm k}_L={\bm k}_1$, whereas $\zeta_S$'s 
are the small-scale modes with the 
wavenumbers ${\bm k}_S={\bm k}$ and $-{\bm k}_1-{\bm k}\simeq -{\bm k}_S$. As we mentioned in Eq.~\eqref{krange}, we are concerned 
with one-loop corrections from the wavenumber in the range 
$k\geq k_{\rm min}\gg k_L\,(=k_1)$, which means that the loop momentum 
$k$ always corresponds to a short mode in the above sense.
We are interested in showing whether the following 
sets of consistency relations hold
\ba
&&
\llangle \tilde{\zeta}(\tau_{*},\bm{k}_L)\tilde{\zeta}'(\tau,\bm{k}_S)
\tilde{\zeta}(\tau,-\bm{k}_L-\bm{k}_S) \rrangle
\overset{?}{\simeq}-\llangle \tilde{\zeta}(\tau_*,\bm{k}_L)\tilde{\zeta}
(\tau_*,-\bm{k}_L) \rrangle
\frac{1}{k_S^3}\frac{{\rm d}\left(k_S^3\llangle \tilde{\zeta}'(\tau,\bm{k}_S)\tilde{\zeta}(\tau,-\bm{k}_S) \rrangle\right)}{{\rm d}\ln k_S}\,,
\label{zetazetaprimezeta}
\\
&&
\llangle \tilde{\zeta}(\tau_*,\bm{k}_L)D_2\tilde{\zeta}(\tau,\bm{k}_S)\tilde{\zeta}(\tau,-\bm{k}_L-\bm{k}_S) \rrangle
\overset{?}{\simeq}-\llangle \tilde{\zeta}(\tau_*,\bm{k}_L)
\tilde{\zeta}(\tau_*,-\bm{k}_L) \rrangle
\frac{1}{k_S^3}\frac{{\rm d}\left(k_S^3\llangle D_2\tilde{\zeta}(\tau,\bm{k}_S)\tilde{\zeta}(\tau,-\bm{k}_S) 
\rrangle\right)}{{\rm d}\ln k_S}\,,
\label{zetaD2zetazeta}
\\
&&
\llangle \tilde{\zeta}( \tau_*,\bm{k}_L)D_2\tilde{\zeta}(\tau,\bm{k}_S)\tilde{\zeta}'(\tau,-\bm{k}_L-\bm{k}_S)\rrangle
\overset{?}{\simeq}-\llangle \tilde{\zeta}(\tau_*,\bm{k}_L)
\tilde{\zeta}( \tau_*,-\bm{k}_L) \rrangle
\frac{1}{k_S^3}\frac{{\rm d}\left(k_S^3\llangle D_2\tilde{\zeta}(\tau,\bm{k}_S)\tilde{\zeta}'(\tau,-\bm{k}_S) \rrangle\right)}
{{\rm d}\ln k_S}\,,
\label{zetaD2zetazetaprime}
\ea
where $\tau_0 \le \tau \le \tau_*$ and $\tau_0$ approximately 
corresponds to the time at which the USR period starts 
($\tau_0 \simeq \tau_s$).
In the following, we will prove that the
consistency relations hold, with some modifications 
to Eqs.~\eqref{zetazetaprimezeta}, \eqref{zetaD2zetazeta}, 
and \eqref{zetaD2zetazetaprime}.
For the first relation associated with Eq.~(\ref{zetazetaprimezeta}), we perform straightforward calculations including all the contributions from the cubic interactions~\eqref{cubicL}. As for Eqs.~(\ref{zetaD2zetazeta}) and 
(\ref{zetaD2zetazetaprime}), we will resort to the Scwhinger-Dyson equation, which enables us 
to show the consistency relations simpler than the brute force calculations. 
In the latter two cases, the quartic interactions~\eqref{reducedS4} become important. 
Indeed, the last two correlation functions satisfy the consistency relations only after the compensation by quartic interactions 
due to spatial derivatives.
By proving the leading-order expression for one-loop corrections 
to the power spectrum~\eqref{ratio-one-loop-tree}, it is possible to show that one-loop corrections are sufficiently small and that the perturbative analysis is under control within the transient USR model, 
which will be studied in Sec.~\ref{oneloopresult}. 

Notice also that Eqs.~\eqref{zetazetaprimezeta}-\eqref{zetaD2zetazetaprime} take unusual forms in the sense that the time argument of the long-wavelength mode differs from others. However, we will resort to the fact that the effective constancy of large-scale curvature perturbations holds within the time range we consider. Nevertheless, we will not exploit such a property 
to the very end of the computations for consistency 
relations or loop corrections.

\subsection{Consistency relations for $\ev{\zeta\zeta\zeta}$ 
and $\ev{\zeta\zeta'\zeta}$}

In Sec.~\ref{AppMaldacena}, we showed the consistency relation of 
$\llangle \zeta^3\rrangle$ for the equal time $\tau_*$. 
However, the three-point function appearing in the master formula \eqref{ratio-one-loop-tree} is of the form
\begin{align}
\llangle \tilde{\zeta}^+(\tau_*,\bm k_L)\tilde{\zeta}^\pm(\tau,\bm k_S)\tilde{\zeta}^\pm(\tau,-\bm k_L-\bm k_S)\rrangle\,,
\end{align}
namely, the time argument of the large-scale mode 
differs from those of small-scale modes. 
Since the tree-level super-Hubble curvature perturbation is effectively 
constant, we have that $\tilde{\zeta}^+(\tau_*,\bm k_L)\simeq \tilde{\zeta}^+(\tau,\bm k_L)$ for $\tau \in [\tau_0,\tau_*]$ (see Sec.~\ref{effconst}). 
By using the consistency relation (\ref{zeta3consistency}) shown in 
Sec.~\ref{AppMaldacena}, it follows that 
\begin{align}
\llangle \tilde{\zeta}^+(\tau_*,\bm k_L)\tilde{\zeta}^\pm(\tau,\bm k_S)\tilde{\zeta}^\pm(\tau,-\bm k_L-\bm k_S)\rrangle\simeq -\llangle\tilde{\zeta}^+(\tau_*,\bm k_L)\tilde{\zeta}^+(\tau_*,-\bm k_L)\rrangle \frac{1}{k_S^3}\frac{{\rm d}\left(k_S^3\llangle \tilde{\zeta}^\pm(\tau,\bm{k}_S)\tilde{\zeta}^\pm(\tau,-\bm{k}_S) \rrangle\right)}{{\rm d}\ln k_S}\,.\label{zeta3consistency2}
\end{align}
Thus, we proved the consistency relation of $\llangle \zeta^3\rrangle$ 
associated with the master formula (\ref{ratio-one-loop-tree}).

Let us now discuss the three-point correlation function containing 
one time derivative $\ev{\zeta\zeta'\zeta}$. 
We consider a particular linear combination of such 
three-point functions that can be written as the time 
derivative of $\llangle\zeta\zeta\zeta \rrangle$, as 
\begin{align}
\llangle \tilde{\zeta}^+(\tau_*,\bm{k}_L)
\{\tilde{\zeta}^+(\tau,\bm{k}_S)\tilde{\zeta}^+(\tau,-\bm{k}_L-\bm{k}_S)\}'\rrangle
=\llangle \tilde{\zeta}^+(\tau_*,\bm{k}_L)\tilde{\zeta}^+(\tau,\bm{k}_S)\tilde{\zeta}^+(\tau,-\bm{k}_L-\bm{k}_S)
\rrangle'\,,
\label{con2m}
\end{align}
where $\tilde{\zeta}^+$ here corresponds to 
a Heisenberg operator rather than the interaction picture ones 
(see the comments at the end of Sec.~\ref{ininpath1}).\footnote{In the operator formalism, we need to be careful since the derivative 
acting on the Heisenberg picture field leads to nontrivial additional corrections as quoted in Ref.~\cite{Pimentel:2012tw}.}

Substituting the consistency relation \eqref{zeta3consistency2} 
into the right-hand side of Eq.~(\ref{con2m}) and taking 
the $\tau$ derivative of the left-hand side of 
Eq.~(\ref{con2m}) further, we obtain
\begin{align}
&\left(\llangle \tilde{\zeta}^+(\tau_*,\bm{k}_L)
\tilde{\zeta}^+{}'(\tau,\bm{k}_S)\tilde{\zeta}^+(\tau,-\bm{k}_L-\bm{k}_S)\rrangle
+\llangle \tilde{\zeta}^+(\tau_*,\bm{k}_L)\tilde{\zeta}^+(\tau,\bm{k}_S) \tilde{\zeta}^+{}' (\tau,-\bm{k}_L-\bm{k}_S) \rrangle\right)\nonumber\\
&\simeq-\llangle \tilde{\zeta}^+(\tau_*,\bm{k}_L)\tilde{\zeta}^+(\tau_*,-\bm{k}_L) \rrangle
\frac{1}{k_S^3}\left[\frac{{\rm d}\left(k_S^3\llangle 
\tilde{\zeta}^+{}' (\tau,\bm{k}_S)\tilde{\zeta}^+(\tau,-\bm{k}_S) \rrangle\right)}{{\rm d}\ln k_S}+\frac{{\rm d}\left(k_S^3\llangle \tilde{\zeta}^+(\tau,\bm{k}_S)
\tilde{\zeta}^+{}' (\tau,-\bm{k}_S) 
\rrangle\right)}{{\rm d}\ln k_S}\right]\,,
\label{MaldacenaCR2}  
\end{align}
where we used $\tilde{\zeta}(\tau_*,\bm k_L) \simeq 
\tilde{\zeta}(\tau,\bm k_L)$ and the effective constancy of 
$\llangle \tilde{\zeta}^+(\tau_*,\bm{k}_L)
\tilde{\zeta}^+(\tau_*,-\bm{k}_L) \rrangle$ within the time range 
we are concerned with. 
This consistency relation differs from Eq.~(\ref{zetazetaprimezeta}), but such a consistency relation is sufficient as the above form precisely appears in the master formula~\eqref{ratio-one-loop-tree}. 

We should note that the above discussion is based on the three-point function of the fields $\langle\zeta^+\zeta^+\zeta^+\rangle$. Nevertheless, we can confirm that $\langle\zeta^+\zeta^-\zeta^-\rangle$ yields the same result despite possible subtleties due to the sign difference. We have illustrated the possible subtlety in Appendix~\ref{explicitform}. 
Since the subtlety appears only at the (measure zero) point $\tau=\tau_*$, it turns out that our result does not suffer from any subtlety.

\subsection{Schwinger-Dyson equation}

We will use the Schwinger-Dyson equation to prove the consistency 
relations (\ref{zetaD2zetazeta}) and (\ref{zetaD2zetazetaprime}) 
containing the EOM derivative operator. 
Here we briefly discuss its derivation within the in-in formalism. 
Let us consider the generalized form of path integral,
\be
Y=\int{\cal{D}}\zeta^+{\cal{D}}\zeta^- 
F[\zeta^+,\zeta^-]\exp\left[\ri\left(
{\cal S}[\zeta^+]-{\cal S}[\zeta^-]\right)\right]\,,
\label{Y}
\ee
where $F$ is an arbitrary functional of $\zeta^+$ and $\zeta^-$.
By changing the integration variable $\zeta^+(x)$ to $\zeta^{+}(x)+\delta\zeta^+(x)$, where $\delta\zeta^+(x)$ is an arbitrary infinitesimal parameter, we obtain
\ba
Y &=& 
\int{\cal{D}}\zeta^+{\cal{D}}\zeta^- F[\zeta^+ + \delta\zeta^+(x),\zeta^-]\exp\left[\ri\left({\cal{S}}[\zeta^+ 
+\delta\zeta^+(x)]-{\cal{S}}[\zeta^-]\right)\right]
\notag\\
&=& \int{\cal{D}}\zeta^+{\cal{D}}\zeta^- F[\zeta^+,\zeta^-]\exp\left[\ri\left({\cal{S}}[\zeta^+]
-{\cal{S}}[\zeta^-]\right)\right]
\notag\\
&&
+\int{\cal{D}}\zeta^+{\cal{D}}\zeta^- \int{\rm d}\tau{\rm d}^3x
\left(\frac{\delta F[\zeta^+,\zeta^-]}{\delta\zeta^+(x)}+\ri F[\zeta^+,\zeta^-]\frac{\delta {\cal{S}}[\zeta^+]}{\delta \zeta^+(x)}\right)\exp\left[\ri\left({\cal{S}}[\zeta^+]-S[\zeta^-]\right)\right] \delta\zeta^+(x)\,.
\ea
Note that the integration measure ${\cal{D}}\zeta^+$ is unchanged 
under this transformation as $\delta\zeta^+(x)$ is 
independent of $\zeta^+$.
Since the first line of the above equation is just $Y$ in Eq.~\eqref{Y}, 
we find the general form of the Schwinger-Dyson equation 
in the in-in formalism,
\be
\int{\cal{D}}\zeta^+{\cal{D}}\zeta^- 
\left(\frac{\delta F[\zeta^+,\zeta^-]}{\delta\zeta^+(x)}+\ri F[\zeta^+,\zeta^-]\frac{\delta {\cal{S}}[\zeta^+]}{\delta \zeta^+(x)}\right)\exp\left[\ri\left({\cal{S}}[\zeta^+]-S[\zeta^-]\right)\right]=0\,.\label{+SDeq}
\ee
The same should hold when considering the infinitesimal shift of $\zeta^-$. By choosing an appropriate form of $F[\zeta^+,\zeta^-]$, 
we can rewrite the correlation functions. 
In particular, notice that
\be
\frac{\delta{\cal{S}}}{\delta\zeta}
=-2\Mpl^2 D_2\zeta+\frac{\delta{\cal{S}}^{(3)}}{\delta\zeta}
=-2\Mpl^2 D_2\zeta+\sum_{A=1}^{n}\tilde{\lambda}_3^{(A)}(\tau){\cal{D}}^{(A)}_1\zeta{\cal{D}}^{(A)}_{1'}\zeta\,.
\ee
Here we have omitted the quartic interactions since we would consider $F$ to be two of $\zeta$ and cubic contributions from quartic interactions cannot give non-zero values.\footnote{The roles of quartic interactions appear 
differently as will be shown later.} The first term contains the EOM derivative operator $D_2$, which appears on the left-hand side of 
both Eqs.~\eqref{zetaD2zetazeta} and \eqref{zetaD2zetazetaprime}. 
We apply the Schwinger-Dyson equations to rewrite them and prove the consistency relations. 

\subsection{$\ev{\zeta D_2\zeta\zeta}$ and $\ev{\zeta D_2\zeta\zeta'}$}\label{D2consistency}

To rewrite Eq.~\eqref{zetaD2zetazeta}, we substitute $F=\zeta^+(y)\zeta^+(z)$ to the Schwinger-Dyson equation~\eqref{+SDeq}, which yields
\be
\ev{\frac{\delta {\cal{S}}[\zeta^+]}{\delta\zeta^+(x)}\zeta^+(y)\zeta^+(z)}=\ri\ev{\zeta^+(y)}\delta^{(4)}(x-z)+\ri\ev{\zeta^+(z)}\delta^{(4)}(x-y)=0\,,
\ee
provided that the tadpole is properly eliminated by renormalization and $\langle\zeta\rangle=0$.\footnote{As we quoted, there are tadpole UV divergent terms from the contraction of cubic interactions, which contribute to the non-1PI diagrams. 
We assume that it is properly removed. 
See also~Ref.~\cite{Pimentel:2012tw}.}
Similarly, we obtain
\be
\int{\cal{D}}\zeta^+{\cal{D}}\zeta^- 
\left(\frac{\delta F[\zeta^+,\zeta^-]}{\delta\zeta^-(x)}-\ri F[\zeta^+,\zeta^-]\frac{\delta {\cal{S}}[\zeta^-]}{\delta \zeta^-(x)}\right)\exp\left[\ri\left({\cal{S}}[\zeta^+]-S[\zeta^-]\right)\right]=0\,,
\ee
and 
\be
\ev{\frac{\delta {\cal{S}}[\zeta^-]}
{\delta\zeta^-(x)}\zeta^+(y)\zeta^-(z)}
=-\ri\ev{\zeta^+(y)}\delta^{(4)}(x-z)-\ri\ev{\zeta^-(z)}\delta^{(4)}(x-y)=0\,,
\ee
where we assumed that the tadpole is removed by appropriate counterterms.

\begin{figure}[h]
\centering
\includegraphics[width=15cm]{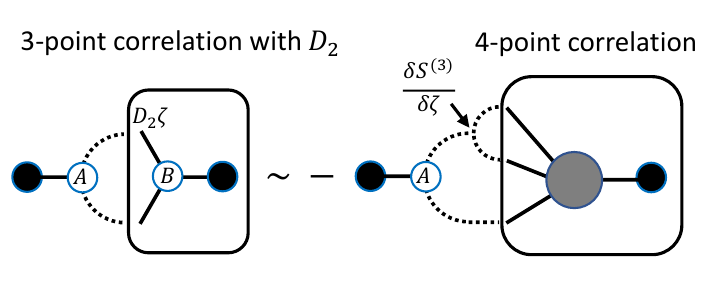}
\caption{\label{figShwingerDyson}
Schematic figure of Eq.~\eqref{3to4}.
The dotted lines denote uncontracted 
legs of vertex $A$.
}
\end{figure}

From the Schwinger-Dyson equation, we can rewrite the three-point correlation function with the EOM derivative operator 
as the effective four-point correlation functions,
\begin{align}
&\ev{\tilde{\zeta}^+(\tau_*,\bm{k}_L)D_2\tilde{\zeta}^{\pm}(\tau,\bm{k}_S)\tilde{\zeta}^{\pm}(\tau,-\bm{k}_L-\bm{k}_S)}\nonumber\\
&=
\frac{1}{2\Mpl^2}\sum_{A=1}^{\tilde{n}}\tilde{\lambda}_3^{(A)}(\tau)\int\frac{{\rm d}^3 q}{(2\pi)^3}\ev{\tilde{\zeta}^+(\tau_*,\bm{k}_L){\cal{D}}^{(A)}_1\tilde{\zeta}^{\pm}(\tau,\bm{q}){\cal{D}}^{(A)}_{1'}\tilde{\zeta}^{\pm}(\tau,\bm{k}_S-\bm{q})\tilde{\zeta}^{\pm}(\tau,-\bm{k}_L-\bm{k}_S)}\,.
\label{3to4}  
\end{align}
This operation is schematically illustrated 
in Fig.~\ref{figShwingerDyson}. 
The leg from vertex $A$ with the EOM 
derivative operator 
is replaced by $\delta{\cal{S}}^{(3)}/\delta\zeta$ 
and thus branches into two uncontracted legs.
It is then contracted by these two new legs and the two 
remaining legs (one originating from the same vertex $A$ 
and the other from the external source), so that it is 
effectively regarded as a set of four-point correlation functions.

As a next step, we evaluate the right-hand side of Eq.~\eqref{3to4}. Notice that we consider the one-loop corrections made of two cubic vertices and 
that the consistency relation to be proven is at tree level. 
Indeed, Eq.~\eqref{one-loop-cut-Fourier} has 
 a coupling $\lambda_3^{(A)}$ explicitly and the expectation value of the three-point function implicitly has another one. Now, the right-hand side of Eq.~\eqref{3to4} has a coupling $\tilde{\lambda}_3^{(A)}$, and therefore the contraction of the four-point function is 
given only by the products of two-point functions up 
to one-loop level.
Therefore, the effective four-point correlations 
in Eq.~\eqref{3to4} can be written down as 
\ba
&&
\llangle\tilde{\zeta}^+(\tau_*,\bm{k}_L)
D_2\tilde{\zeta}^\pm(\tau,\bm{k}_S)\tilde{\zeta}^\pm(\tau,-\bm{k}_L-\bm{k}_S)\rrangle
\notag\\
&&
=\frac{1}{2\Mpl^2}\sum_{A=1}^{\tilde{n}}\tilde{\lambda}_3^{(A)}(\tau)
\Biggl[ \llangle\tilde{\zeta}^+(\tau_*,\bm{k}_L){\cal{D}}^{(A)}_1\tilde{\zeta}^\pm(\tau,{-{\bm k}_L})\rrangle
\llangle {\cal{D}}^{(A)}_{1'}\tilde{\zeta}^\pm(\tau,\bm{k}_S)\tilde{\zeta}^\pm(\tau,-\bm{k}_S)\rrangle+(1\leftrightarrow1')\Biggr] \,.
\label{leading-order-SD}
\ea
Thus, we have found that the three-point correlation functions are 
expressed by the products of two-point functions even before 
taking the squeezing limit.

Let us evaluate Eq.~\eqref{leading-order-SD} more concretely 
for the transient USR model.
From the cubic action~\eqref{cubicL}, we obtain
\ba
\frac{\delta{\tilde{\cal{S}}}^{(3)}}{\delta\zeta} 
&\simeq&
\Mpl^2\left(-6\zeta D_2\zeta-4a^2\epsilon\zeta\partial^2\zeta\right)
\nonumber \\
& &
+\frac{a\epsilon\Mpl^2}{H}\left[-2\zeta'\partial^2\zeta
-4(\partial\zeta')(\partial\zeta)+6\zeta'\zeta''\right]
+a^2\epsilon\Mpl^2\left[ 3\eta\zeta'^2-\eta(\partial\zeta)^2\right]\,,
\label{dS3dzeta}
\ea
up to leading order in the slow-roll approximation of $\epsilon$, 
with $\eta$ of order 1 during the USR regime.
Furthermore, the terms on the second line of Eq.~\eqref{dS3dzeta} 
contain either time or spatial derivatives for both $\zeta$. 
Therefore, as $\zeta_{k_1}'$ and $k_1\zeta_{k_1}$ are suppressed for the super-Hubble Fourier mode, we can neglect those terms 
in Eq.~(\ref{dS3dzeta})\footnote{
The term containing $\zeta''$ in the second line of Eq.~\eqref{dS3dzeta} may lead to contributions proportional to $\delta(\tau-\tau_*)$ in the 
right-hand side of Eq.~\eqref{leading-order-SD}. 
Notice that the time-dependent couplings in Eq.~\eqref{cubicL}, which explicitly appear in Eq.~\eqref{premaster}, are negligibly small at $\tau=\tau_*$. Therefore, such contributions do not affect one-loop corrections and 
hence we neglect them in the following.}.
Therefore, substituting the terms on the first line of Eq.~\eqref{dS3dzeta} 
into Eq.~\eqref{leading-order-SD}, we obtain
\ba
\llangle\tilde{\zeta}^+(\tau_*,\bm{k}_L)
D_2\tilde{\zeta}^\pm(\tau,\bm{k}_S)
\tilde{\zeta}^\pm(\tau, -{\bm k}_L-\bm{k}_S)\rrangle
&\simeq&
-3\llangle\tilde{\zeta}^+(\tau_*,\bm{k}_L)
\tilde{\zeta}^\pm(\tau,-\bm{k}_L)\rrangle
\llangle D_2\tilde{\zeta}^\pm(\tau, 
{\bm k}_S+\bm{k}_L)\tilde{\zeta}^\pm
(\tau,-{\bm k}_S-\bm{k}_L)\rrangle
\notag\\
&&
-2a^2(\tau)\epsilon(\tau)
\llangle\tilde{\zeta}^+(\tau_*,\bm{k}_L)\tilde{\zeta}^\pm
(\tau,-\bm{k}_L)\rrangle
\llangle \partial^2\tilde{\zeta}^\pm(\tau,
\bm{k}_S+\bm{k}_L)\tilde{\zeta}^\pm(\tau,-\bm{k}_S-\bm{k}_L)\rrangle
\notag\\
&\simeq& -\llangle\tilde{\zeta}^+(\tau_*,\bm{k}_L)\tilde{\zeta}^\pm(\tau,-\bm{k}_L)\rrangle
\frac{1}{k^3}\frac{{\rm d}\left(k^3\llangle D_2\tilde{\zeta}^\pm(\tau,\bm{k}_S)\tilde{\zeta}^\pm(\tau,-\bm{k}_S)\rrangle\right)}{{\rm d}\ln k}
\notag\\
&&
-2a^2(\tau)\epsilon(\tau)
\llangle\tilde{\zeta}^+(\tau_*,\bm{k}_L)\tilde{\zeta}^\pm(\tau,-\bm{k}_L)\rrangle
\llangle\partial^2\tilde{\zeta}^\pm(\tau,\bm{k}_S)
\tilde{\zeta}^\pm(\tau,-\bm{k}_S)\rrangle\,,
\label{D2zetazeta}
\ea
where we have used the fact that $\llangle D_2\tilde{\zeta}^\pm(\tau,\bm{k})\tilde{\zeta}^\pm(\tau,-\bm{k})\rrangle$ 
is a $k$-independent variable and taken the limit $k_1\ll k$. 
As we will show in Sec.~\ref{oneloopresult}, 
the two-point function including $D_2$ is 
a singular object that can be removed by the mechanism discussed in Appendix~\ref{mostgeneralpathintegral}. 
Nevertheless, we leave it to prove the consistency relation. 
The above result is equivalent to the statement
\ba
&&
\llangle \tilde{\zeta}^+(\tau_*,\bm{k}_L)
D_2\tilde{\zeta}^\pm(\tau,\bm{k}_S)\tilde{\zeta}^\pm(\tau,-\bm{k}_L-\bm k_S) \rrangle
+2a^2(\tau)\epsilon(\tau)\llangle \tilde{\zeta}^+(\tau_*,\bm{k}_L)
\tilde{\zeta}^\pm(\tau,-\bm{k}_L)\rrangle\llangle\partial^2\tilde{\zeta}^\pm(\tau,\bm{k}_S)\tilde{\zeta}^\pm(\tau,-\bm{k}_S)\rrangle
\notag\\
&&
\simeq
-\llangle \tilde{\zeta}^+(\tau,\bm{k}_L)\tilde{\zeta}^\pm(\tau,-\bm{k}_L) \rrangle
\frac{1}{k_S^3}\frac{{\rm d}\left(k_S^3\llangle D_2\tilde{\zeta}^\pm(\tau,\bm{k}_S)\tilde{\zeta}^\pm(\tau,-\bm{k}_S) \rrangle\right)}{{\rm d}\ln k_S}\,.
\label{MaldacenaCR3}
\ea
This does not prove the consistency relation for 
$\langle\zeta^2 D_2\zeta\rangle$ yet. 
It shows that unless the second line of Eq.~\eqref{D2zetazeta} 
is canceled, the consistency relation does not hold.

This is the place where the quartic interactions \eqref{reducedS4} 
play the role. Consider the one-loop correction from the first term of \eqref{reducedS4} and take the limit $k_1 \to 0$. 
Namely, we drop the term proportional to $k_1$, which reads
\begin{align}
&\llangle\tilde{\zeta}(\tau_*,\bm{k}_L)\tilde{\zeta}(\tau_*,-\bm{k}_L)\rrangle^{(4)}_{\text{1-loop,1PI}}|_{\rm 1st}\nonumber\\
&\simeq 
\ri\int \rd\tau\int \frac{\rd^3k_S}{(2\pi)^3} 2a^2(\tau)\epsilon(\tau) \eta(\tau)\Biggl[G_{++}(k_L;\tau,\tau_*)\llangle \tilde{\zeta}^+(\tau,\bm k_L)\tilde{\zeta}^+(\tau,-\bm k_L)\rrangle\llangle\partial^2\tilde\zeta^+(\tau,\bm k_S)\tilde\zeta^+(\tau,-\bm k_S)\rrangle\nonumber\\
&\hspace{5.1cm} 
-G_{+-}(k_L;\tau,\tau_*)\llangle \tilde{\zeta}^+(\tau,\bm k_L)
\tilde{\zeta}^-(\tau,-\bm k_L)\rrangle\llangle\partial^2\tilde\zeta^-(\tau,\bm k_S)\tilde\zeta^-(\tau,-\bm k_S)\rrangle\Biggr]\,,
\end{align}
where the abbreviation 
$\int_{k>k_{\rm min}} \rd^3 k\to \int \rd^3k_S$ was used.
Notice also that the three-point function 
above appears in the one-loop correction as
\begin{align}
&\llangle\tilde{\zeta}(\tau_*,\bm{k}_L)\tilde{\zeta}(\tau_*,-\bm{k}_L)\rrangle_{\text{1-loop,1PI}}^{(3)}\nonumber\\
&\supset\ri\int{\rm d}\tau\int
\frac{{\rm d}^3 k_S}{(2\pi)^3}\eta(\tau)\Biggl[ G_{++}(k_L;\tau,\tau_*)
\llangle
\tilde{\zeta}^+(\tau_*,\bm{k}_L)D_2\tilde{\zeta}^+(\tau,\bm{k}_S)\tilde{\zeta}^+(\tau,-\bm{k}_L-\bm{k}_S)\rrangle_{\rm tree}\nonumber\\
&\hspace{3.5cm}
-G_{-+}(k_L;\tau,\tau_*)
\llangle
\tilde{\zeta}^+(\tau_*,\bm{k}_L)D_2\tilde{\zeta}^-(\tau,\bm{k}_S)\tilde{\zeta}^-(\tau,-\bm{k}_L-\bm{k}_S)\rrangle_{\rm tree}\Biggr]\,\nonumber\\
&\simeq \ri\int{\rm d}\tau\int\frac{{\rm d}^3 k_S}{(2\pi)^3}\eta(\tau)\Biggl[ G_{++}(k_L;\tau,\tau_*)
\Biggl(-\frac{1}{k^3_S}\llangle \tilde{\zeta}^+(\tau_*,\bm{k}_L)\tilde{\zeta}^+(\tau,-\bm{k}_L) \rrangle\frac{{\rm d}\left(k_S^3\llangle D_2\tilde{\zeta}^+(\tau,\bm{k}_S)\tilde{\zeta}^+(\tau,-\bm{k}_S)\rrangle\right)}{{\rm d}\ln k_S}\nonumber\\
&\hspace{6.1cm}-2a^2(\tau) \epsilon(\tau)
\llangle\tilde{\zeta}^+(\tau,\bm{k}_L)\tilde{\zeta}^+(\tau,-\bm{k}_L)\rrangle
\llangle\partial^2\tilde{\zeta}^+(\tau,\bm{k}_S)\tilde{\zeta}^+(\tau,-\bm{k}_S)\rrangle\Biggr)\nonumber\\
&\hspace{3.5cm}-G_{-+}(k_L;\tau,\tau_*)
\Biggl(-\frac{1}{k_S^3}\llangle \tilde{\zeta}^+(\tau_*,\bm{k}_L)\tilde{\zeta}^-(\tau,-\bm{k}_L) \rrangle\frac{{\rm d}\left(k_S^3\llangle D_2\tilde{\zeta}^-(\tau,\bm{k}_S)\tilde{\zeta}^-(\tau,-\bm{k}_S)\rrangle\right)}{{\rm d}\ln k_S}\nonumber\\
&\hspace{6.2cm}-2a^2(\tau)\epsilon(\tau)
\llangle\tilde{\zeta}^+(\tau_*,\bm{k}_L)\tilde{\zeta}^-(\tau,-\bm{k}_L)\rrangle
\llangle\partial^2\tilde{\zeta}^-(\tau,\bm{k}_S)\tilde{\zeta}^-(\tau,-\bm{k}_S)\rrangle\Biggr)\Biggr]\,,\label{zetaD2zetaloop}
\end{align}
where we exploited Eq.~\eqref{D2zetazeta} in the approximate 
equality on the second line. 
The sum of the 1PI contribution from the cubic and quartic interactions leaves only the terms that take the desired consistency relation forms of the three-point function $\langle\zeta^2D_2\zeta\rangle$ as expected. 
Thus, with the aid of the quartic interaction 
in Eq.~\eqref{reducedS4}, we have proven that 
the consistency relation (\ref{zetaD2zetazeta}) 
correctly holds after the cancellation of the second line 
in Eq.~\eqref{D2zetazeta}. 

Now, let us discuss the consistency relation of 
$\ev{\zeta D_2\zeta\zeta'}$ in the same way as the previous case. 
From the Schwinger-Dyson equation, we have
\ba
&&
\llangle\tilde{\zeta}(\tau_*,\bm{k}_L)D_2\tilde{\zeta}^\pm(\tau,\bm{k}_S)
\tilde{\zeta}^{\pm}{}'(\tau,-\bm{k}_L-\bm{k}_S)\rrangle
\notag\\
&&
\simeq
\frac{1}{2\Mpl^2}\sum_{A=1}^{\tilde{n}}\tilde{\lambda}_3^{(A)}(\tau)
\Biggl(\llangle\tilde{\zeta}^+(\tau_*,\bm{k}_L){\cal{D}}^{(A)}_1\tilde{\zeta}^\pm(\tau,-\bm{k}_L)\rrangle
\llangle {\cal{D}}^{(A)}_{1'}\tilde{\zeta}^\pm(\tau,\bm{k}_S+\bm{k}_L)
\tilde{\zeta}^{\pm}{}' (\tau,-\bm{k}_L-\bm{k}_S)\rrangle+(1\leftrightarrow1')\Biggr)\,
\notag\\
&&
\simeq-3\llangle\tilde{\zeta}^+(\tau_*,\bm{k}_L)\tilde{\zeta}^\pm(\tau,-\bm{k}_L)\rrangle
\llangle  D_2\tilde{\zeta}^\pm(\tau_,\bm{k}_S)
\tilde{\zeta}^{\pm}{}'(\tau,-\bm{k}_S)\rrangle \nonumber \\
&&~~\,
-2a^2(\tau) \epsilon(\tau)
\llangle\tilde{\zeta}^+(\tau_*,\bm{k}_L)\tilde{\zeta}^\pm(\tau,-\bm{k}_L)\rrangle
\llangle \partial^2\tilde{\zeta}^\pm(\tau_,\bm{k}_S)
\tilde{\zeta}^{\pm}{}' (\tau,-\bm{k}_S)\rrangle\,,\label{D2prime}
\ea
where we used Eq.~\eqref{dS3dzeta}. 
This is equivalent to 
\ba
&&
\llangle \tilde{\zeta}^+(\tau_*,\bm{k}_L)D_2\tilde{\zeta}^\pm(\tau,\bm{k}_S)
\tilde{\zeta}^{\pm}{}' (\tau,-\bm{k}_L-\bm k_S) \rrangle
+2a^2(\tau)\epsilon(\tau)\llangle \tilde{\zeta}^+(\tau_*,\bm{k}_L)\tilde{\zeta}^\pm(\tau,-\bm{k}_L)\rrangle \llangle\partial^2\tilde{\zeta}^\pm(\tau,\bm{k}_S)
\tilde{\zeta}^{\pm}{}'(\tau,-\bm{k}_S)\rrangle
\notag\\
&&
\simeq 
-\llangle \tilde{\zeta}^+(\tau_*,\bm{k}_L)\tilde{\zeta}^\pm(\tau,-\bm{k}_L) \rrangle
\frac{1}{k_S^3}\frac{{\rm d}\left(k_S^3\llangle D_2\tilde{\zeta}^\pm(\tau,\bm{k}_S)
\tilde{\zeta}^{\pm}{}'(\tau,-\bm{k}_S) \rrangle\right)}{{\rm d}\ln k_S}\,.
\label{MaldacenaCR4}
\ea
Again, we need to consider the quartic interaction. 
In this case, we focus on the second term in Eq.~\eqref{reducedS4}, 
which yields the one-loop contribution
\begin{align}
& \llangle\tilde{\zeta}(\tau_*,\bm{k}_L)\tilde{\zeta}(\tau_*,-\bm{k}_L)\rrangle^{(4)}_{\text{1-loop,1PI}}|_{\rm 2nd}\nonumber\\
& \simeq 
\ri\int \rd\tau\int \frac{\rd^3k_S}{(2\pi)^3} \frac{4a(\tau)\epsilon(\tau)}{H}\Biggl[G_{++}(k_L;\tau,\tau_*)\llangle \tilde{\zeta}^+(\tau_*,\bm k_L)\tilde{\zeta}^+(\tau,-\bm k_L)\rrangle\llangle\partial^2\tilde\zeta^+(\tau,\bm k_S)\tilde\zeta^+{}'(\tau,-\bm k_S)\rrangle\nonumber\\
   &\hspace{4.3cm}\,-G_{-+}(k_L;\tau,\tau_*)\llangle \tilde{\zeta}^+(\tau_*,\bm k_L)\tilde{\zeta}^-(\tau,-\bm k_L)\rrangle\llangle\partial^2\tilde\zeta^-(\tau,\bm k_S)\tilde\zeta^-{}'(\tau,-\bm k_S)\rrangle\Biggr]\,.
   \label{quarticD2prime}
\end{align}
Note that we dropped contributions including 
$\zeta'(\tau(\tau_*),\pm\bm k_1)$ and $k_1$, 
which only leaves the pair of two-point functions here.
The 1PI one-loop corrections from the cubic 
interaction include
\begin{align}
&\llangle\tilde{\zeta}(\tau_*,\bm{k}_L)
\tilde{\zeta}(\tau_*,-\bm{k}_L)\rrangle_{\text{1-loop,1PI}}^{(3)}\nonumber\\
&\supset\ri\int{\rm d}\tau\int\frac{{\rm d}^3 k_S}{(2\pi)^3}\frac{2}{a(\tau)H}\Biggl[ G_{++}(k_L;\tau,\tau_*)
\llangle
\tilde{\zeta}^+(\tau_*,\bm{k}_L)D_2\tilde{\zeta}^+(\tau,\bm{k}_S) \tilde{\zeta}^{+}{}'(\tau,-\bm{k}_L-\bm{k}_S)\rrangle_{\rm tree}\nonumber\\
&\hspace{4cm}-G_{-+}(k_L;\tau,\tau_*)
\llangle
\tilde{\zeta}^+(\tau_*,\bm{k}_L)D_2\tilde{\zeta}^-(\tau,\bm{k}_S)\tilde{\zeta}^{-}{}'(\tau,-\bm{k}_L-\bm{k}_S)\rrangle_{\rm tree}
\Biggr]\,\nonumber\\
&\simeq \ri\int{\rm d}\tau\int\frac{{\rm d}^3 k_S}{(2\pi)^3}\frac{2}{a(\tau)H}\Biggl[ G_{++}(k_L;\tau,\tau_*)
\Biggl(-\llangle \tilde{\zeta}^+(\tau_*,\bm{k}_L)\tilde{\zeta}^+(\tau,-\bm{k}_L) \rrangle
\frac{1}{k_S^3}\frac{{\rm d}\left(k^3_S\llangle D_2\tilde{\zeta}^+(\tau,\bm{k}_S)
\tilde{\zeta}^{+}{}'(\tau,-\bm{k}_S) \rrangle\right)}{{\rm d}\ln k_S}\nonumber\\
&\hspace{6.4cm}-2a^2(\tau)\epsilon(\tau)\llangle \tilde{\zeta}^+(\tau_*,\bm{k}_L)\tilde{\zeta}^+(\tau,-\bm{k}_L)\rrangle \llangle\partial^2\tilde{\zeta}^+(\tau,\bm{k}_S)
\tilde{\zeta}^{+}{}'(\tau,-\bm{k}_S)\rrangle\Biggr)\nonumber\\
&\hspace{4cm}-G_{-+}(k_L;\tau,\tau_*)
\Biggl(-\llangle \tilde{\zeta}^+(\tau_*,\bm{k}_L)
\tilde{\zeta}^-(\tau,-\bm{k}_L) \rrangle
\frac{1}{k_S^3}\frac{{\rm d}\left(k_S^3\llangle D_2\tilde{\zeta}^-(\tau,\bm{k}_S)\tilde{\zeta}^{-}{}'(\tau,-\bm{k}_S) 
\rrangle\right)}{{\rm d}\ln k_S}\nonumber\\
&\hspace{6.8cm}-2a^2(\tau)\epsilon(\tau)\llangle \tilde{\zeta}^+(\tau_*,\bm{k}_L)\tilde{\zeta}^-(\tau,-\bm{k}_L)\rrangle \llangle\partial^2\tilde{\zeta}^-(\tau,\bm{k}_S)
\tilde{\zeta}^{-}{}'(\tau,-\bm{k}_S)\rrangle\Biggr)\Biggr]\,,\label{zetaprimeD2zetaloop}
\end{align}
where we used Eq.~\eqref{D2prime} in the approximate equality. 
As in the previous case, we find the precise cancellation of 
Eq.~\eqref{quarticD2prime} with
the second and the fourth lines of the last approximate equality of \eqref{zetaprimeD2zetaloop}. 
This proves the consistency relation (\ref{zetaD2zetazetaprime}).

Since we showed all the consistency relations, the master 
formula \eqref{ratio-one-loop-tree} for one-loop corrections 
to the tree-level power spectrum holds within our reduced set of interactions in a self-consistent manner. We expect that  
the consistency relations should hold even by incorporating  
all the interactions, but the proof by explicit 
computations would be rather cumbersome. 
In the next section, we use the proven 
formula~\eqref{ratio-one-loop-tree} to evaluate one-loop 
corrections to the power spectrum in the transient USR model.

\section{One-loop corrections to the power spectrum 
in transient USR inflation}
\label{oneloopresult}

In the previous section, we proved the four consistency 
relations, with which the formula of one-loop 
corrections \eqref{ratio-one-loop-tree} can be used. 
In this section, we show that large one-loop corrections argued in Ref.~\cite{Kristiano:2022maq} 
are absent by using the formula (\ref{ratio-one-loop-tree}). 
The contributions to Eq.~(\ref{ratio-one-loop-tree}) contain the expectation values of cubic interactions, where, in our case, 
we have picked up the (self-consistent) set of leading terms~\eqref{cubicL}. Therefore, we separately evaluate 
the contributions from each cubic interaction in the following. 
As we have seen, the terms with spatial derivatives should be 
accompanied by quartic interactions, and we also take 
such terms into account.

\begin{itemize}
\item Interaction 1: $\displaystyle {\cal L}_1=\Mpl^2\int{\rm d}\tau{\rm d}^3x\frac{a^2}{2}\epsilon\eta'\zeta^2\zeta'$

In this case, the coupling $\lambda=\lambda_3^{(A)}(\tau)$ and 
derivative operators ${\cal{D}}$'s in Eq.~(\ref{cubicgen}) 
are given by
\be
\lambda=\Mpl^2\frac{a^2}{2}\epsilon\eta',
\hspace{0.3cm}
\text{and}
\hspace{0.3cm}
{\cal{D}}=\partial_\tau,1,1\,.
\ee
Substituting them into the master formula~\eqref{ratio-one-loop-tree}, 
the ratio between the one-loop correction and the tree-level power 
spectrum is evaluated as
\ba
&&
\frac{P_{\cal{\zeta}}^{(\text{1-loop})}(\tau_*,k_L)}{P_{\cal{\zeta}}^{(\rm {tree})}(\tau_*,k_L)}
\simeq
\left[\Mpl^2a^2\epsilon\Delta\eta\left({\rm Im}\left(\zeta_{k_L}'(\tau)\bar{\zeta}_{k_L}(\tau_*)\right){\cal{P}}_{\cal{\zeta}}^{\rm (tree)}(\tau,k_{\rm min})+2{\rm Im}\left(\zeta_{k_L}(\tau)\bar{\zeta}_{k_L}(\tau_*)\right)\partial_\tau{\cal{P}}_{\cal{\zeta}}^{\rm (tree)}
(\tau,k_{\rm min})\right)\right]_{\tau_s}^{\tau_e}
\notag
\\
&&
\hspace{0.2cm}
\simeq
\left[\Mpl^2a^2\epsilon\Delta\eta\left(
\underset{=-(4a^2\epsilon \Mpl^2)^{-1}}{\underline{{\rm Im}\left(\zeta_{k_L}'(\tau)\bar{\zeta}_{k_L}(\tau)\right)}}{\cal{P}}_{\cal{\zeta}}^{\rm (tree)}(\tau,k_{\rm min})+2\underset{=0}{\underline{{\rm Im}\left(\zeta_{k_L}(\tau)\bar{\zeta}_{k_L}(\tau)\right)}}\partial_\tau{\cal{P}}_{\cal{\zeta}}^{\rm (tree)}(\tau,k_{\rm min})\right)\right]_{\tau_s}^{\tau_e}
\notag
\\
&&
\hspace{0.2cm}
\simeq-\frac{\Delta\eta}{4}\left[
{\cal{P}}_{\cal{\zeta}}^{\rm (tree)}(\tau_s,k_{\rm min})-{\cal{P}}_{\cal{\zeta}}^{\rm (tree)}(\tau_e,k_{\rm min})\right]
\simeq 0\,,\label{bulkoneloop}
\ea
where we used $\zeta_{k_L}(\tau_*)\simeq\zeta_{k_L}(\tau)$ and the normalization condition of the mode function~\eqref{normalizationzeta}. 
Note that the power spectrum on the right-hand side is the dimensionless one $\mathcal{P}_\zeta^{(\rm tree)}(\tau,k)=
[k^3/(2\pi^2)]\llangle\zeta(\tau,\bm k)\zeta(\tau,-\bm k)\rrangle$, not the one in Eq.~\eqref{Pzetadef}.
The last equality is justified since $k_{\rm min}$ is chosen to be far outside the horizon at the beginning of the USR phase, i.e., in the range 
(\ref{krange}). Since $k_{\rm min} \ll k_s$, the effective constancy of 
curvature perturbations during the USR period 
holds for the mode $k_{\rm min}$. 
Note that the final approximate equality in Eq.~(\ref{bulkoneloop}) denotes 
the equality up to the corrections by non-constant parts of the mode function, 
which would be proportional to $r_{k_{\rm min}}(\tau_*)$. 
However, they are supposed to be negligibly small as long as we are concerned with the mode that exits the Hubble horizon 
well before the USR phase.

\item Interactions 2, 3: $\displaystyle {\cal L}_2=\Mpl^2\frac{\eta}{2}\zeta^2D_2\zeta+\Mpl^2\frac{a^2\epsilon\eta}{3}\zeta^3\partial^2\zeta$ and $\displaystyle {\cal L}_3=\Mpl^2\frac{2}{aH}\zeta\zeta'D_2\zeta+\Mpl^2\frac{2a\epsilon}{H}\zeta^2\zeta'\partial^2\zeta$

One can readily substitute the coupling and derivative operators into \eqref{ratio-one-loop-tree}. However, as we showed in Eqs.~\eqref{MaldacenaCR3} and \eqref{MaldacenaCR4}, the resultant 
three-point function becomes the product of the large-scale mode 
power spectrum and the two-point function of 
operators including $D_2$. 
This is achieved only after including the quartic 
interactions that compensate the terms containing spatial derivatives, 
to prove the correct consistency relations as we have explicitly shown. The resultant correlation function takes the form 
$\llangle D_2\zeta^{\pm}(\tau,\bm k_S)\zeta^\pm(\tau,-\bm k_S)\rrangle=\mp \ri\delta(0)(2\Mpl^2)^{-1}$, which 
follows from Eq.~\eqref{EOMtimeprop}. We can generally remove 
such a singular term by the mechanism discussed in Appendix~\ref{mostgeneralpathintegral}. Therefore, the terms proportional to $\llangle D_2\zeta^{\pm}(\tau,\bm k_S)\zeta^\pm(\tau,-\bm k_S)\rrangle$ give no contributions to (\ref{ratio-one-loop-tree}). 
It is also possible that $D_2$ acts on the propagator 
that explicitly appears in Eq.~(\ref{ratio-one-loop-tree}), but it leads to $\delta(\tau-\tau_*)$ and the time-dependent couplings become negligibly small since $\tau_*$ is sufficiently far from $\tau_e$.
Thus, we can simply neglect such contributions.
\end{itemize}

Thus, we showed that leading-order one-loop 
corrections to the tree-level power spectrum are suppressed 
in transient USR inflation. We recall that there exists 
a correction to the large-scale mode roughly proportional to 
the factor $r_k$, see Eq.~\eqref{defrk}. 
Since $r_k$ is very much smaller than 1, this correction 
is strongly suppressed even in the presence of the USR period 
during which the curvature perturbation can be enhanced. 
However, the enhancement of the CMB mode is negligibly small in realistic transient USR models. Therefore, as we have seen, the proof of the 
time-independence of $\zeta$ given in Ref.~\cite{Pimentel:2012tw} applies 
more or less in the same manner. Notice that our result is practically independent of whether $\eta$ is continuous in time and of 
replacing the USR phase with a constant roll phase. 
Such a rather background-independent result arises as a consequence of 
the symmetry that inflationary spacetime has. 
For completeness, we explicitly present all the one-loop contributions from the reduced Lagrangian~\eqref{cubicL} in Appendix~\ref{explicitform}.

\section{Conclusion}\label{conclusion}

In this paper, we have shown that one-loop corrections to 
the tree-level power spectrum of large-scale curvature 
perturbations induced by the growth of small-scale perturbations 
are suppressed to be small in the context of transient 
USR inflation. As emphasized in Introduction, this result 
is based on the work by Pimentel, Senatore, and Zaldarriaga~\cite{Pimentel:2012tw}, where the consistency relation 
plays a crucial role in proving the constancy of $\zeta$ against one-loop corrections. The consistency relation is expected to hold as 
it originates from the symmetry of curvature perturbations trivially present on the FLRW background. 
The absence of large one-loop corrections is rather natural from 
physical intuition that effective zero modes would be unphysical 
for local observers inside the horizon. 

Using the master formula shown in Ref.~\cite{Pimentel:2012tw} achieves 
the independence of the form of an interaction Lagrangian (Hamiltonian). 
The key to applying the formula is the consistency relation and 
the constancy of $\zeta$ for long-wavelength 
perturbations at tree level, which we discussed in Sec.~\ref{effconst}. 
As we showed explicitly, the CMB-scale curvature perturbation 
stays as an effective constant during the USR phase, 
despite the enhancement of $\zeta$ for small-scale 
modes relevant to the seed 
generation of PBHs. Indeed, the effective constancy of $\zeta$
for CMB perturbations after the Hubble radius crossing is
necessary to show that the consistency relations hold. 
In other words, if we consider one-loop corrections to the power spectrum 
of a mode that is not constant during the USR phase, our result does not apply.\footnote{Such a situation is exceptional also in the all-order-proof of the $\zeta$ constancy~\cite{Senatore:2012ya,Assassi:2012et}, 
since both proofs assume the tree-level conservation of $\zeta$. 
Furthermore, as explicitly shown in Ref.~\cite{Motohashi:2023syh} (Figure 9), the consistency relation does not hold for the mode that exits 
the Hubble horizon slightly before the end of 
the USR stage $-k_L\tau_e=0.1$. Our result cannot be applied to such momentum modes since neither the consistency relations nor 
the effective constancy hold.} 
In this sense, we are not sure whether the perturbation 
theory of sub-Hubble modes during the USR phase transition 
leads to large one-loop corrections, which is beyond the scope of this work.\footnote{Recently, 
non-perturbative approaches using lattice simulations have also been 
studied in inflation models with small-scale features~\cite{Caravano:2021pgc,Caravano:2024tlp,Mizuguchi:2024kbl}.}

At technical levels, the boundary and EOM terms may cause the subtlety 
of one-loop corrections, and one of the lessons from this work is that 
we need to reduce the effective Lagrangian in such a way that the 
resulting Lagrangian effectively reproduces the property of symmetry 
of the system. Namely, the consistency relations hold not just for $\ev{\zeta^3}$ but for all the three-point functions that appear in the interactions, as we have proven explicitly.

Although we have used the path-integral formalism to derive 
the master formula (\ref{ratio-one-loop-tree}),   
one can apply the formula in Ref.~\cite{Pimentel:2012tw} 
based on the operator formalism, but we expect no differences between them. Unfortunately, it seems difficult to find the origin of the discrepancy between ours and that in Ref.~\cite{Kristiano:2022maq} due to multiple differences in formulations and approximations besides the ``leading-order'' interactions taken into account. However, we emphasize again that the generality of the master formula requires only the tree-level constancy of $\zeta$ and the consistency relations, both of which are satisfied within the transient USR model. In other words, if our result were somehow falsified, at least one of the requirements 
would be false. 

It is expected that the master formula also holds for more general single-field inflation with second-order equations of motion 
(dubbed Horndeski's theories) \cite{Horndeski,Kobayashi:2011nu}.
Indeed, the seed generation of PBHs was studied in the context of  
Higgs inflation \cite{Ezquiaga:2017fvi}, $f(R)$ gravity \cite{Frolovsky:2022ewg}, 
nonminimal derivative couplings to gravity \cite{Fu:2019ttf}, 
Galileons \cite{Choudhury:2023hvf}, scalar Gauss-Bonnet 
couplings \cite{Kawai:2021edk}, 
Higgs Gauss-Bonnet inflation \cite{Kawaguchi:2022nku} and so on, 
which all belong to the subclasses of Horndeski's theories. 
Upon using the cubic-order action 
derived in Refs.~\cite{Gao:2011qe,DeFelice:2011uc} for 
full Horndeski's theories, it would be of interest to study whether 
one-loop corrections to the large-scale power spectrum 
remain small by proving more general consistency relations 
possibly arising from additional operators.

Throughout this work, we have considered one-loop corrections to 
the two-point function where only cubic and quartic interactions contribute. One may wonder e.g., whether higher loop corrections yield large corrections. Our result cannot be applied and another way of investigation would be necessary. Nevertheless, the all-order proof of the constancy of $\zeta$ shown in Ref.~\cite{Senatore:2012ya} only assumes the constancy 
at the tree level, and the constancy against higher-order corrections is shown by induction. Therefore, we expect that the approximate constancy of super-Hubble modes within the transient USR model would suffice the assumptions used in Ref.~\cite{Senatore:2012ya}. 
Then, the absence of enhancement of CMB modes should be true 
in all-order in perturbation series. We will discuss it elsewhere.

\section*{Note added}

After the initial submission of our paper, the reference \cite{Fumagalli:2024jzz} appeared on arXiv. In Ref.~\cite{Fumagalli:2024jzz}, the author also applies the master formula 
in Ref.~\cite{Pimentel:2012tw} to the transient USR model as ours but uses the operator formalism. Due to the differences between path integral and operator formalisms, such as the roles of boundary and EOM terms, it is not easy to compare its intermediate results and technical points directly with ours. Nevertheless, the final results of Ref.~\cite{Fumagalli:2024jzz} are consistent 
with ours. Furthermore, in Ref.~\cite{Fumagalli:2024jzz}, 
the removal of tadpole contributions is 
explicitly discussed.

\section*{Acknowledgements}
We thank Jacopo Fumagalli, Mohammad Ali Gorji, Keisuke Inomata, Jason Kristiano, 
Yusuke Mikura, Yuichiro Tada, and Shuichiro Yokoyama for 
useful discussions. 
RK is supported by JSPS KAKENHI Grants No.~24KJ2108. 
ST is supported by the Grant-in-Aid for Scientific 
Research Fund of the JSPS No.~22K03642 and 
Waseda University Special Research Project 
No.~2024C-474.

\appendix

\section{Path-integral formalism for general models 
with derivative couplings}
\label{mostgeneralpathintegral}

We review how to treat the system with derivative 
interactions based on the analysis in Ref.~\cite{Abolhasani:2022twf}.
In particular, we consider the terms up to quartic order as they are relevant to this work.\footnote{The reduced Lagrangian~\eqref{cubicL} discussed 
in the main text includes the second derivative of $\zeta$ 
with respect to 
time in the EOM terms. However, the original Lagrangian consists of polynomials of at most first-order derivatives of the fields, 
and the second-order derivative terms such as EOM terms appear 
due to the technique of integration-by-parts. 
In this sense, Eq.~\eqref{mostgenerallagrangian} 
is the most general case we would have.} 
We consider the most general Lagrangian 
up to quartic order given by 
\be
{\cal{L}}(\zeta,\zeta')=\sum_{n=2}^{4}\sum_{m=0}^{n}\lambda_{n-m}^{(n)}\zeta^{n-m}(\zeta')^{m}\,,
\label{mostgenerallagrangian}
\ee
where the couplings $\lambda_{n-m}^{(n)}$'s are generally 
time-dependent and may contain spatial derivatives acting 
on each $\zeta$.\footnote{The spatial derivatives can be treated as constant parameters in Fourier space, and therefore, we deal with the spatial derivatives as if they were constant and implicitly include 
them in the couplings.} 
From the above Lagrangian, the canonical conjugate momentum 
of $\zeta$ is 
\be
\pi_\zeta=\frac{\partial{\cal{L}}}{\partial\zeta'}=\sum_{n=2}^{4}
\sum_{m=1}^{n}m\lambda_{n-m}^{(n)}\zeta^{n-m}(\zeta')^{m-1}\,.
\label{pizeta}
\ee
From Eq.~\eqref{pizeta}, we can express $\zeta'$ as a series in $\zeta$ and $\pi_\zeta$, as
\ba
&&
\zeta'(\zeta,\pi_\zeta)=\frac{1}{2\lambda_0^{(2)}}\left(\pi_\zeta-\lambda_1^{(2)}\zeta\right)
\notag\\
&&
-\frac{3\lambda_0^{(3)}}{8(\lambda_0^{(2)})^3}\pi_\zeta^2-\frac{\zeta\pi_\zeta}{4(\lambda_0^{(2)})^3}\left(3\lambda_1^{(2)}\lambda_0^{(3)}-2\lambda_0^{(2)}\lambda_1^{(3)}\right)-\frac{\zeta^2}{8(\lambda_0^{(2)})^3}\left(3(\lambda_1^{(2)})^2\lambda_0^{(3)}-4\lambda_0^{(2)}\lambda_1^{(2)}\lambda_1^{(3)}
+4(\lambda_0^{(2)})^2\lambda_2^{(3)}\right)
\notag\\
&&
+\frac{\pi_\zeta^3}{16(\lambda_0^{(2)})^5}\left(9(\lambda_0^{(3)})^2-4\lambda_0^{(2)}\lambda_0^{(4)}\right)
-\frac{3\zeta\pi_\zeta^2}{16(\lambda_0^{(2)})^5}
\left[ \lambda_1^{(2)}\left(9(\lambda_0^{(3)})^2-4\lambda_0^{(2)}\lambda_0^{(4)}\right)+2\lambda_0^{(2)}\left(-3\lambda_0^{(3)}\lambda_1^{(3)}+\lambda_0^{(2)}\lambda_1^{(4)}\right)\right]
\notag\\
&&
+\frac{\zeta^2\pi_\zeta}{16(\lambda_0^{(2)})^5}\biggl[ 
3(\lambda_1^{(2)})^2\left(9(\lambda_0^{(3)})^2-4\lambda_0^{(2)}\lambda_0^{(4)}\right)+12\lambda_0^{(2)}\lambda_1^{(2)}\left(-3\lambda_0^{(3)}\lambda_1^{(3)}+\lambda_0^{(2)}\lambda_1^{(4)}\right)
\notag\\
&&
\hspace{1.8cm}
+4(\lambda_0^{(2)})^2\left(2(\lambda_1^{(3)})^2+3\lambda_0^{(3)}
\lambda_2^{(3)}-2\lambda_0^{(2)}\lambda_2^{(4)}\right)\biggr]
\notag\\
&&
-\frac{\zeta^3}{16(\lambda_0^{(2)})^5}
\biggl[ (\lambda_1^{(2)})^3\left(9(\lambda_0^{(3)})^2 -4\lambda_0^{(2)}\lambda_0^{(4)}\right)+6\lambda_0^{(2)}(\lambda_1^{(2)})^2\left(-3\lambda_0^{(3)}\lambda_1^{(3)}+\lambda_0^{(2)}\lambda_1^{(4)}\right)
\notag\\
&&
\hspace{1.8cm}
+4(\lambda_0^{(2)})^2\lambda_1^{(2)}
\left( 2(\lambda_1^{(3)})^2+3\lambda_0^{(3)}\lambda_2^{(3)}
-2\lambda_0^{(2)}\lambda_2^{(4)}\right)
+8(\lambda_0^{(2)})^3\left(-\lambda_1^{(3)}\lambda_2^{(3)}+\lambda_0^{(2)}\lambda_3^{(4)}\right)
\biggr]\,.
\ea
The Legendre transform of the Lagrangian 
yields the Hamiltonian in the form 
\be
{\cal{H}}(\zeta,\pi_\zeta)=\pi_\zeta \zeta'(\zeta,\pi_\zeta)-{\cal{L}}(\zeta,\zeta'(\zeta,\pi_\zeta))=\sum_{n=2}^{4}\sum_{m=0}^{n}
\tilde{\lambda}_{n-m}^{(n)}\zeta^{n-m}\pi_\zeta^{m}\,,
\ee
where
\ba
&&
\tilde{\lambda}_0^{(2)}=\frac{1}{4\lambda_0^{(2)}},
\hspace{1cm}
\tilde{\lambda}_1^{(2)}=-\frac{\lambda_1^{(2)}}{2\lambda_0^{(2)}},
\hspace{1cm}
\tilde{\lambda}_2^{(2)}=\frac{1}{4\lambda_0^{(2)}}\left((\lambda_1^{(2)})^2-4\lambda_0^{(2)}\lambda_2^{(2)}\right),
\notag\\
&&
\tilde{\lambda}_0^{(3)}=-\frac{\lambda_0^{(3)}}{8(\lambda_0^{(2)})^3},
\hspace{1cm}
\tilde{\lambda}_1^{(3)}=\frac{1}{8(\lambda_0^{(2)})^3}\left(3\lambda_1^{(2)}\lambda_0^{(3)}-2\lambda_0^{(2)}\lambda_1^{(3)}\right),
\notag\\
&&
\tilde{\lambda}_2^{(3)}=-\frac{1}{8(\lambda_0^{(2)})^3}\left(3(\lambda_1^{(2)})^2\lambda_0^{(3)}-4\lambda_0^{(2)}\lambda_1^{(2)}\lambda_1^{(3)}+4(\lambda_0^{(2)})^2\lambda_2^{(3)}\right)\,,
\notag\\
&&
\tilde{\lambda}_3^{(3)}=\frac{1}{8(\lambda_0^{(2)})^3}\left((\lambda_1^{(2)})^3\lambda_0^{(3)}-2\lambda_0^{(2)}(\lambda_1^{(2)})^2\lambda_1^{(3)}+4(\lambda_0^{(2)})^2\lambda_1^{(2)}\lambda_2^{(3)}-8(\lambda_0^{(2)})^3\lambda_3^{(3)}\right)\,,
\notag\\
&&
\tilde{\lambda}_0^{(4)}=\frac{1}{64(\lambda_0^{(2)})^5}\left(9(\lambda_0^{(3)})^2-4\lambda_0^{(2)}\lambda_0^{(4)}\right)\,,
\notag\\
&&
\tilde{\lambda}_1^{(4)}=\frac{1}{16(\lambda_0^{(2)})^5}\left(-9\lambda_1^{(2)}(\lambda_0^{(3)})^2+6\lambda_0^{(2)}\lambda_0^{(3)}\lambda_1^{(3)}+4\lambda_0^{(2)}\lambda_1^{(2)}\lambda_0^{(4)}-2(\lambda_0^{(2)})^2\lambda_1^{(4)}\right)\,,
\notag\\
&&
\tilde{\lambda}_2^{(4)}=\frac{1}{32(\lambda_0^{(2)})^5}\Bigl(27(\lambda_1^{(2)})^2(\lambda_0^{(3)})^2-36\lambda_0^{(2)}\lambda_1^{(2)}\lambda_0^{(3)}\lambda_1^{(3)}+8(\lambda_0^{(2)})^2(\lambda_1^{(3)})^2+12(\lambda_0^{(2)})^2\lambda_0^{(3)}\lambda_2^{(3)}
\notag\\
&&
\hspace{2.7cm}
-12\lambda_0^{(2)}(\lambda_1^{(2)})^2\lambda_0^{(4)}+12(\lambda_0^{(2)})^2\lambda_1^{(2)}\lambda_1^{(4)}-8(\lambda_0^{(2)})^3\lambda_2^{(4)}\Bigr),
\notag\\
&&
\tilde{\lambda}_3^{(4)}=\frac{1}{32(\lambda_0^{(2)})^5}\Bigl(-18(\lambda_1^{(2)})^3(\lambda_0^{(3)})^2+36\lambda_0^{(2)}(\lambda_1^{(2)})^2\lambda_0^{(3)}\lambda_1^{(3)}-16(\lambda_0^{(2)})^2\lambda_1^{(2)}(\lambda_1^{(3)})^2-24(\lambda_0^{(2)})^2\lambda_1^{(2)}\lambda_0^{(3)}\lambda_2^{(3)}
\notag\\
&&
\hspace{2.7cm}
+16(\lambda_0^{(2)})^3\lambda_1^{(3)}\lambda_2^{(3)}+8\lambda_0^{(2)}(\lambda_1^{(2)})^3\lambda_0^{(4)}-12(\lambda_0^{(2)})^2(\lambda_1^{(2)})^2\lambda_1^{(4)}+16(\lambda_0^{(2)})^3\lambda_1^{(2)}\lambda_2^{(4)}-16(\lambda_0^{(2)})^4\lambda_3^{(4)}\Bigr),
\notag
\\
&&
\tilde{\lambda}_4^{(4)}=\frac{1}{64(\lambda_0^{(2)})^5}\Bigl(9(\lambda_1^{(2)})^4(\lambda_0^{(3)})^2-24\lambda_0^{(2)}(\lambda_1^{(2)})^3\lambda_0^{(3)}\lambda_1^{(3)}+16(\lambda_0^{(2)})^2(\lambda_1^{(2)})^2(\lambda_1^{(3)})^2+24(\lambda_0^{(2)})^2(\lambda_1^{(2)})^2\lambda_0^{(3)}\lambda_2^{(3)}
\notag\\
&&
\hspace{2.7cm}
-32(\lambda_0^{(2)})^3\lambda_1^{(2)}\lambda_1^{(3)}\lambda_2^{(3)}+16(\lambda_0^{(2)})^4(\lambda_2^{(3)})^2-4\lambda_0^{(2)}(\lambda_1^{(2)})^4\lambda_0^{(4)}+8(\lambda_0^{(2)})^2(\lambda_1^{(2)})^3\lambda_1^{(4)}
\notag\\
&&
\hspace{2.7cm}
-16(\lambda_0^{(2)})^3(\lambda_1^{(2)})^2\lambda_2^{(4)}+32(\lambda_0^{(2)})^4\lambda_1^{(2)}\lambda_3^{(4)}-64(\lambda_0^{(2)})^5\lambda_4^{(4)}\Bigr).
\ea
Let us consider sourced generating functional of $\zeta$ and its canonical momentum $\pi_\zeta$. 
For now, we consider the generating functional for time-ordered 
products, which can be straightforwardly generalized to 
the in-in formalism. 
The generating functional $Z[J,K]$ of $\zeta$ 
and $\pi_\zeta$ is given by
\be
Z[J,K]=\int{\cal{D}}\zeta{\cal{D}}\pi_\zeta
\exp\left[\frac{\ri}{\hbar}\int{\rm d}\tau{\rm d}^3x 
\left(\pi_\zeta\zeta'-{\cal{H}}(\zeta,\pi_\zeta)+J\zeta+K\pi_\zeta\right)\right]\,,
\label{Zzetapizeta}
\ee
where $J$ and $K$ are external source fields.
The Hamiltonian can be decomposed into the free 
and interaction parts, ${\cal{H}}_{\rm free}$ 
and ${\cal{H}}_{\rm int}$.
Here, we assign quadratic terms in the Hamiltonian 
as the free part and higher-order terms as the interaction part.
Substituting ${\cal{H}}={\cal{H}}_{\rm free}+{\cal{H}}_{\rm int}$ into Eq.~\eqref{Zzetapizeta} and performing the Gauss integral 
in terms of $\pi_\zeta$, we can rewrite the generating 
functional $Z[J,K]$, as
\be
Z[J,K]=N_0 \exp\left[-\frac{\ri}{\hbar}\int{\rm d}\tau
{\rm d}^3x\,{\cal{H}}_{\rm int}\left(\frac{\hbar}{\ri}\frac{\delta}{\delta J(x)},\frac{\hbar}{\ri}\frac{\delta}{\delta K(x)}\right)\right]Z_0[J,K]\,,
\label{Zzeta}
\ee
where $N_0$ is the normalization factor and $\delta/\delta X$ 
is the functional derivative with respect to $X$.
$Z_0$ is a free part of the generating functional 
given by 
\be
Z_0[J,K]=\int {\cal{D}}\zeta \exp\left[\frac{\ri}{\hbar}\int{\rm d}\tau{\rm d}^3x\left({\cal{L}}_{\rm free}+\pi_{\zeta,{\rm free}}^{(\rm cl)}K+\lambda_0^{(2)}K^2+J\zeta\right)\right]\,,
\ee
where ${\cal{L}}_{\rm free}$ and $\pi_{\zeta,\rm free}^{(\rm cl)}$ are free parts of the Lagrangian and classical canonical momentum, i.e.,
\ba
{\cal{L}}_{\rm free}
&=& \lambda_0^{(2)}(\zeta')^2+\lambda_1^{(2)}\zeta(\zeta')
+\lambda_2^{(2)}\zeta^2\,,\\
\pi_{\zeta,\rm free}^{(\rm cl)} 
&=&\frac{\partial{\cal{L}}_{\rm free}}{\partial\zeta'}=2\lambda_0^{(2)}\zeta'+\lambda_1^{(2)}\zeta\,.
\ea

Due to the presence of derivative couplings, we cannot easily pull ${\cal{H}}_{\rm int}$ back into the integral, which is evident from the fact that Eq.~\eqref{Zzeta} contains $K^2$ in the integrand.
Let us look at a specific example. 
A term $\lambda_1^{(3)}\zeta(\zeta')^2$ in the interaction 
Lagrangian leads to 
\ba
&&
{\cal{H}}_{\rm int}=-\frac{\lambda_1^{(3)}}{4(\lambda_0^{(2)})^2}\zeta\pi_\zeta^2
+\frac{\lambda_1^{(2)}\lambda_1^{(3)}}{2(\lambda_0^{(2)})^2}
\zeta^2\pi_\zeta-\frac{(\lambda_1^{(2)})^2\lambda_1^{(3)}}
{4(\lambda_0^{(2)})^2}\zeta^3
\nonumber \\
&&
\hspace{1.1cm}
+\frac{(\lambda_1^{(3)})^2}{4(\lambda_0^{(2)})^3}\zeta^2\pi_\zeta^2-\frac{\lambda_1^{(2)}(\lambda_1^{(3)})^2}{2(\lambda_0^{(2)})^3}\zeta^3\pi_\zeta+\frac{(\lambda_1^{(2)})^2(\lambda_1^{(3)})^2}{4(\lambda_0^{(2)})^3}\zeta^4\,.
\label{A12}
\ea
The first and second lines show the cubic and quartic 
Hamiltonians, respectively.
Let us focus on the first term in the cubic part,
\be
{\cal{H}}_{\rm int}\supset\tilde{\lambda}_1^{(3)}\zeta\pi_\zeta^2
\rightarrow\tilde{\lambda}_1^{(3)}\left(\frac{\hbar}{\ri}\frac{\delta}{\delta J}\right)\left(\frac{\hbar}{\ri}\frac{\delta}{\delta K}\right)^2\,.
\ee
Substituting it into Eq.~\eqref{Zzeta}, we obtain 
\ba
\hspace{-1.0cm}
Z[J,K]\Bigr|_{J,K=0}
&\supset&
\left[-\frac{\ri}{\hbar}\int {\rm d}^4 y\tilde{\lambda}_1^{(3)}(y)\left(\frac{\hbar}{\ri}\frac{\delta}{\delta J(y)}\right)\left(\frac{\hbar}{\ri}\frac{\delta}{\delta K(y)}\right)^2\right]Z_0[J,K] \biggr|_{J,K=0}
\notag \\
\hspace{-1.0cm}
&=& \int {\cal{D}}\zeta
\left[ -\frac{\ri}{\hbar}\int{\rm d}^4 y\tilde{\lambda}_1^{(3)}(y)\zeta(y)\left((\pi_{\zeta,{\rm free}}^{(\rm cl)}(y))^2-2\ri\hbar\lambda_0^{(2)}(y)\delta^{(4)}(0)\right)
\right] \exp\left(\frac{\ri}{\hbar}
\int{\rm d}^4 x\,{\cal{L}}_{\rm free}\right)
\notag \\
\hspace{-1.0cm}
&=&\int {\cal{D}}\zeta\left[-\frac{\ri}{\hbar}\int{\rm d}^4 y\left( {\cal{H}}_{\rm int}(\zeta(y),\pi_{\zeta,{\rm free}}^{(\rm cl)}(y))-2\ri\hbar\lambda_0^{(2)}(y)\tilde{\lambda}_1^{(3)}(y)\zeta(y)\delta^{(4)}(0)\right)\right]\exp\left(\frac{\ri}{\hbar}
\int{\rm d}^4 x\,{\cal{L}}_{\rm free}\right),
\label{A14}
\ea
at linear order in $\tilde{\lambda}_1^{(3)}$, and
\ba
Z[J,K]\Bigr|_{J,K=0}
&\supset&
\frac{1}{2}\left[-\frac{\ri}{\hbar}\int {\rm d}^4 y\tilde{\lambda}_1^{(3)}(y)\left(\frac{\hbar}{\ri}\frac{\delta}{\delta J(y)}\right)\left(\frac{\hbar}{\ri}\frac{\delta}{\delta K(y)}\right)^2\right]^2Z_0[J,K] \biggr|_{J,K=0}
\notag\\
&=&\int {\cal{D}}\zeta\,\frac{1}{2}\Biggl[ 
\left(-\frac{\ri}{\hbar}\int{\rm d}^4 y\tilde{\lambda}_1^{(3)}(y)\zeta(y)\left((\pi_{\zeta,{\rm free}}^{(\rm cl)}(y))^2-2\ri\hbar\lambda_0^{(2)}(y)\delta^{(4)}(0)\right)\right)^2
\notag\\
&&
+\frac{8\ri}{\hbar}\int{\rm d}^4 y \lambda_0^{(2)}(y)(\tilde{\lambda}_1^{(3)}(y))^2\zeta^2(y)(\pi_{\zeta,{\rm free}}^{(\rm cl)}(y))^2\Biggr]
\exp\left(\frac{\ri}{\hbar}\int{\rm d}^4 x\,{\cal{L}}_{\rm free}\right)
\notag\\
&=&
\int {\cal{D}}\zeta\Biggl[ \frac{1}{2}\left(-\frac{\ri}{\hbar}\int{\rm d}^4 y \left({\cal{H}}_{\rm int}(\zeta(y),\pi_{\zeta,{\rm free}}^{\rm (cl)}(y))-2\ri\hbar\lambda_0^{(2)}(y)\tilde{\lambda}_1^{(3)}(y)\zeta(y)\delta^{(4)}(0)\right)\right)^2
\notag\\
&&
+\frac{4\ri}{\hbar}\int{\rm d}^4 y \lambda_0^{(2)}(y)(\tilde{\lambda}_1^{(3)}(y))^2\zeta^2(y)(\pi_{\zeta,{\rm free}}^{(\rm cl)}(y))^2\Biggr]
\exp\left(\frac{\ri}{\hbar}\int{\rm d}^4 x\,{\cal{L}}_{\rm free}\right)\,,
\label{A15}
\ea
at quadratic order in $\tilde{\lambda}_1^{(3)}$.\footnote{Here, we drop $\hbar^2$ terms because it is irrelevant up to the one-loop level.}
It can be seen that two types of non-trivial terms appear in Eqs.~\eqref{A14} and \eqref{A15} due to the $K^2$ term in $Z_0$.
One is the term containing $\delta^{(4)}(0)$, which arises when one of the two $K$-functional derivatives in an interaction acts on $K^2$ in $Z_0$ and the other on $K$, which has appeared outside $Z_0$ due to the previous operation.
As this term has a factor $\hbar$, it can be 
expected that it only gives contributions 
at the loop level and not at the tree level.
As our purpose is the loop calculation, such additional terms may 
also need to be taken into account.
We will discuss more details later.
Another is the effective quartic term appearing in Eq.~\eqref{A15}.
It arises when one $K$-functional derivative in one of the interactions acts on $K^2$ in $Z_0$ and one $K$-functional derivative in another of 
the interactions acts on $K$, which has appeared outside $Z_0$ 
due to the previous operation.
This effective quartic term exactly cancels the contribution from the original quartic Hamiltonian, the first term in Eq.~\eqref{A12}. 
Similarly, there are effective quartic terms from the 
second-order contributions of the cubic Hamiltonian, as 
\ba
&&
\left(\tilde{\lambda}_1^{(3)}\zeta\pi_\zeta^2\right)\times\left(\tilde{\lambda}_2^{(3)}\zeta^2\pi_\zeta\right)\rightarrow\lambda_0^{(2)}\tilde{\lambda}_1^{(3)}\tilde{\lambda}_2^{(3)}\zeta^3\pi_{\zeta,{\rm free}}^{(\rm cl)}\,,
\notag\\
&&
\left(\tilde{\lambda}_2^{(3)}\zeta^2\pi_\zeta\right)
\times\left(\tilde{\lambda}_2^{(3)}\zeta^2\pi_\zeta\right)\rightarrow\lambda_0^{(2)}(\tilde{\lambda}_2^{(3)})^2\zeta^4\,,
\ea
and they cancel the contributions from the original quartic Hamiltonian, 
the second and third terms in Eq.~\eqref{A12}.
This process results in all contributions from 
the quartic Hamiltonian generated from the cubic Lagrangian vanishing.
Therefore, the path integral is given by
\ba
&&
Z[J,K]\Bigr|_{J,K=0}=\int {\cal{D}}\zeta 
\exp \left[\frac{\ri}{\hbar}\int {\rm d}\tau{\rm d}^3x\left({\cal{L}}_{\rm free}-{\cal{H}}_{\rm int}^{(\rm cubic)}(\zeta,\pi_{\zeta,{\rm free}}^{\rm (cl)})+\ri\hbar\delta^{(4)}(0)\alpha(\zeta,\zeta')\right)\right]
\notag\\
&&
\hspace{2.2cm}
=\int {\cal{D}}\zeta \exp
\left[\frac{\ri}{\hbar}\int {\rm d}\tau{\rm d}^3x\left({\cal{L}}_{\rm free}+\lambda_1^{(3)}\zeta(\zeta')^2+\ri\hbar\delta^{(4)}(0)\alpha(\zeta,\zeta')\right)\right]\,,
\ea
where $\alpha(\zeta,\zeta')$ is a function of 
$\zeta$ and $\zeta'$, and we used 
\be
{\cal{H}}_{\rm int}^{(\rm cubic)}=-\frac{\lambda_1^{(3)}}{4(\lambda_0^{(2)})^2}\zeta\pi_\zeta^2
+\frac{\lambda_1^{(2)}\lambda_1^{(3)}}{2(\lambda_0^{(2)})^2}\zeta^2\pi_\zeta-\frac{(\lambda_1^{(2)})^2\lambda_1^{(3)}}{4(\lambda_0^{(2)})^2}\zeta^3\xrightarrow
[\pi_\zeta\to\pi_{\zeta,{\rm free}}^{\rm (cl)}]{}-\lambda_1^{(3)}\zeta(\zeta')^2\,.
\ee
Thus, there are terms proportional to $\delta^{(4)}(0)$ 
in addition to the original Lagrangian.

Let us start with the more general Lagrangian~\eqref{mostgenerallagrangian} and specify the form of $\alpha(\zeta,\zeta')$ (see Ref.~\cite{Abolhasani:2022twf} 
for detailed discussion). 
We define the ``effective Lagrangian'' $\mathcal{L}_{\rm eff}$ by
\be
\exp\left(\frac{\ri}{\hbar}\int{\rm d}\tau{\rm d}^3x{\cal{L}}_{\rm eff}\right)
\equiv N\int{\cal{D}}\pi_\zeta\exp\left[\frac{\ri}{\hbar}\int{\rm d}\tau{\rm d}^3x\left(\pi_\zeta \zeta'-{\cal{H}}(\zeta,\pi_\zeta)\right)\right]\,,
\label{def_effLagrangian}
\ee
where $N$ is a normalization factor. 
To perform the integration of $\pi_\zeta$, we need to expand the Lagrangian terms around a ``stationary point'', namely, 
around a classical solution
\be
\pi_\zeta\zeta'-{\cal{H}}={\cal{L}}-\frac{1}{2}\frac{\partial^2{\cal{H}}}{\partial\pi_\zeta^2}\biggr|_{\rm cl}
\left(\pi_\zeta-\pi_\zeta^{(\rm cl)}\right)^2+\cdots\,,
\label{clexpansion}
\ee
where $\cal{L}$ is the classical Lagrangian given by Eq.~\eqref{mostgenerallagrangian}.
The classical solution satisfies 
$\zeta'=\partial{\cal{H}}/\partial\pi_\zeta$ and is given by $\pi_\zeta=\pi_\zeta^{(\rm cl)}=\partial{\cal{L}}/\partial\zeta'$.
The ellipses denote the terms that give rise to 
$\mathcal{O}(\hbar^2)$ contributions in the effective Lagrangian, which we ignore at the one-loop order. 
Substituting Eq.~\eqref{clexpansion} into 
Eq.~\eqref{def_effLagrangian}, we obtain
\ba
&&
\int{\cal{D}}\pi_\zeta\exp
\left[\frac{\ri}{\hbar}\int{\rm d}\tau{\rm d}^3x\left(\pi_\zeta \zeta'-{\cal{H}}(\zeta,\pi_\zeta)\right)\right]
\notag\\
&&
\simeq\prod_{j}\exp\left(\ri\frac{\Omega}{\hbar}{\cal{L}}\right)\int{\rm d}\pi_{\zeta,j}\exp\left[-\ri\frac{\Omega}{2\hbar}\frac{\partial^2{\cal{H}}}{\partial\pi_\zeta^2}\Biggr|_{{\rm cl},j}\left(\pi_{\zeta,j}-\pi_{\zeta,j}^{(\rm cl)}\right)^2\right]
\notag\\
&&
=\prod_{j}\left(\frac{\ri\pi\hbar}{\Omega\,
\partial^2{\cal{H}}_0/\partial\pi_\zeta^2} \right)^{1/2}
\exp\left( 
\ri\frac{\Omega}{\hbar}{\cal{L}}\right)
\left[1+\frac{\partial^2{\cal{H}}_{\rm int}/\partial\pi_\zeta^2}{\partial^2{\cal{H}}_{0}/\partial\pi_\zeta^2}\Biggr|_{{\rm cl},j}
\right]^{-1/2}
\notag\\
&&
\simeq
\prod_{j}\left( \frac{2\ri\pi\hbar\lambda_0^{(2)}}{\Omega}\right)^{1/2}
\exp\left[ \ri\frac{\Omega}{\hbar}{\cal{L}}-\lambda_0^{(2)}\frac{\partial^2{\cal{H}}}{\partial\pi_\zeta^2}\Biggr|_{{\rm cl},j}+(\lambda_0^{(2)})^2\left(\frac{\partial^2{\cal{H}}}{\partial\pi_\zeta^2}\Biggr|_{{\rm cl},j}\right)^2\right]\,.
\ea
Here, we have taken the discrete limit of spacetime with an 
infinitesimal 4-volume $\Omega$, approximated the exponent 
by a quadratic term in the first equality, performed 
the Gaussian integration in the second equality, and used $\partial^2{\cal{H}}_0/\partial\pi_\zeta^2=(1/2)\lambda_0^{(2)}$ and $(1+x)^{-1/2}=\exp(-x/2+x^2/4)+{\cal{O}}(x^3)$ in the third equality. 
In the continuous limit $\Omega\to 0$, we identify the exponent as the effective Lagrangian given by
\be
{\cal{L}}_{\rm eff}={\cal{L}}+\ri\hbar\delta^{(4)}(0)\alpha+{\cal{O}}\left(\hbar^2\right)\,,
\ee
where
\ba
&&
\alpha=\lambda_0^{(2)}\left[ 
\frac{\partial^2{\cal{H}}}{\partial\pi_\zeta^2}\Biggr|_{{\rm cl},j}-\lambda_0^{(2)}\left(\frac{\partial^2{\cal{H}}}{\partial\pi_\zeta^2}\Biggr|_{{\rm cl},j}\right)^2\right]
\notag\\
&&
\hspace{0.33cm}
=-\frac{\lambda_1^{(3)}}{2\lambda_0^{(2)}}\zeta-\frac{3\lambda_0^{(3)}}{2\lambda_0^{(2)}}\zeta'+\left(\frac{(\lambda_1^{(3)})^2}{4(\lambda_0^{(2)})^2}-\frac{\lambda_2^{(4)}}{2\lambda_0^{(2)}}\right)\zeta^2
-\frac{3\lambda_1^{(4)}}{2\lambda_0^{(2)}}\zeta\zeta'+\left(\frac{9(\lambda_0^{(3)})^2}{4(\lambda_0^{(2)})^2}-\frac{3\lambda_0^{(4)}}{\lambda_0^{(2)}}\right)(\zeta')^2+{\cal{O}}(\zeta^3)\,.
\label{alphaterm}
\ea

Finally, we discuss the role of $\alpha$-terms 
for the computation of correlation functions. 
The $\alpha$ terms in the effective Lagrangian contain $\hbar$, 
which implies their relevance to one-loop corrections. 
The divergences including $\delta^{(4)}(0)$ arise from one-loop corrections consisting of vertices generated from interactions containing two or more $\zeta'$. Such unphysical divergences can be properly removed by taking into account the $\alpha$-terms. 
Hereafter, we take the natural unit $\hbar=1$. 

Let us consider the same term in the interaction part of Lagrangian, 
$\lambda_1^{(3)}\zeta(\zeta')^2$, as before.
In this case, the effective Lagrangian is
\be
{\cal{L}}_{\rm eff}=\lambda_0^{(2)}(\zeta')^2+\lambda_1^{(2)}\zeta\zeta'
+\lambda_2^{(2)}\zeta^2+\lambda_1^{(3)}\zeta(\zeta')^2
-\ri \delta^{(4)}(0)\frac{\lambda_1^{(3)}}{2\lambda_0^{(2)}}\zeta+\ri\delta^{(4)}(0)\frac{(\lambda_1^{(3)})^2}{4(\lambda_0^{(2)})^2}\zeta^2\,.
\label{ex_effLagrangian}
\ee
\begin{figure}[h]
\centering
\includegraphics[width=15cm]{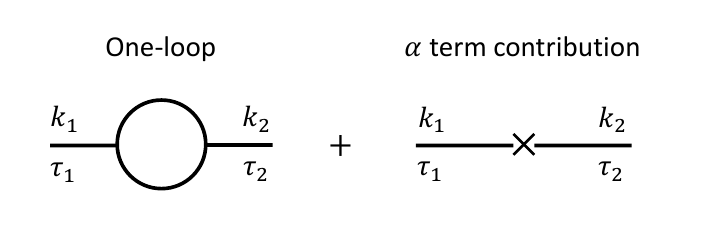}
\caption{\label{fig_alphaterm}
Feynman diagrams of the one loop and $\alpha$-term contributions 
to the two-point function.}
\end{figure}

As an illustration, let us consider a one-loop correction to 
the two-point function diagrammatically shown in Fig.~\ref{fig_alphaterm}.
The first diagram in the figure 
is the 1PI one-loop diagram consisting of two cubic interactions, 
whereas the second one consists of the last quadratic term 
in Eq.~\eqref{ex_effLagrangian}, which has the divergent 
factor $\delta^{(4)}(0)$.

Focusing on the cubic interaction $\zeta(\zeta')^2$ with the coupling $\lambda_{1}^{(3)}$, we find the following contribution to 
the left diagram in Fig.~\ref{fig_alphaterm}, 
\ba
&&
(\text{Left figure})\supset\frac{1}{2}2^2 \ri^2\int\frac{{\rm d}^3k}{(2\pi)^3}\int{\rm d}\tau{\rm d}\tau'\lambda_1^{(3)}(\tau)\lambda_1^{(3)}(\tau')G(k_1;\tau,\tau_1)\partial_{\tau\tau'}G(k;\tau,\tau')\partial_{\tau\tau'}G(|\bm{k}_1+\bm{k}|;\tau,\tau')G(k_2;\tau',\tau_2)
\notag\\
&&
\hspace{1.95cm}
\supset\frac{1}{2}\int\frac{{\rm d}^3k}{(2\pi)^3}\int{\rm d}\tau \frac{(\lambda_1^{(3)}(\tau))^2}{(\lambda_0^{(2)}(\tau))^2}G(k_1;\tau,\tau_1)G(k_2;\tau,\tau_2)\delta(0)
\notag\\
&&
\hspace{1.95cm}
=\frac{1}{2}\delta^{(4)}(0)\int{\rm d}\tau \frac{(\lambda_1^{(3)}(\tau))^2}{(\lambda_0^{(2)}(\tau))^2}G(k_1;\tau,\tau_1)G(k_2;\tau,\tau_2)\,,
\ea
where a factor $1/2$ in the first line is a symmetry factor, $\partial_{\tau\tau'}\equiv\partial_{\tau}\partial_{\tau'}$, 
and as noted before, the coupling may contain spatial derivatives 
which just become constant factors.
In the second equality, we used $\partial_{\tau\tau'}G(k;\tau,\tau')\supset{\ri}\delta(\tau-\tau')/{2\lambda_0^{(2)}}$.
On the other hand, from the right diagram 
in Fig.~\ref{fig_alphaterm}, we have
\ba
&&
(\text{Right figure})=2\ri\int{\rm d}\tau\,\ri\delta^{(4)}(0)\frac{(\lambda_1^{(3)}(\tau))^2}{4(\lambda_0^{(2)}(\tau))^2}G(k_1;\tau,\tau_1)G(k_2;\tau,\tau_2)
\notag\\
&&
\hspace{2.19cm}
=-\frac{1}{2}\delta^{(4)}(0)\int{\rm d}\tau \frac{(\lambda_1^{(3)}(\tau))^2}{(\lambda_0^{(2)}(\tau))^2}G(k_1;\tau,\tau_1)G(k_2;\tau,\tau_2)\,.
\ea
As a result, the singular term due to the two time-derivatives acting on the time-ordered propagator is canceled out by the 
contribution from the $\alpha$ term.
Thus, the unphysical divergences that differ from the UV divergence due to high-energy modes can be simply canceled by the divergent pieces appearing in the effective Lagrangian. In practice, keeping the interactions proportional to $\delta^{(4)}(0)$ is cumbersome. Therefore, we simply omit the unphysical divergent contributions both in the effective Lagrangian as well as in the computation of correlation functions on the basis of the above observation.

\section{Boundary term interactions 
in the in-in path-integral formalism}
\label{Bterminpathintegral}

This Appendix will show that the temporal total derivative terms do not contribute to the correlation functions within the path-integral formalism under an appropriate prescription.
Instead of proving it in general, we present some explicit computations 
relevant to the ones in the main text including the diagrams 
with an internal line and a loop.
The prescription is based on our previous work~\cite{Kawaguchi:2024lsw}, 
where only the tree-level contribution without internal lines 
has been discussed.
Here, we extend it to the case with internal lines and a loop. 

\subsection{Three-point correlation function}

\begin{figure}[h]
\centering
\includegraphics[width=15cm]{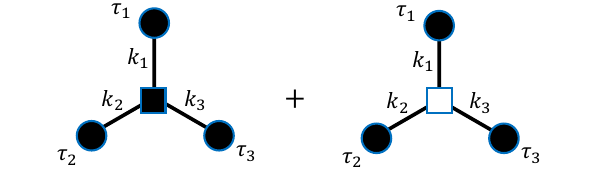}
\caption{\label{fig_3point_boundary}
Feynman diagrams for the three-point correlation function from 
boundary term interactions.
The square vertices denote contributions from boundary terms.
Black and white represent the $+$ and $-$ types, respectively.
}
\end{figure}

Let us consider a general form of ``boundary'' interactions,
\be
{\cal{L}}_{\partial}=\frac{\rm d}{{\rm d}\tau}\left(\lambda^{(\partial)}{\cal{D}}_1^{(\partial)}\zeta{\cal{D}}_2^{(\partial)}
\zeta{\cal{D}}_3^{(\partial)}\zeta\right)\,,
\label{boundary_cubic_lagrangian}
\ee
where ${\cal{D}}_i^{(\partial)}$'s represent either a time-derivative 
operator or an identity operator,\footnote{This means that $\zeta$ without time derivatives may appear.} and $\lambda^{(\partial)}$ is (generally) 
a time-dependent coupling.
We calculate the first-order correction from the boundary interaction to a three-point correlation function $\ev{\zeta_1\zeta_2\zeta_3}$ with possibly different time arguments $\tau_i$ ($i=1,2,3$), which is represented in Fig.~\ref{fig_3point_boundary}~\cite{Kawaguchi:2024lsw}. 
Here, we take the prescription $\tau_f>\tau_i$, namely, 
the temporal boundary is at the future of operators we are concerned with.
The black and white squares denote the $+$ and $-$ types of interaction~\eqref{boundary_cubic_lagrangian}, respectively.
From these figures, we have
\ba
&&
(\text{Left figure})=\ri\int^{\tau_f}_{-\infty}{\rm d}\tau\frac{\rm d}{{\rm d}\tau}\left(\lambda^{(\partial)}(\tau){\cal{D}}_1^{(\partial)}G_{++}(k_1;\tau,\tau_1){\cal{D}}_2^{(\partial)}G_{++}(k_2;\tau,\tau_2){\cal{D}}_3^{(\partial)}G_{++}(k_3;\tau,\tau_3)\right)+(\text{5 perms})
\notag\\
&&
\hspace{1.95cm}
=\ri\lambda^{(\partial)}(\tau_f){\cal{D}}_1^{(\partial)}G_{++}(k_1;\tau_f,\tau_1){\cal{D}}_2^{(\partial)}G_{++}(k_2;\tau_f,\tau_2){\cal{D}}_3^{(\partial)}G_{++}(k_3;\tau_f,\tau_3)+(\text{5 perms})\,,
\notag\\
&&
(\text{Right figure})=-\ri\int^{\tau_f}_{-\infty}{\rm d}\tau\frac{\rm d}{{\rm d}\tau}\left(\lambda^{(\partial)}(\tau){\cal{D}}_1^{(\partial)}G_{-+}(k_1;\tau,\tau_1){\cal{D}}_2^{(\partial)}G_{-+}(k_2;\tau,\tau_2){\cal{D}}_3^{(\partial)}G_{-+}(k_3;\tau,\tau_3)\right)+(\text{5 perms})
\notag\\
&&
\hspace{2.2cm}
=-\ri\lambda^{(\partial)}(\tau_f){\cal{D}}_1^{(\partial)}G_{-+}(k_1;\tau_f,\tau_1){\cal{D}}_2^{(\partial)}G_{-+}(k_2;\tau_f,\tau_2){\cal{D}}_3^{(\partial)}G_{-+}(k_3;\tau_f,\tau_3)+(\text{5 perms})\,.
\ea
Notice that, so long as the Lagrangian has at most second-order time derivatives, the operators ${\cal{D}}_i^{\partial}$ include at most 
a single time derivative as there is an overall time derivative. Therefore,  ${\cal{D}}_i^{\partial} G_{++}(k;\tau_f,\tau_i)={\cal{D}}_i^{\partial} G_{-+}(k;\tau_f,\tau_i)$ is satisfied by setting the final slice time $\tau_f$ 
to a time later than all of $\tau_i$'s ($i=1,2,3$).
The contributions from the left and right figures therefore cancel each other thanks to our prescription that the final time is taken to be the 
future of any operator insertions we consider.\footnote{In the case 
where the external operators of which we take the expectation value have time derivatives, the $\delta$-function contribution may emerge when the time derivatives act on the (anti-) time-ordered propagator. However, the 
$\delta$-function is of the form $\delta(\tau_f-\tau_i)$, 
which vanishes by our prescription since $\forall \tau_i<\tau_f$ and $\tau_i$ being the time argument of the external field. 
Thus, the cancellation is true even in such a case.} 

%

\subsection{Four-point correlation function}

\begin{figure}[h]
\centering
\includegraphics[width=15cm]{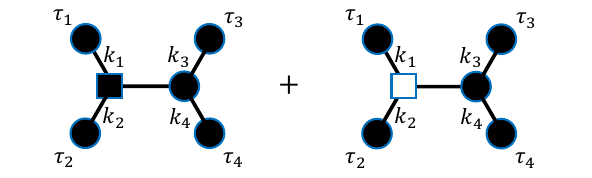}
\\
\includegraphics[width=15cm]{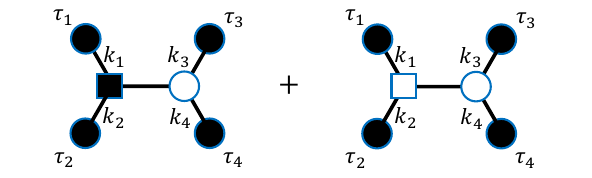}
\caption{\label{fig_4point_boundary}
Feynman diagrams for the four-point correlation function 
arising from the interactions of one boundary term and one bulk term.
}
\end{figure}

Next, let us consider a bit non-trivial case, the second-order 
correction from the cubic interaction to a four-point correlation function, which contains a single internal line. 
We consider corrections to a four-point correlation function from two cubic vertices, and one of the vertices to be the boundary interaction~\eqref{boundary_cubic_lagrangian}, 
while the other to be a bulk interaction.
Notice that, by distributing the total time derivative 
in the boundary interaction, we may regard the boundary term as 
a set of bulk terms, and therefore, the following discussions 
apply to the case where two vertices are the boundary interaction. 
Here, we formally assume the bulk interaction
\be
{\cal{L}}_{b}=\lambda^{(b)}{\cal{D}}_1^{(b)}\zeta{\cal{D}}_2^{(b)}\zeta{\cal{D}}_3^{(b)}\zeta\,,
\label{bulk_cubic_lagrangian}
\ee
where $\lambda^{(b)}$ is a coupling constant and $\mathcal{D}_i^{(b)}$'s 
are the time derivative operators (or identity) as is the case of the boundary term. In this case, the four diagrams differ in the choice of either $+$ or $-$ vertex for each interaction as shown in Fig.~\ref{fig_4point_boundary}. Looking at the two diagrams in the upper part of Fig.~\ref{fig_4point_boundary} while focusing on the cubic vertices in Fig.~\ref{fig_3point_boundary} as sub-diagrams, we notice that the sum of the two cancels with each other as the case of the previous one, and similarly for the lower two diagrams.
However, there is a subtlety: When more than two derivatives act on the internal line of the type either $(++)$ or $(--)$, 
a $\delta$-function from the (anti-)time-ordered propagator appears. 
Note that both the $(+-)$ and $(-+)$ type propagators 
do not give rise to the $\delta$-function.
Therefore, only one diagram is non-vanishing among the upper and lower diagrams, respectively. 
Suppose that $\partial_{\tau\tau'}$ acts on 
internal propagators. Then, from the upper one, we find
\ba
&&
(\text{Upper figures})\supset \ri^2\int^{\tau_f}_{-\infty}{\rm d}\tau\cdots\partial_{\tau\tau'}G_{++}(|\bm{k}_1+\bm{k}_2|;\tau,\tau')|_{\tau'=\tau_f}\cdots
\notag\\
&&
\hspace{2.45cm}
\supset -\int^{\tau_f}_{-\infty}{\rm d}\tau\cdots\frac{\ri}{2a^2\epsilon\Mpl^2}\delta(\tau-\tau_f)\cdots\,,
\ea
where ellipses denote the other components in the integrand.
The omitted terms are canceled by the above argument, 
but the $\delta$-function term remains, namely, 
the $\delta$-function contribution cannot be canceled among 
the sum of the upper diagrams.
Notice however that a similar thing happens to 
the lower diagrams and we are left with
\ba
&&
(\text{Lower figures})\supset (-\ri)^2\int^{\tau_f}_{-\infty}{\rm d}\tau\cdots\partial_{\tau\tau'}G_{--}(|\bm{k}_1+\bm{k}_2|;\tau,\tau')|_{\tau'=\tau_f}\cdots
\notag\\
&&
\hspace{2.4cm}
\supset -\int^{\tau_f}_{-\infty}{\rm d}\tau\cdots\frac{-\ri}{2a^2\epsilon\Mpl^2}\delta(\tau-\tau_f)\cdots\,.
\ea
Thus, we find that the remaining terms cancel between 
the contributions from the upper and lower diagrams 
due to the opposite sign, which originates from the difference 
of time-ordered and anti-time-ordered propagators. 
Although we have checked only the simplest case,
it is straightforward to extend it to the case with 
additional interactions by replacing the external point 
with the interaction vertices.

%

\subsection{Comment on the one-loop case}

\begin{figure}[h]
\centering
\includegraphics[width=15cm]{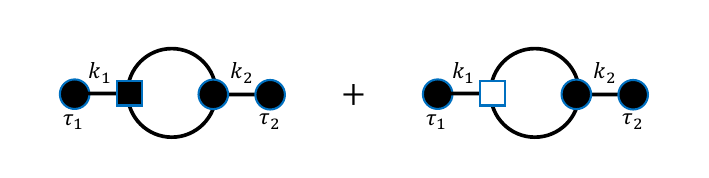}
\\
\includegraphics[width=15cm]{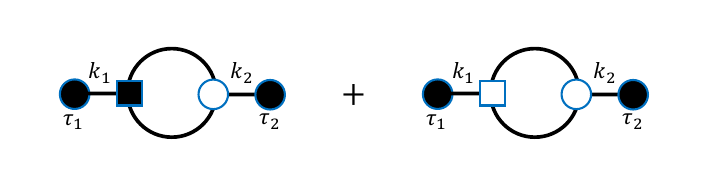}
\caption{\label{fig_oneloop_boundary}
Feynman diagrams for one-loop correction to the power spectrum 
from the interactions of one boundary term and one bulk term.
}
\end{figure}

We finally comment on the case with loops. 
We consider one-loop diagrams with two cubic vertices and can take one of 
the vertices to be the boundary term and the other to 
be the bulk one without loss of generality, 
as we explained previously.
Once again, we have four types of diagrams as shown in Fig.~\ref{fig_oneloop_boundary}. 
In most of the terms, the cancellation mechanism works 
in the same way as the tree-level cases. 
Only the nontrivial case we see in the one-loop diagrams 
is that multiple time derivatives act both of the two 
internal propagators, which results in two 
$\delta$-functions with the same argument.
The $\delta$-function with the same argument leads to an 
unphysical divergent factor $\delta^{(4)}(0)$. 
Notice however that such an unphysical contribution should be 
eliminated by the $\alpha$-terms discussed in the previous section.
As we emphasized, the total derivative terms can become a set of bulk terms. This is so since the total derivative terms appear to reorganize derivatives by performing integration by parts. Therefore, according to the general discussions about the derivative interactions in Appendix~\ref{mostgeneralpathintegral}, there should be induced 
counterterms ($\alpha$-terms) that precisely cancel the 
unphysical divergent terms.
Thus, the problematic divergent term that is not canceled 
by the reasons mentioned above can be properly removed by incorporating the $\alpha$-terms. Hence there are no contributions from the boundary interactions, which justifies that we deal with only the bulk and EOM terms in the main text. Although we have discussed just a one-loop case, we expect that a similar argument should hold due to the causality: 
If the boundary terms appear only in the asymptotic future, 
they cannot contribute to changing the physical quantities 
at any finite time.

\section{Proof of the identity~\eqref{one-loop-cut-Fourier}}
\label{proofid}

Here, we will show some details for the derivation of 
the equality in Eq.~\eqref{one-loop-cut-Fourier}. 
Recall that the first equality of 
Eq.~\eqref{one-loop-cut-Fourier} is 
\ba
&&
P_\zeta^{(\text{1-loop})}(\tau_*,k_1) 
\notag\\
&&
=\ri\int^{\tau_f}_{-\infty}{\rm d}\tau\Biggl[\sum_{A=1}^{n}\lambda_3^{(A)}\sum_{(i_1,i_2,i_3)}\Biggr({\cal{D}}_{i_1}^{(A)}G_{++}(k_1;\tau,\tau_*)\int\frac{{\rm d}^3 k}{(2\pi)^3}\llangle
\tilde{\zeta}^+(\tau_*,\bm{k}_1){\cal{D}}_{i_2}^{(A)}\tilde{\zeta}^+(\tau,\bm{k}){\cal{D}}_{i_3}^{(A)}\tilde{\zeta}^+(\tau,-\bm{k}_1-\bm{k})}\rrangle_{{\rm tree}\notag\\
&&
\hspace{2.0cm}
-{\cal{D}}_{i_1}^{(A)}G_{-+}(k_1;\tau,\tau_*)\int\frac{{\rm d}^3 k}{(2\pi)^3}\llangle {\cal{D}}_{i_2}^{(A)}\tilde{\zeta}^-(\tau,\bm{k}){\cal{D}}_{i_3}^{(A)}\tilde{\zeta}^-(\tau,-\bm{k}_1-\bm{k})\tilde{\zeta}^+(\tau_*,\bm{k}_1)\rrangle_{{\rm tree}}\Biggr)\Biggr]\,.
\ea

Notice that there is a single vertex to evaluate 
$\llangle \cdots \rrangle$, which is formally 
given by $\int_{-\infty}^{\tau_f} \rd \tau_1 \sum_{A=1}^n\lambda_{3}^{(A)}\mathcal{L}^{(A)}_3(\tau_1)$. 
From now on, we neglect the derivative operators, 
momentum integration, and couplings for notational simplicity, 
and focus on the propagators and time integration. 
Then, we obtain 
\ba
\hspace{-0.7cm}
& &P_\zeta^{(\text{1-loop})}(\tau_*,k_1)\nonumber\\
\hspace{-0.7cm}
& &\simeq- \int^{\tau_f}_{-\infty} \rd\tau 
\int^{\tau_f}_{-\infty} \rd \tau_1 \nonumber\\
\hspace{-0.7cm}
& &\times 
\Biggl[G_{++}(k_1;\tau,\tau_*)\Bigl\{G_{++}(k_1;\tau_*,\tau_1)G_{++}(k,\tau,\tau_1)G_{++}(\tilde{k};\tau,\tau_1)-G_{+-}(k_1;\tau_*,\tau_1)G_{+-}(k,\tau,\tau_1)G_{+-}(\tilde{k};\tau,\tau_1)\Bigr\}\nonumber\\
\hspace{-0.7cm}
& &~~~-G_{-+}(k_1;\tau,\tau_*)\Bigl\{G_{++}(k_1;\tau_*,\tau_1)G_{-+}(k,\tau,\tau_1)G_{-+}(\tilde{k};\tau,\tau_1)-G_{+-}(k_1;\tau_*,\tau_1)G_{--}(k,\tau,\tau_1)G_{--}(\tilde{k};\tau,\tau_1)\Bigr\}\Biggr]\,,
\label{Pzetalo}
\ea
where $\tilde{k}=|\bm k_1-\bm k|$. 
To simplify this equation further, 
we decompose the time integration as 
\begin{align}
\int_{-\infty}^{\tau_f} \rd \tau 
\int_{-\infty}^{\tau_f} \rd\tau_1=
\underbrace{\int_{-\infty}^{\tau_*} 
\rd\tau\int_{-\infty}^{\tau_*} \rd \tau_1}_{\text{(i)}}
+\underbrace{\int_{-
\infty}^{\tau_*}\rd\tau\int_{\tau_*}^{\tau_f} 
\rd\tau_1}_{\text{(ii)}}+\underbrace{\int_{\tau_*}^{\tau_f}
\rd \tau\int_{-\infty}^{\tau}\rd \tau_1}_{\text{(iii)}}+
\underbrace{\int_{\tau_*}^{\tau_f}\rd\tau\int_{\tau}^{\tau_f}
\rd\tau_1}_{\text{(iv)}}\,.
\end{align}
We will show that the integrals (ii), (iii), and (iv) vanish, 
as expected from causality. 
To see this, notice the following. 
For (ii), since $\tau_1>\tau_*>\tau$, 
we have $G_{++}(k_1;\tau_*,\tau_1)=G_{+-}(k_1;\tau_*,\tau_1)$ and $G_{++}(k(\tilde{k}),\tau,\tau_1)=G_{+-}(k(\tilde{k}),\tau,\tau_1)$. 
Then, the terms on the first line of the curly bracket in 
Eq.~(\ref{Pzetalo}) cancel each other. 
For a similar reason, the terms on the second line also cancel out. 
For (iii), we have $\tau>\tau_*$ and $G_{++}(k_2;\tau,\tau_*)=G_{-+}(k_2;\tau,\tau_*)$, and since $\tau>\tau_1$, there is also 
the relation $G_{++}(k(\tilde{k}),\tau,\tau_1)=G_{-+}(k(\tilde{k}),\tau,\tau_1)$. 
Therefore, the terms on the first line are exactly canceled 
by the corresponding terms on the second line. 
For (iv), it follows that $\tau_1>\tau>\tau_*$, 
and then we have $G_{++}(k;\tau_*,\tau_1)=G_{+-}(k;\tau_*,\tau_1)$ 
and $G_{++}(k(\tilde{k});\tau,\tau_1)=G_{+-}(k(\tilde{k});\tau,\tau_1)$. 
Hence the terms on the first line cancel out and a similar argument 
leads to the cancellation of terms on the second line. 
Thus we are left with (i) only, 
so that we reached the expression~\eqref{one-loop-cut-Fourier}.

However, the above manipulation has a subtlety. 
Although we have omitted the derivative operators from interactions, the second-order time derivative operator acting on the propagators may lead to the $\delta$-function with respect to time. 
However possible $\delta$-functions are the following three types, $\delta(\tau-\tau_*)$ from $G_{++}(k_2;\tau,\tau_*)$, $\delta(\tau_*-\tau_1)$ from $G_{++(--)}(k_1;\tau_*,\tau_1)$, and $\delta(\tau-\tau_1)$ from $G_{++(--)}(k(\tilde{k});\tau,\tau_1)$. 
For either choice, we notice that the integral over 
$\tau\in[\tau_*,\tau_f]$ cancels out, and therefore, the upper limit of the $\tau$ integration can be taken as $\tau_*$. Note that the double 
$\delta$-function can appear with the same argument, which causes a singular factor $\delta(0)$. However, such an unphysical divergence should be canceled out by the mechanism discussed in Appendix~\ref{mostgeneralpathintegral}. So, in general, we may assume the $\tau$ integration to be $\tau_*$. Notice also that the time integration appears from the interaction vertices, and the two integrations should be symmetric. Therefore, we may also take the upper limit of $\tau_1$ integration to be $\tau_*$. This can be simply understood as the consequence of causality since we are concerned with the quantum corrections to the operator inserted at $\tau=\tau_*\,(<\tau_f)$ 
and future events should not affect their expectation values.

\section{Note on renormalization}\label{renormalization}

We give a few comments on the renormalization issue in the transient USR model. As we already argued, there are no UV divergences for the 1PI contributions from cubic vertices. This is expected from the general result~\cite{Pimentel:2012tw}, where the background dynamics 
is not specified. In general, a non-smooth background leads to 
unphysical behavior of physical quantities due to the excitation of 
short wavelength modes by discontinuity. 
Indeed, in our model, the non-smoothness causes the quantity $\mathcal{B}_k$ in Eq.~\eqref{2ndSRAB} decaying only some power in $k$, but on a smooth background, we expect the exponential 
decay of ${\cal B}_k$ instead. 
In our discussion, thanks to the symmetric property of $\zeta$, 
the unphysical behavior does not occur but in general, 
one should be careful when dealing with the non-smooth background.

Furthermore, if the background is replaced by a smooth one, 
namely, the unphysical behavior is properly removed, we expect that the renormalization of various quantities should be done in the same way 
as the usual slow-roll inflation even in the transient USR model. 
This is because the structure of the UV divergence is independent 
of the background field configuration. 
More concretely, we can examine formal heat kernel expansion (see e.g.~\cite{Vassilevich:2003xt}) or adiabatic regularization~\cite{Zeldovich:1971mw,Parker:1974qw} in computing the 
UV divergent part of expectation values of the operators. 
As long as the state under consideration is physically a reasonable one, or more specifically, the class of Hadamard states whose two-point functions have a singularity structure similar to that of 
the Minkowski vacuum state\footnote{See 
e.g., Ref.~\cite{Khavkine:2014mta} for a review of the Hadamard states.},
the renormalization can be done without specifying field profiles. 
Then, since the difference between a slow-roll model and the transient USR model comes from the time dependence of $a(t)$, the counterterm we need should be the same in both cases. Such a property originates from the fact that for Hadamard states the UV modes contributing to the divergence do not feel 
e.g., the background spacetime curvature, and therefore, the UV divergence can be formally written without specifying the background field configuration. In the literature, one-loop corrections to the power spectrum from the leading-order cubic interaction has explicit UV-cutoff dependence, which requires renormalization and the finite part is dealt with by implicitly or explicitly assuming the renormalization of the corrections~\cite{Kristiano:2022maq,Choudhury:2023vuj,Firouzjahi:2023aum}. 
We expect that it is an artifact of the technical complications. 
If there were, counterterms would be 
necessary, which implies 
the time-dependence to the super-Hubble modes~\cite{Pimentel:2012tw}. 
If such a divergence were to appear in the transient USR model, 
this implies that we would have slow-roll models breaking 
the constancy of $\zeta$.
As we have shown, this is not the case. An exception is the tadpole term, but it can be removed by an appropriate renormalization condition~\cite{Pimentel:2012tw}.

Finally, we add a few details about dimensional regularization applied in the main text. 
Let us define the right-hand side 
of Eq.~\eqref{ratio-one-loop-tree} as 
$F(\delta)$, while taking 
$k_{\rm max}\to \infty$. 
$F(\delta)$ is ill-defined for some $\delta$, namely it is not a convergent integral. Nevertheless, there would be some parameter range where $F(\delta)$ is finite, and we define 
the original integral as the limit $F(\delta)|_{\delta\to 0}$. 
If the limit exists, we would find no singular terms. However, this is generally not the case 
and we find singular terms proportional to $1/\delta$ and so on. 
In the presence of such singular terms, 
we need to add counterterms to remove 
the singular contribution.

It is not obvious whether the divergent integral can be finite by appropriately chosen $\delta$ since one needs modification of the mode function~\cite{Senatore:2009cf}, and $\delta$-dependence from both mode functions and the integration may cancel with each other. 
If this is the case, the dimensional regularization cannot make the divergent integral convergent. Let us look into a bit more details. As mentioned above, dimensional regularization requires us to change the mode function for consistency. Indeed, the scalar curvature perturbation $\zeta$ in $p+1$ ($p=3+\delta$) spacetime dimensions is expressed as
\begin{align}
\zeta(\tau, \bm x)=\int \frac{{\rm d}^p k}{(2\pi)^p}\left[\hat{a}_{\bm k}\zeta^{(p)}_k(\tau)e^{\ri \bm k \cdot \bm x}
+{\rm h.c.}\right]\,,
\end{align}
where
\begin{align}
\zeta^{(p)}_k(\tau)=\frac{\sqrt{-\pi\tau}}{2\sqrt{2\epsilon}\,a^{(p-1)/2}(\tau)
\mu^{(p-3)/2}}e^{\ri\frac{(2\nu_p+1)\pi}{4}}H_{\nu_p}^{(1)}(-k\tau)\,,
\end{align}
and $\nu_p \simeq (p+\eta)/2$, where we 
have neglected contributions to $\nu_p$ having first or third slow-roll parameters. 
As we are concerned with the convergence 
of the momentum integration particularly from the contribution $k\to\infty$, the asymptotic behavior of the mode function becomes important, 
which reads
\begin{align}
\zeta^{(p)}_k(\tau)=\frac{\sqrt{-\pi\tau}}{2\sqrt{2\epsilon}\,a^{(p-1)/2}(\tau)
\mu^{(p-3)/2}}e^{\ri\frac{(2\nu_p+1)\pi}{4}}
\left(-\frac{2}{\pi k\tau}\right)^{\frac12}e^{\ri(- k\tau-\frac12\pi\nu_p-\frac14\pi)}\left[1+\mathcal{O}((k\tau)^{-1})\right]\,.
\end{align}
The modification of spatial dimension $\delta$ does not affect the momentum dependence for such high momentum modes. Note that this is not 
the case for small $k$ modes since the asymptotic behavior changes for $-k\tau \ll 1$. 
Now, the integrand of 
Eq.~\eqref{ratio-one-loop-tree} with respect to momentum integration contains a two-point function of $\zeta$ with some derivative operators, which may increase the power of momentum. Schematically, the momentum dependence of the integrand becomes $k^{p-2+n_d}$, where $n_d$ corresponds to the number of spatial derivatives or the time derivative that acts to increase the momentum power. So, if $\delta<-2-n_d^{\rm max}$, the integral becomes finite. Thus, we would define $F(\delta)$ in such a region, and consider analytic continuation of $F(\delta)$ to 
$\delta \to 0$. As explained in the main 
text, the limit does not give rise to singular terms, and $F(0)$ turns out to be finite. 
Thus, we have confirmed that using dimensional regularization leads to a finite result.\footnote{One may still wonder how the apparently divergent quantity could become finite without singular contribution. A similar example is the zeta function regularization: 
The sum of positive integers is $\zeta(-1)=-1/12$. There are no singular contributions, but the original sum is divergent. What happens here? 
In general, we cannot define divergent quantity and such a quantity can be defined as a limit of finite quantity. Crudely speaking, each regularization contains (implicit) counterterms that remove divergent quantities and allow us to give a finite value to the originally divergent quantity. The implicit counterterms depend on the choice of regularization, and therefore the regularized quantity depends on the regularization, but such unphysical differences can be removed by (explicit) counterterms that we will add to the Lagrangian. As a result, the physical value only depends on the renormalization condition we would impose.} 
Finally, we note that, if one applies the cut-off regularization, the time integration has to be carefully modified as discussed 
in Ref.~\cite{Senatore:2009cf}. 
We will not discuss the cut-off regularization here.

\section{Explicit computations of the master formula}
\label{explicitform}

We show the explicit form of one-loop corrections 
for the reduced Lagrangian (\ref{cubicL}). 
Using the formula~\eqref{premaster}, we find
\begin{align}
{P}_\zeta^{(\text{1-loop})}(\tau_*,k_1)
\simeq \frac{\ri}{2} \Mpl^2\int^{\tau_*}_{\tau_0}{\rm d}\tau\frac{{\rm d}^3 k}{(2\pi)^3}\left[\frac{a^2(\tau)\epsilon(\tau)\eta'(\tau)}{2}F_1
+\frac{\eta(\tau)}{2}F_2+\frac{2}{a(\tau)H}F_3\right]\,,
\end{align}
where $F_{1,2,3}$ are given by
\begin{align}
F_1=&+2G_{++}(k_1;\tau,\tau_*)\llangle\tilde{\zeta}^+(\tau_*,\bm k_1)\tilde{\zeta}^+(\tau,\bm k)
{\tilde\zeta^+}{}'(\tau,-\bm k-\bm k_1)\rrangle+2G_{++}(k_1;\tau,\tau_*)\llangle\tilde{\zeta}^+(\tau_*,\bm k_1)
{\tilde\zeta^+}{}'(\tau,\bm k)\tilde{\zeta}^+(\tau,-\bm k-\bm k_1)\rrangle\nonumber\\
&+2\partial_\tau G_{++}(k_1;\tau,\tau_*)\llangle\tilde{\zeta}^+(\tau_*,\bm k_1)\tilde{\zeta}^+(\tau,\bm k)\tilde{\zeta}^+(\tau,-\bm k-\bm k_1)\rrangle
-2G_{-+}(k_1;\tau,\tau_*)\llangle\tilde{\zeta}^+(\tau_*,\bm k_1)
\tilde{\zeta}^-(\tau,\bm k){\tilde\zeta^-}{}'(\tau,-\bm k-\bm k_1)\rrangle\nonumber\\
&-2G_{-+}(k_1;\tau,\tau_*)\llangle\tilde{\zeta}^+(\tau_*,\bm k_1)
{\tilde\zeta^-}{}'(\tau,\bm k)\tilde{\zeta}^-(\tau,-\bm k-\bm k_1)\rrangle
-2\partial_\tau G_{-+} (k_1;\tau,\tau_*)\llangle\tilde{\zeta}^+(\tau_*,\bm k_1)
\tilde{\zeta}^-(\tau,\bm k)\tilde{\zeta}^-(\tau,-\bm k-\bm k_1)\rrangle,\\
F_2=&+2G_{++}(k_1;\tau,\tau_*)\llangle\tilde{\zeta}^+(\tau_*,\bm k_1)\tilde{\zeta}^+(\tau,\bm k)D_2\tilde{\zeta}^+(\tau,-\bm k-\bm k_1)\rrangle+2G_{++}(k_1;\tau,\tau_*)\llangle\tilde{\zeta}^+(\tau_*,\bm k_1)D_2\tilde{\zeta}^+(\tau,\bm k)\tilde{\zeta}^+(\tau,-\bm k-\bm k_1)\rrangle\nonumber\\
&-\frac{\ri\delta(\tau-\tau_*)}{\Mpl^2}\llangle\tilde{\zeta}^+(\tau_*,\bm k_1)\tilde{\zeta}^+(\tau,\bm k)\tilde{\zeta}^+(\tau,-\bm k-\bm k_1)\rrangle-2G_{-+}(k_1;\tau,\tau_*)\llangle\tilde{\zeta}^+(\tau_*,\bm k_1)\tilde{\zeta}^-(\tau,\bm k)D_2\tilde{\zeta}^-(\tau,-\bm k-\bm k_1)\rrangle\nonumber\\
&-2G_{-+}(k_1;\tau,\tau_*)\llangle\tilde{\zeta}^+(\tau_*,\bm k_1)D_2\tilde{\zeta}^-(\tau,\bm k)\tilde{\zeta}^-(\tau,-\bm k-\bm k_1)\rrangle,\\
F_3=&+G_{++}(k_1;\tau,\tau_*)\llangle\tilde{\zeta}^+(\tau_*,\bm k_1)
{\tilde\zeta^+}{}' (\tau,\bm k)D_2\tilde{\zeta}^+(\tau,-\bm k-\bm k_1)\rrangle+G_{++}(k_1;\tau,\tau_*)\llangle\tilde{\zeta}^+(\tau_*,\bm k_1)D_2\tilde{\zeta}^+(\tau,\bm k)
{\tilde\zeta^+}{}'(\tau,-\bm k-\bm k_1)\rrangle\nonumber\\
&+\partial_\tau G_{++}(k_1;\tau,\tau_*)\llangle\tilde{\zeta}^+
(\tau_*,\bm k_1)D_2\tilde{\zeta}^+(\tau,\bm k)\tilde{\zeta}^+(\tau,-\bm k-\bm k_1)\rrangle+\partial_\tau G_{++}(k_1;\tau,\tau_*)\llangle\tilde{\zeta}^+(\tau_*,\bm k_1)\tilde{\zeta}^+(\tau,\bm k)D_2\tilde{\zeta}^+(\tau,-\bm k-\bm k_1)\rrangle\nonumber\\
&-\frac{\ri\delta(\tau-\tau_*)}{2\Mpl^2}\llangle\tilde{\zeta}^+(\tau_*,\bm k_1)
{\tilde\zeta^+}{}' (\tau,\bm k)\tilde{\zeta}^+(\tau,-\bm k-\bm k_1)\rrangle-\frac{\ri\delta(\tau-\tau_*)}{2\Mpl^2}\llangle\tilde{\zeta}^+(\tau_*,\bm k_1)\tilde{\zeta}^+(\tau,\bm k)
{\tilde\zeta^+}{}' (\tau,-\bm k-\bm k_1)\rrangle\nonumber\\
&-G_{-+}(k_1;\tau,\tau_*)\llangle\tilde{\zeta}^+(\tau_*,\bm k_1)
{\tilde\zeta^-}{}' (\tau,\bm k)D_2\tilde{\zeta}^-(\tau,-\bm k-\bm k_1)\rrangle-G_{-+}(k_1;\tau,\tau_*)\llangle\tilde{\zeta}^+(\tau_*,\bm k_1)D_2\tilde{\zeta}^-(\tau,\bm k){\tilde\zeta^-}{}'(\tau,-\bm k-\bm k_1)\rrangle\nonumber\\
&-\partial_\tau G_{-+}(k_1;\tau,\tau_*)\llangle\tilde{\zeta}^+(\tau_*,\bm k_1)D_2\tilde{\zeta}^-(\tau,\bm k)\tilde{\zeta}^-(\tau,-\bm k-\bm k_1)\rrangle
-\partial_\tau G_{-+}(k_1;\tau,\tau_*)\llangle\tilde{\zeta}^+(\tau_*,\bm k_1)
\tilde{\zeta}^-(\tau,\bm k)D_2\tilde{\zeta}^-(\tau,-\bm k-\bm k_1)\rrangle.
\end{align}
Note that a prime represents the partial derivative 
with respect to $\tau$.
Following the results in Sec.~\ref{D2consistency}, 
we may neglect the three-point functions including the $D_2$ operator since they give terms proportional to $\delta^{(4)}(0)$ and should be canceled by the $\alpha$-terms following from the general 
discussion given in Appendix~\ref{mostgeneralpathintegral}.
Dropping the terms that contain three-point functions including $D_2$, 
the leading-order terms of $F_{1,2,3}$ yield
\begin{align}
F_1^{\rm LO}= &+2G_{++}(k_1;\tau,\tau_*)\llangle\tilde{\zeta}^+(\tau_*,\bm k_1)
\{\tilde{\zeta}^+(\tau,\bm k)\tilde{\zeta}^+(\tau,-\bm k-\bm k_1)\}'\rrangle
+2\partial_\tau G_{++}(k_1;\tau,\tau_*)\llangle\tilde{\zeta}^+
(\tau_*,\bm k_1)\tilde{\zeta}^+(\tau,\bm k)\tilde{\zeta}^+(\tau,-\bm k-\bm k_1)\rrangle\nonumber\\
&-2G_{-+}(k_1;\tau,\tau_*)\llangle\tilde{\zeta}^+(\tau_*,\bm k_1)
\{\tilde{\zeta}^-(\tau,\bm k)\tilde{\zeta}^-(\tau,-\bm k-\bm k_1)\}'\rrangle-2\partial_\tau G_{-+}(k_1;\tau,\tau_*)\llangle\tilde{\zeta}^+(\tau_*,\bm k_1)\tilde{\zeta}^-(\tau,\bm k)\tilde{\zeta}^-(\tau,-\bm k-\bm k_1)\rrangle\,,\\
\label{F1LO}
F_2^{\rm LO}=&-\frac{\ri\delta(\tau-\tau_*)}{\Mpl^2}\llangle\tilde{\zeta}^+(\tau_*,\bm k_1)\tilde{\zeta}^+(\tau,\bm k)\tilde{\zeta}^+(\tau,-\bm k-\bm k_1)\rrangle\,,\\
F_3^{\rm LO}=&-\frac{\ri\delta(\tau-\tau_*)}{2\Mpl^2}\llangle\tilde{\zeta}^+(\tau_*,\bm k_1){\tilde\zeta^+}{}'(\tau,\bm k)\tilde{\zeta}^+(\tau,-\bm k-\bm k_1)\rrangle
-\frac{\ri\delta(\tau-\tau_*)}{2\Mpl^2}\llangle\tilde{\zeta}^+(\tau_*,\bm k_1)\tilde{\zeta}^+(\tau,\bm k){\tilde\zeta^+}{}'(\tau,-\bm k-\bm k_1)\rrangle\,.
\end{align}
By using these results, we obtain
\begin{align}
P_\zeta^{(\text{1-loop})}(\tau_*,k_1)
\simeq&\frac{\ri}{2}\Mpl^2\int^{\tau_*}_{\tau_0}{\rm d}\tau\frac{{\rm d}^3 k}{(2\pi)^3}\left[\frac{a^2(\tau)\epsilon(\tau)\eta'(\tau)}{2}
F_1^{\rm LO}\right]+\int
\frac{{\rm d}^3 k}{(2\pi)^3}
\frac{\eta(\tau_*)}{4}
\langle\tilde{\zeta}^+(\tau_*,\bm k_1)\tilde{\zeta}^+(\tau_*,\bm k)\tilde{\zeta}^+(\tau_*,-\bm k-\bm k_1)\rangle  \nonumber\\
&+\int\frac{{\rm d}^3 k}{(2\pi)^3}
\frac{1}{2a(\tau_*)H} \left[\llangle\tilde{\zeta}^+(\tau_*,\bm k_1){\tilde\zeta^+}{}'(\tau,\bm k)\tilde{\zeta}^+(\tau,-\bm k-\bm k_1)\rrangle+\llangle\tilde{\zeta}^+(\tau_*,\bm k_1)\tilde{\zeta}^+(\tau,\bm k){\tilde\zeta^+}{}'(\tau,-\bm k-\bm k_1) \rrangle \right]|_{\tau\to\tau_*}.
\label{oneloopexplicit}
\end{align}
Notice that the first term is the one included in Eq.~\eqref{bulkoneloop}. 
On the other hand, the rest of terms are evaluated at a future time $\tau_*\,(>\tau_e)$. By assumption, the slow-roll parameter $\eta$  becomes negligibly small in the secondary SR phase within our model, which implies that the second integral in Eq.~(\ref{oneloopexplicit}) is negligible. Furthermore, since $|\tau_*|\ll |\tau_e|$, the third integral in Eq.~(\ref{oneloopexplicit}), which is proportional to $(a(\tau_*)H)^{-1}\simeq -\tau_*$, should be also small. 
Therefore, only the first integral gives the leading-order contribution to Eq.~(\ref{oneloopexplicit}). Note that the momentum integration of these terms can be also evaluated 
as Eq.~\eqref{ratio-one-loop-tree} since the consistency relation holds. 
Thus, we have confirmed the validity of our discussion in Sec.~\ref{oneloopresult}. 

We also comment on a technical issue associated with the relation 
between the correlation functions of $+$ and $-$ fields. 
The origin of these fields is the presence of the time-evolution operator and its Hermitian conjugate, and formally, the difference between $\zeta^+(\tau,\bm x_1)$ and $\zeta^-(\tau,\bm x_1)$ would disappear if we consider correlation functions at the equal time. This argument is correct up to the possible $\delta$-functions associated with the canonical commutation relations when the external fields have time derivatives. As an illustration, we consider
\begin{align}
\lim_{\tau_1\to \tau_2}\langle0|\zeta^+(\tau_1,\bm x)
\zeta^+{}'(\tau_2,\bm y)|0\rangle
=&\lim_{\tau_1\to \tau_2}  \langle 0|
T\{\zeta(\tau_1,\bm x)\zeta'(\tau_2,\bm y) \} |0\rangle
=-\ri\lim_{\tau_1\to\tau_2}\partial_{\tau_2}\Delta_{++}(\tau_1,\bm x;\tau_2,\bm y)\nonumber\\
=&\frac12\int\frac{\rd^3 k}{(2\pi)^3}\partial_{\tau_2}|\zeta_k(\tau_2)|^2
e^{-\ri\bm k \cdot (\bm x-\bm y)}\,.
\label{ppex}
\end{align}
On the other hand, we have 
\be
\lim_{\tau_1\to \tau_2}\langle0|\zeta^+(\tau_1,\bm x)
\zeta^{-}{}'(\tau_2,\bm y)|0\rangle=-\ri 
\partial_{\tau_2}\Delta_{+-}(\tau_1,\bm x;\tau_2,\bm y)
=\int\frac{\rd^3k}{(2\pi)^3}\zeta_k'(\tau_2)
\bar\zeta_k(\tau_2)e^{-\ri\bm k \cdot (\bm x-\bm y)}\,,
\label{pmex}
\ee
which differs from the former. 
This is due to our prescription of the Heaviside 
step function, and not the difference between 
the path-integral and operator formalisms, 
since, even for the operator formalism, we would have
\begin{align}
\lim_{\tau_1\to \tau_2} \langle 0|T\{\zeta(\tau_1,\bm x)\zeta'(\tau_2,\bm y) \}|0\rangle=\frac12 \langle 0|\zeta(\tau_2,\bm x)\zeta'(\tau_2,\bm y)+\zeta'(\tau_2,\bm y)\zeta(\tau_2,\bm x)|0\rangle\,,
\end{align}
which reproduces the above result. 
Now, the difference between (\ref{ppex}) and 
(\ref{pmex}) is given by
\begin{align}
&\lim_{\tau_1\to \tau_2}\langle0|\zeta^+(\tau_1,\bm x)\zeta^{+}{}'(\tau_2,\bm y)|0\rangle-\lim_{\tau_1\to \tau_2}\langle0|\zeta^+(\tau_1,\bm x)\zeta^{-}{}'(\tau_2,\bm y)|0\rangle\nonumber\\
&=\frac12 \int\frac{\rd^3k}{(2\pi)^3}
\left[ \bar{\zeta}'_k(\tau_2)\zeta_k(\tau_2)-\zeta_k'(\tau_2)\bar\zeta_k(\tau_2)\right] 
e^{-\ri\bm k \cdot (\bm x-\bm y)}\nonumber\\
&=\frac12 \int\frac{\rd^3k}{(2\pi)^3}\frac{\ri}{2a^2\epsilon \Mpl^2}e^{-\ri\bm k \cdot (\bm x-\bm y)}=\frac{\ri}{4a^2\epsilon \Mpl^2}\delta^{(3)}(\bm x-\bm y)\,,
\end{align}
where we used the normalization condition \eqref{normalizationzeta}.
This is also expected from the operator formalism. 
Hence, the equal time limit of the (anti-)time-ordered products may 
lead to some subtlety depending on the definition of the Heaviside step function.
Notice however that e.g., in the evaluation of the connected 
three-point functions, we take the contraction of ``external fields'' 
with the fields in interaction terms, which form three propagators. 
Then, the equal time limit of the external fields causes no subtlety since 
the external fields are contracted only with the fields in interaction terms. 
In other words, for connected diagrams, the ordering 
ambiguity does not provide any contributions. Nevertheless, this argument 
has a possible loophole discussed below. 

The possible loophole of the above argument is the case that 
some operators have two (or more) time derivatives such as 
$D_2$ since such derivatives give rise to a time-dependent 
$\delta$-function, 
which enforces the equal time limit of some operators. 
To illustrate the subtlety and its resolution, we consider 
a simple interacting system
\begin{align}
\mathcal{L}_{\rm int}=\lambda\zeta^2D_2\zeta\,,
\end{align}
where $\lambda$ is a coupling. 
Now, let us evaluate the correlation function
\begin{align}
\langle0|\zeta^+(\tau,\bm x_1)
\zeta^+(\tau,\bm x_2)\zeta^+(\tau,\bm x_3)|0\rangle\,,
\end{align}
and
\begin{align}
\langle0|\zeta^+(\tau,\bm x_1)\zeta^-(\tau,\bm x_2)
\zeta^-(\tau,\bm x_3)|0\rangle\,,
\end{align}
at first order in $\lambda$. 
Notice that we have two interactions consisting of either 
$+$ or $-$ fields. Summing up the contributions from 
the $\pm$-vertices for the former, we find
\begin{align}
\ri\,2 \lambda \int \rd\tau_1 \rd^3\bm{y}(-\ri\Delta_{++}(\tau,\bm x_1;\tau_1,\bm y))(-\ri\Delta_{++}(\tau,\bm x_2;\tau_1,\bm y))\frac{-\ri}{2\Mpl^2}\delta(\tau_1-\tau)\delta^{(3)}(\bm x_3-\bm y)+(3\leftrightarrow 1)+(3\leftrightarrow2)\,,
\end{align}
where the first $\ri$ comes from the $+$ vertex, and we used 
$D_2(-\ri\Delta_{++}(\tau,\bm x;\tau_1,\bm y))
=-\ri(2\Mpl^2)^{-1}\delta(\tau_1-\tau)\delta^{(3)}(\bm x_3-\bm y)$.
The latter is evaluated as
\begin{align}
&\ri\, 2\lambda \int \rd\tau_1 \rd^3\bm{y}
(-\ri\Delta_{+-}(\tau,\bm x_2;\tau_1,\bm y))(-\ri\Delta_{+-}(\tau,\bm x_3;\tau_1,\bm y))\frac{-\ri}{2\Mpl^2}\delta(\tau_1-\tau)\delta^{(3)}(\bm x_1-\bm y)\nonumber\\
&-\ri\, 2\lambda \int \rd\tau_1 \rd^3\bm{y}(-\ri\Delta_{+-}(\tau,\bm x_1;\tau_1,\bm y))(-\ri\Delta_{--}(\tau,\bm x_3;\tau_1,\bm y))\frac{+\ri}{2\Mpl^2}\delta(\tau_1-\tau)\delta^{(3)}(\bm x_2-\bm y)+(2\leftrightarrow 3)\,,
\end{align}
where the first and second lines arise from the $+$ 
and $-$ vertices, respectively, and we used 
$D_2(-\ri\Delta_{--}(\tau,\bm x;\tau_1,\bm y))=+\ri(2\Mpl^2)^{-1}
\delta(\tau_1-\tau)\delta^{(3)}(\bm x_3-\bm y)$. 
After evaluating the $\delta$-function, we obtain the same result 
from both. In this sense, we may choose either of the signs for the external fields. Thus, even though the intermediate results differ from each other, we have confirmed that the equal time limit of $\langle \zeta^+\zeta^+\zeta^+\rangle$ and $\langle\zeta^+\zeta^-\zeta^-\rangle$ coincide in this case. As we will see below, this case is special since the external operators have no time derivatives.

Actual subtlety arises when the external operators have time derivatives. 
Let us consider 
\begin{align}
\langle0|\zeta^+(\tau,\bm x_1)
{\zeta^+}{}'(\tau,\bm x_2)\zeta^+(\tau,\bm x_3)|0\rangle\,,
\end{align}
and
\begin{align}
\langle0|\zeta^+(\tau,\bm x_1)
\zeta^-{}'(\tau,\bm x_2)\zeta^-(\tau,\bm x_3)|0\rangle\,.
\end{align}
Here, we need to be more careful how to define 
the coincident time limit 
of these operators. The appropriate way is to define them as
\begin{align}
 \langle0|\zeta^+(\tau,\bm x_1)
\zeta^+{}'(\tau,\bm x_2)\zeta^+(\tau,\bm x_3)|0 \rangle
\equiv
\langle0|\zeta^+(\tau_1,\bm x_1)
\zeta^+{}'(\tau_2,\bm x_2)\zeta^+(\tau_3,\bm x_3)|0 \rangle|_{\tau_{1,2,3}\to\tau}\,.
\end{align}
The former is evaluated as
\begin{align}
2\ri \lambda\int \rd \tau_4 \rd^3\bm y\Biggl[&(-\ri\Delta_{++}(\tau_1,\bm x_1;\tau_4,\bm y))(-\ri\partial_{\tau_2}\Delta_{++}(\tau_2,\bm x_2;\tau_4,\bm y))\frac{-\ri}{2\Mpl^2}\delta(\tau_4-\tau_3)\delta^{(3)}(\bm x_3-\bm y)\nonumber\\
&+(-\ri\Delta_{++}(\tau_3,\bm x_3;\tau_4,\bm y))(-\ri\partial_{\tau_2}\Delta_{++}(\tau_2,\bm x_2;\tau_4,\bm y))\frac{-\ri}{2\Mpl^2}\delta(\tau_4-\tau_1)\delta^{(3)}(\bm x_1-\bm y)\nonumber\\
&+(-\ri\Delta_{++}(\tau_1,\bm x_1;\tau_4,\bm y))(-\ri\Delta_{++}(\tau_3,\bm x_3;\tau_4,\bm y))\frac{-\ri}{2\Mpl^2}\partial_{\tau_2}\delta(\tau_4-\tau_2)\delta^{(3)}(\bm x_2-\bm y)\Biggr]_{\tau_{1,2,3}\to\tau}\nonumber\\
=\frac{\lambda}{\Mpl^2}\int\frac{\rd^3 k_1}{(2\pi)^3}& \int\frac{\rd^3 k_2}{(2\pi)^3}\Biggl[\frac12|\zeta_{k_1}(\tau)|^2\frac{\rd}{\rd \tau}|\zeta_{k_2}(\tau)|^2e^{\ri \bm k_1\cdot(\bm x_1-\bm x_3)+\ri \bm k_2\cdot(\bm x_2-\bm x_3)}+\frac12|\zeta_{k_1}(\tau)|^2\frac{\rd}{\rd\tau}|\zeta_{k_2}(\tau)|^2e^{\ri \bm k_1\cdot(\bm x_1-\bm x_3)+\ri \bm k_2\cdot(\bm x_2-\bm x_1)}\nonumber\\
&\hspace{1.4cm}+\frac12\frac{\rd}{\rd \tau}
\left(|\zeta_{k_1}(\tau)|^2|\zeta_{k_2}(\tau)|^2\right)e^{\ri \bm k_1\cdot(\bm x_1-\bm x_2)+\ri \bm k_2\cdot(\bm x_3-\bm x_2)}\Biggr]\,,
\end{align}
whereas the latter gives
\begin{align}
2\ri \lambda\int \rd \tau_4 \rd^3\bm y\Biggl[&-(-\ri\Delta_{+-}(\tau_1,\bm x_1;\tau_4,\bm y))(-\ri\partial_{\tau_2}\Delta_{--}(\tau_2,\bm x_2;\tau_4,\bm y))\frac{+\ri}{2\Mpl^2}\delta(\tau_4-\tau_3)\delta^{(3)}(\bm x_3-\bm y)\nonumber\\
&+(-\ri\Delta_{-+}(\tau_3,\bm x_3;\tau_4,\bm y))(-\ri\partial_{\tau_2}\Delta_{-+}(\tau_2,\bm x_2;\tau_4,\bm y))\frac{-\ri}{2\Mpl^2}\delta(\tau_4-\tau_1)\delta^{(3)}(\bm x_1-\bm y)\nonumber\\ 
&-(-\ri\Delta_{+-}(\tau_1,\bm x_1;\tau_4,\bm y))(-\ri\Delta_{--}(\tau_3,\bm x_3;\tau_4,\bm y))\frac{+\ri}{2\Mpl^2}\partial_{\tau_2}\delta(\tau_4-\tau_2)\delta^{(3)}(\bm x_2-\bm y)\Biggr]_{\tau_{1,2,3}\to\tau}\nonumber\\
=\frac{\lambda}{\Mpl^2}\int \frac{\rd^3k_1}{(2\pi)^3}&\frac{\rd^3k_2}{(2\pi)^3}\Biggl[\frac12|\zeta_{k_1}(\tau)|^2\frac{\rd}{\rd \tau}|\zeta_{k_2}(\tau)|^2e^{\ri \bm k_1\cdot(\bm x_1-\bm x_3)+\ri \bm k_2\cdot(\bm x_2-\bm x_3)}+|\zeta_{k_1}(\tau)|^2\zeta_{k_2}'(\tau)\bar\zeta_{k_2}(\tau)e^{\ri \bm k_1\cdot(\bm x_1-\bm x_3)+\ri \bm k_2\cdot (\bm x_2-\bm x_1)}\nonumber\\  
&\hspace{1cm}+\left\{|\zeta_{k_2}(\tau)|^2\zeta_{k_1}'(\tau)\bar\zeta_{k_1}(\tau)+\frac12|\zeta_{k_1}(\tau)|^2\frac{\rd}{\rd\tau}|\zeta_{k_2}(\tau)|^2\right\}e^{\ri\bm k_1\cdot(\bm x_1-\bm x_2)+\ri\bm k_2\cdot(\bm x_3-\bm x_2)}\Biggr]\,.
\end{align}
One can take the differences between them and find
\begin{align}
&\langle0|\zeta^+(\tau,\bm x_1)
\zeta^+{}'(\tau,\bm x_2)\zeta^+(\tau,\bm x_3)|0\rangle-  \langle0|\zeta^+(\tau,\bm x_1)
{\zeta^-}{}'(\tau,\bm x_2)\zeta^-(\tau,\bm x_3)|0\rangle\nonumber\\
&=\frac{\ri\lambda}{4a^2(\tau)\epsilon(\tau)\Mpl^4}\int \frac{\rd^3 k_1}{(2\pi)^3}|\zeta_{k_1}(\tau)|^2\left[e^{\ri\bm k_1\cdot(\bm x_1-\bm x_3)}\delta^{(3)}(\bm x_1-\bm x_2)
+e^{\ri\bm k_1\cdot(\bm x_2-\bm x_3)}\delta^{(3)}(\bm x_1-\bm x_2)\right]\,.
\end{align}
Once again, we find the contact contribution associated with the ordering ambiguity. 
Notice that like the case of the previous argument, as long as we are concerned with 
connected correlation functions at different space points, the ambiguity does not 
contribute to the correlation function since the spatial $\delta$-function trivially 
vanishes. Nevertheless, we have found the subtlety associated with the EOM operators: 
They may give contact contributions proportional to the spatial $\delta$-function,
which may give rise to ambiguities of correlation functions unless we consider the 
expectation values of operators at different points. Since we are concerned with 
loop contributions, one may wonder if the above-mentioned subtlety causes problems 
in our discussion. Fortunately, this is not the case as discussed below.

We find that despite possible subtlety, it turns out that our computations
do not suffer from the above ordering issue. To illustrate this point, 
we consider $F_1^{\rm LO}$ appearing in 
Eq.~\eqref{oneloopexplicit}, which have
\begin{align}
F_1^{\rm LO}\supset&+2\partial_\tau G_{++}(k_1;\tau,\tau_*)
\llangle\zeta^+(\tau_*,\bm k_1)\zeta^+(\tau,\bm k)\zeta^+
(\tau,-\bm k-\bm k_1)\rrangle\nonumber\\
&-2\partial_\tau G_{-+}(k_1;\tau,\tau_*)\llangle\zeta^+(\tau_*,\bm k_1)
\zeta^-(\tau,\bm k)\zeta^-(\tau,-\bm k-\bm k_1)\rrangle\,.
\end{align}
From the contraction with the third term in Eq.~\eqref{cubicL}, 
we have 
\begin{align}
\llangle\zeta^+(\tau_*,\bm k_1)\zeta^+(\tau,\bm k)\zeta^+(\tau,-\bm k-\bm k_1)\rrangle\supset \frac{4}{a(\tau)H}\partial_{\tau_1}\left[G_{++}(k_1;\tau_*,\tau_1)\left(G_{++}(k;\tau,\tau_1)+G_{++}(\tilde{k};\tau,\tau_1)\right)\right]_{\tau_1\to\tau}\,,
\end{align}
whereas
\begin{align}
\llangle\zeta^+(\tau_*,\bm k_1)\zeta^-(\tau,\bm k)\zeta^-(\tau,-\bm k-\bm k_1)\rrangle\supset \frac{4}{a(\tau)H}\partial_{\tau_1}\left[G_{+-}(k_1;\tau_*,\tau_1)\left(G_{--}(k;\tau,\tau_1)+G_{--}(\tilde{k};\tau,\tau_1)\right)\right]_{\tau_1\to\tau},
\end{align}
where we defined $|\bm k+\bm k_1|=\tilde{k}$.
Thus, the integrand contains
\begin{align}
F_1^{\rm LO}\supset\frac{8}{a(\tau)H}
\left[\{(\partial_\tau G_{++}(k_1,\tau,\tau_*))^2-(\partial_\tau 
G_{-+}(k_1,\tau,\tau_*))^2\} \{ G_{++}(k;\tau,\tau)+G_{++}(\tilde{k};\tau,\tau) \}
\right]\,,
\end{align}
where we used 
$G_{++}(k(\tilde{k});\tau,\tau)=G_{--}(k(\tilde{k});\tau,\tau)$. 
Strictly speaking, this quantity is not real at $\tau=\tau_*$ but in our case, it is just a point with measure zero. Except that point, 
$G_{++}(k_1,\tau,\tau_*)=\bar{G}_{-+}(k_1,\tau,\tau_*)$ holds for $\tau\in [\tau_0,\tau_*)$, which is the region of integration. Then, the integrand becomes pure imaginary, which yields a real correction to the power spectrum with the overall imaginary unit $\ri$. One can confirm similar results 
for other contributions. Therefore, there is no ordering issue within our model.

Although the ordering subtlety does not appear within our model, we give a few comments 
related to the solution to the above ordering issue in general. 
We would have several options: One is to use the Keldysh basis 
$\zeta_{c}=(\zeta_{+}+\zeta_{-})/2$ and $\zeta_q=(\zeta_{+}-\zeta_{-})$ 
(see e.g., Ref.~\cite{Berges:2015kfa}) as basic variables and to compute 
the expectation values of $\zeta_c$. 
Such a computation corresponds to taking the average over 
possible sign assignments on the fields. We may also employ the wave functional approach~\cite{Maldacena:2002vr,Harlow:2011ke}. 
These approaches would avoid the above-mentioned issue.

\bibliographystyle{mybibstyle}
\bibliography{bib}

\end{document}